\documentclass[12pt,a4paper]{article}
\pdfoutput=1
\usepackage{amssymb,amsmath}
\usepackage{graphicx}
\usepackage{subfig}
\usepackage{color}
\usepackage{slashed}
\usepackage{enumerate}
\usepackage{dsfont}
\usepackage{hyperref}
\usepackage{bm}
\usepackage{multirow}
\usepackage{pifont}
\usepackage{makecell}
\usepackage{gensymb}
\usepackage{lipsum}
\usepackage{cite}
\usepackage{setspace}
\definecolor{nicered}{rgb}{0.7,0.1,0.1}
\definecolor{nicegreen}{rgb}{0.1,0.5,0.1}
\hypersetup{colorlinks,citecolor= nicegreen,linkcolor= black%nicered
}
\allowdisplaybreaks
\begin{document}
\begin{titlepage}
% -------------------------------------------------
% Uncomment to add preprint numbers when necessary
% -------------------------------------------------
% \begin{flushright}

% \end{flushright}
% -------------------------------
% Author addresses abbreviations
% -------------------------------
\newcommand{\AddrLiege}{{\sl \small IFPA, Dep. AGO, Universite de
      Liege,\\
      \small \sl Bat B5, Sart Tilman B-4000 Liege 1, Belgium}}
\newcommand{\AddrValencia}{{\sl \small AHEP Group,
      Instituto de Fisica Corpuscular-C.S.I.C./Universitat de
      Valencia,\\
      \sl \small Edificio Institutos de Paterna, Apt 22085,
      E-46071 Valencia, Spain}}
\vspace*{1.5cm}
% ----------------
% Front page
% ----------------
\hfill{Preprint: IFIC/14-61}

\begin{center}
  \textbf{\large Systematic classification of two-loop realizations\\[3mm]% 
    of the Weinberg operator}\\[15mm]
  D. Aristizabal Sierra%
  \footnote{e-mail address: {\tt daristizabal@ulg.ac.be}},
   A. Degee%
  \footnote{e-mail address: {\tt audrey.degee@ulg.ac.be}}%
  \vspace{0.3cm}\\
  \AddrLiege.\vspace{0.4cm}
  \vspace{0.8cm}\\
  L. Dorame%
  \footnote{e-mail address: {\tt dorame@ific.uv.es}},
   M. Hirsch%
  \footnote{e-mail address: {\tt mahirsch@ific.uv.es}}%
  \vspace{0.3cm}\\
  \AddrValencia.\vspace{0.4cm}\\
\end{center}
\vspace*{0.5cm}
\begin{abstract}
  \onehalfspacing
  We systematically analyze the $d=5$ Weinberg operator at 2-loop
  order.  Using a diagrammatic approach, we identify two different
  interesting categories of neutrino mass models: (i) Genuine 2-loop
  models for which both, tree-level and 1-loop contributions, are
  guaranteed to be absent. And (ii) finite 2-loop diagrams, which
  correspond to the 1-loop generation of some particular vertex
  appearing in a given 1-loop neutrino mass model, thus being {\em
    effectively 2-loop.}  From the large list of all possible 2-loop
  diagrams, the vast majority are infinite corrections to lower order
  neutrino mass models and only a moderately small number of diagrams
  fall into these two interesting classes.  Moreover, all diagrams in
  class (i) are just variations of three basic diagrams, with examples
  discussed in the literature before.  Similarly, we also show that
  class (ii) diagrams consists of only variations of these three plus
  two more basic diagrams.  Finally, we show how our results can be
  consistently and readily used in order to construct two-loop
  neutrino mass models.

\end{abstract}
\end{titlepage}
\setcounter{footnote}{0}

\tableofcontents
\section{Introduction}
\label{sec:intro}
Neutrino masses observed in oscillation experiments
\cite{Fukuda:1998mi,Ahmad:2002jz,Eguchi:2002dm} are so far the only
signal for physics beyond the standard model (SM) measured in
laboratories.  However, while we do know now mass squared differences
and mixing angles to a very high precision
\cite{Forero:2014bxa,Capozzi:2013csa,GonzalezGarcia:2012sz}, there is
no experimental data on whether neutrinos are Dirac or Majorana
particles.

From the low energy point of view Majorana neutrino masses
are described by a lepton-number-breaking dimension five effective
operator, known as the Weinberg operator \cite{Weinberg:1980bf}:
\begin{equation}
  \label{eq:weinberg-operator-intro}
  {\cal O}_5=\frac{c_{\alpha\beta}}{\Lambda}
  \left(\overline{L^c_\alpha}\,i\tau_2\,H\right)
  \left(H^T\,i\tau_2\,L_\beta\right)\ .
\end{equation}
The smallness of the observed light neutrino masses can then be
explained from eq.~(\ref{eq:weinberg-operator-intro}) as being due to
either a large scale $\Lambda$ or a small coffecient $c_{\alpha\beta}$
(or both). However, disentangling these possibilities requires going
beyond this effective operator picture.

It is well-known that at tree-level only three UV completions for the
Weinberg operator exist~\cite{Ma:1998dn}: These are usually called
type-I
\cite{Minkowski:1977sc,Yanagida:1979as,GellMann:1980vs,Mohapatra:1979ia},
type-II
\cite{Schechter:1980gr,Magg:1980ut,Mohapatra:1980yp,Cheng:1980qt} and
type-III \cite{Foot:1988aq} seesaw. All of them have in common that
for $c_{\alpha\beta} \simeq {\cal O}(10^{-2}-1)$, $\Lambda \simeq
10^{13-15}$ GeV is needed to produce sub-eV neutrino masses. Thus,
while being an attractive possibility from the theoretical point of
view, experimentally the classical seesaws do not offer any possible
tests---apart from neutrino masses themselves and the fact that
neutrinos are predicted to be Majorana particles, thus a finite rate
for $0\nu\beta\beta$ decay should exist.

However, $c_{\alpha\beta}$ could easily be smaller than ${\cal O}(1)$. 
Essentially there are three possibilities to arrange this:

\begin{enumerate}
\item The neutrino mass is generated at tree level, but an additional
  suppression enters through a small lepton-number-violating (LNV)
  coupling. The so-called ``inverse'' seesaw \cite{Mohapatra:1986bd}
  or ``linear'' seesaw \cite{Akhmedov:1995ip,Akhmedov:1995vm} are
  examples for this approach.

\item The neutrino mass is generated radiatively. The additional
  suppression is guaranteed by a combination of loop integrals and
  sub-EW scale masses (for example SM charged lepton masses) entering
  the diagrams.  At the one- and two-loop level, the Zee
  \cite{Zee:1980ai} \footnote{The minimal Zee model
    \cite{Wolfenstein:1980sy} is ruled out since it predicts maximal
    mixing in the atmospheric as well as in the solar sector. However,
    its non-minimal version is fully consistent with neutrino
    data~\cite{Balaji:2001ex,He:2003ih,AristizabalSierra:2006ri}.}
  and the Cheng-Li-Babu-Zee models stand as benchmark references
  \cite{Cheng:1980qt,Zee:1985id,Babu:1988ki} and probably due to this
  reason they have been the subject of intensive phenomenological
  studies
  \cite{AristizabalSierra:2006ri,Choudhury:1994vr,Kitabayashi:2000nf,
    Babu:2005ia,AristizabalSierra:2006gb,
    Nebot:2007bc,Schmidt:2014zoa,Herrero-Garcia:2014hfa}.

\item The neutrino mass is forbidden at $d=5$, but appears from
  effective operators of higher dimension
  \cite{Babu:1999me,Bonnet:2009ej}.  Such an approach is not feasible
  in models with only the SM Higgs doublet, since $(H^{\dagger}H)$ is
  a complete singlet and can not carry any charges\footnote{Note that
    an exception to this statement does exist. If the UV completion
    involves higher $SU(2)$ representations (fourtuple scalar and
    triplet fermion), then a model generating at the effective level
    the $d=7$ effective operator $LHLHH^\dagger H$ can be written
    \cite{Babu:2009aq}.}. However, in a two-Higgs doublet world (or
  more complicated setups) forbidding the $d=5$ while allowing $d=7$
  could be realized with, for example, the help of some flavor
  symmetry that prevents the direct Yukawa coupling of the SM Higgs
  doublet to the light fermions.
\end{enumerate}

\begin{figure}%[pt]
  \centering
  \includegraphics[scale=0.88]{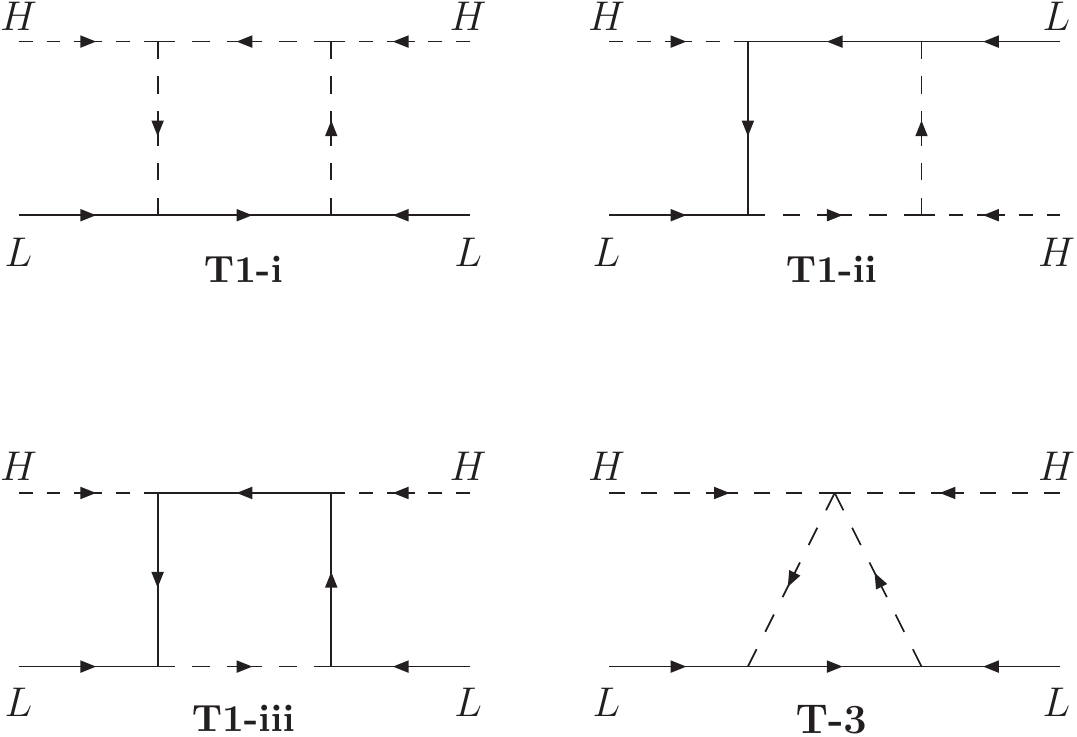}
  \caption{\it The four diagrams leading to genuine 1-loop neutrino
    mass models. The notation of \cite{Bonnet:2012kz} is used to
    classify these diagrams. Just to mention two examples: Diagram
    T1-ii corresponds to the the classical Zee model
    \cite{Zee:1980ai}, while an example for T-3 is the ``scotogenic''
    model of \cite{Ma:2006km}.}
\label{Fig:1lpdiags}
\end{figure}

In this paper we will focus on the second possibility: Loop neutrino
masses. In \cite{Bonnet:2012kz} the Weinberg operator was studied
systematically at the one-loop level. Two topologies (for a total of
four diagrams) were identified to give neutrino masses at the 1-loop
level {\em genuinely} (i.e. without producing neutrino masses at
tree-level), see fig. \ref{Fig:1lpdiags}. Three more diagrams were
found, that can be understood as 1-loop realizations of one of the
known tree-level seesaws and the relation between tree- and 1-loop
diagrams were discussed. In our current work, we extend this analysis
\cite{Bonnet:2012kz} to the 2-loop level, following the same
diagrammatic-based approach.

The systematic decomposition used in \cite{Bonnet:2012kz} allows one
to identify all possible realizations of ${\cal O}_5$ at a given loop
level, in principle. However, while there are only 12 diagrams (out of
which only seven turn out to be of any interest) at the 1-loop level,
at the 2-loop level one can naively expect to find order ${\cal O}
(100)$ diagrams, which need to be studied. We have followed therefore
a sort of ``algorithm'' for ${\cal O}_5$ at the two-loop order: $(i)$
derive all possible two-loop topologies; exclude all 1-loop reducible
topologies from this list. $(ii)$ into the remaining topologies insert
fermions and scalars, to create all possible diagram variations;
exclude from further analysis all those diagrams which need
non-renormalizable vertices. For these first two steps we have
extensively used {\tt FeynArts} \cite{Hahn:2000kx}, in order to ensure
that no topology is missed. $(iii)$ Identify in this list of diagrams
(with renormalizable vertices only) all those, for which no 1-loop
diagram (nor a tree-level neutrino mass) exist. We call these diagrams
``genuine 2-loop diagrams'' and classify them as class-I diagrams
(or models). ($iv$) For all remaining diagrams one can then
distinguish diagrams which lead to finite loop integrals from those
with inifinite integrals. The former cases, which are our
``class-II'' diagrams, can present interesting models of neutrino
mass, even though they are not genuinely 2-loop. The characterization
of class-II diagrams (and their corresponding models) is similar to
the discussion given in \cite{Bonnet:2012kz} for the 1-loop order:
Class-II diagrams can give a theoretical motivation for the
smallness of a particular vertex, generated at 1-loop order. This
particular vertex then appears in one of the four genuine 1-loop
neutrino mass diagrams (see fig. \ref{Fig:1lpdiags}), making the whole
construction {\em effectively} 2-loop. Diagrams with infinite loop
integrals, on the other hand, can never lead to interesting models and
can therefore be discarded.

Surprisingly, the result of the above exercise allows one to show that
in the moderate number of diagrams of class-I {\em all} cases are
variations of only three basic diagrams, two of which have been known
in the literature for a long time: The Cheng-Li-Babu-Zee
\cite{Cheng:1980qt,Zee:1985id,Babu:1988ki} diagram (CLBZ in the
following) and another similar diagram first considered in two
independent papers by Petcov and Toshev \cite{Petcov:1984nz} and by
Babu and Ma \cite{Babu:1988ig} (PTBM in the following).  The third
basic diagram we call the ``rainbow'' diagram (RB in the
following). Similarly, it can be shown that {\em all} diagrams in
class-II can be described by variations of just five basic types of
diagrams: we call them the non-genuine CLBZ and PTBM and RB diagrams
plus two internal scalar correction diagrams (two categories, called
ISC-i and ISC-ii).

Before entering into the details, let us mention that our study
considers only scalar bosons, while, for example, the original papers
on the PTBM diagram \cite{Petcov:1984nz,Babu:1988ig} use the SM
$W$-boson. We decided to concentrate on scalars for essentially two
reasons: (a) From a topological point of view, diagrams with scalar or
vector bosons are equivalent. Thus, from our list of diagrams for
scalars the corresponding diagrams for vectors can be easily
derived.\footnote{Of course, the propagator of a massive vector boson
  is different from that of a scalar. Thus, the expressions for the
  2-loop integrals need to be modified accordingly.} And (b) apart
from the few cases with SM $W$-bosons, new vector-mediated cases require
that the vector should be a gauge boson under a new symmetry, and the
mass should be given by the spontaneous breaking of that
symmetry. This means that the scalar sector of the model needs to be
discussed as well; see \cite{Ma:2012xj} for a recent example.

Our list of diagrams allows us to recover 2-loop models discussed
previously in the literature. Apart from the standard diagrams CLBZ
\cite{Cheng:1980qt,Zee:1985id,Babu:1988ki} and PTBM
\cite{Petcov:1984nz,Babu:1988ig} in their original incarnations
(enumerated as CLBZ-1 and PTBM-1 in the following)\footnote{The
  numbers of the variants quoted correspond to those given in
  figs.~\ref{fig:diagrams-1}, \ref{fig:diagrams-2} and
  figs. \ref{fig:non-genuine-diagrams-i},
  \ref{fig:non-genuine-diagrams-ii},
  \ref{fig:non-genuine-diagrams-iii} and
  \ref{fig:non-genuine-diagrams-3} in appendix
  \ref{sec:non-ren-topo-divergent-diag}.}, we have found a number of
variations of these genuine diagrams and also several realizations of
our class-II diagrams have been discussed in the literature.  For
example, a variant of CLBZ-1 with an additional neutral scalar vev to
generate the lepton number violating triple scalar vertex
$h^-h^-k^{++}$ has been discussed in
\cite{Bamba:2008jq,Lindner:2011it}.  A supersymmetric extension of
CLBZ has been discussed in \cite{Aoki:2010ib}. A new model with a
scalar diquark and a scalar leptoquark has been discussed in
\cite{Kohda:2012sr}, 2-loop neutrino masses are generated by the
CLBZ-1 diagram. Ref. \cite{Ma:2007gq} considers a model with neutrino
masses due to CLBZ-1 and a $\mathbb{Z}_3$ symmetry to eliminate
tree-level seesaw and also explain dark matter.  There are also models
in the literature based on other variants of CLBZ. CLBZ-3 appears in
\cite{delAguila:2011gr}, CLBZ-9 in \cite{Guo:2012ne,Li:2012mu} and
CLBZ-8 and CLBZ-10 appear within the 331-model of \cite{Chang:2006aa}.
A possible connection between two-loop neutrino masses and dark matter
has been explored in\cite{Aoki:2014cja,Okada:2014qsa} in two models
giving each a CLBZ-3 type diagram.  Then there are also models, based
on CLBZ, using vectors instead of scalars
\cite{Chen:2006vn,Chen:2007dc,King:2014uha}. All these models are
realizations of a 2-loop gauge-mediated diagram involving an internal
effective coupling (see e.g. \cite{Gustafsson:2014vpa}). In
refs. \cite{Chen:2006vn,Chen:2007dc} this effective coupling is
generated at the tree level via the mixing of an $SU(2)$ scalar
triplet with a doubly charged singlet, thus resulting in a 2-loop
model (effectively CLBZ-9). Note that this construction requires that
the tree-level coupling between the triplet and the leptons is absent.
Ref. \cite{Gustafsson:2012vj}, instead, discusses the case where the
effective coupling is induced at the 1-loop order, thus leading to a
3-loop gauge-mediated neutrino mass model. Finally, ref.
\cite{King:2014uha} considers a partial UV completion involving an
effective lepton number-violating vertex for $W$ bosons with a doubly
charged singlet. Then there are models based on variants of PTBM such
as \cite{Angel:2013hla}, which uses leptoquarks and a colour octet
fermion. Also in R-parity violation PTBM diagrams appear
\cite{Borzumati:2002bf} and can be used to constrain the R-parity
violating soft SUSY breaking parameters.  Such R-parity violating SUSY
models have not only PTBM diagrams, but also RB type 2-loop diagrams
\cite{Dey:2008ht}. Then there are leptoquark models \cite{Babu:2010vp}
and extensions of the SM with vector-like quarks \cite{Babu:2011vb},
with scalar and SM $W$ boson diagrams. In this case, both CLBZ and
PTBM in various variants contribute to the neutrino mass.  RB diagrams
where considered, for example, in
\cite{Kajiyama:2013rla,Baek:2013fsa}. However, those models contain
RB-diagrams of higher order, $[(L H)^2S^2]/\Lambda^3$, and thus do not
fall into our classification scheme.  The 1-loop diagram {\bf
  T}-$\boldsymbol{3}$, see fig. \ref{Fig:1lpdiags}, contains a quartic
scalar vertex, usually its parameter is called $\lambda_5$.  The
radiative generation of $\lambda_5$ for this diagram, via diagrams of
class ISC-i has been considered in \cite{Ma:2007yx}.  Similarly ISC-i
variant-5 was discussed in \cite{Aoki:2013gzs}.

On top of these ``pure'' 2-loop models, also mixed situations, where
one (or more) neutrinos have tree-level masses, while one neutrino
mass is generated at 2-loop level have been considered.
Ref. \cite{Grimus:2000kv} considers such a situation, with some
neutrinos getting a mass through CLBZ-1.  Similarly,
\cite{Davidson:2006tg} assumes two neutrino masses to be tree-level
and calculates the minimal mass for the remaining neutrino, generated
through diagrams with Higgses of the form PTBM-1 in both SM and MSSM.
Ref. \cite{Joshipura:1999xe} considers a variant with some neutrinos
receiving 1-loop neutrino masses and others are 2-loop.  Also,
\cite{Chang:1999hga} consider models where neutrino mass appear at
1-loop level and also at 2-loop level with CLBZ-1, PTBM-4 and two
variants of the RB diagram\footnote{The diagrams shown for the RB
  are NG-RB-2 and one diagram with an infinite
  integral. The latter is, of course, not an interesting 2-loop model,
  but a correction to the 1-loop diagram.}.

We mention that there are also papers on two-loop models in the
literature, which are not covered by our classification, because they
are of higher dimension than the Weinberg operator.  To quote two more
examples, the papers \cite{Kitabayashi:2000za,Kitabayashi:2001jp,
  Kitabayashi:2001id,Kitabayashi:2000nq} have several variants of CLBZ
at $d=7$, while \cite{Kajiyama:2013zla} has a variant of the RB at
$d=7$. Finally there is the paper \cite{Okada:2014vla}, that discusses
{\em Dirac} neutrino masses at 2-loop.

Our work is, of course, not the first attempt to organize neutrino
mass models systematically. Apart from the above-mentioned paper
\cite{Bonnet:2012kz}, which treats the 1-loop case, a set of ``rules
and recipes'' for neutrino masses at 1-loop, 2-loop and higher orders
has been discussed in \cite{Farzan:2012ev} and our approach has some
overlap with this paper, too. Then, there is the interesting work of
\cite{Babu:2001ex}, which writes down all lepton number violating
operators from $d=5$ to $d=11$. Decomposing these operators, one can
find a list of tree-level, 1-loop, 2-loop etc. diagrams, which allow
to specify neutrino mass models
\cite{Babu:2001ex,Choi:2002bb,degouvea:2007xp,Angel:2012ug,delAguila:2012nu}.
Our study is complementary to the analysis done in these papers in
that it provides further insight for the specific two-loop case, {\em
  exhaustively} listing all possible diagrams. However, different from
\cite{Babu:2001ex,Choi:2002bb,degouvea:2007xp,Angel:2012ug,delAguila:2012nu},
we put quite some emphasis on our classification schemes, which allow
us to distinguish ``genuine'' models, i.e. those for which the absence
of 1-loop masses is {\em guaranteed} from the ``non-genuine'' (or
class-II) models.

The rest of this paper is organized as follows. In section
\ref{sec:generalities} we discuss the ``strategy'' followed in this
paper and introduce our notation. Section
\ref{sec:2-loop-order-realizations} is devoted to the classification
of relevant topologies. ``Genuine'' and ``non-genuine'' diagrams are
discussed next and SM electroweak-sector quantum number assignments
are given. In sec.~\ref{sec:constructing-model-guide} we exemplify the
use of our results by constructing two specific examples of UV
completions. In sec.~\ref{sec:conclusions} we summarize and present
our conclusions. The bulk of the technical details of our calculation
is collected in appendices \ref{sec:non-ren-topo-divergent-diag},
\ref{sec:quantum-numbers-appendix} and \ref{sec:formulas-2-loop},
where we list renormalizable topologies leading to non-genuine finite
and infinite diagrams, non-renormalizable topologies, as well as
non-genuine diagrams, tables with the different quantum number
assignments and relevant formulas for the evaluation of two-loop
integrals.

\section{Two-loop 1PI topologies, diagrams,
  genuine models and quantum numbers}
\label{sec:2-loop-order-realizations}

\subsection{Generalities, strategy and notation}
\label{sec:generalities}

A systematic classification of the two-loop order realizations of
${\cal O}_5$ using the diagrammatic method does so far not exist. The
underlying reason is probably related with the fact that tackling the
problem via the diagram-based method turns out to be challenging, due
to the large number of two-loop diagrams, and unless precise
guidelines are followed such study is not possible. Thus, in this
section we discuss some generic guidelines that will allow us to deal
with the 2-loop classification of ${\cal O}_5$.

At the 2-loop order, the dimension five effective operator consist of
a set of topologies: ${\cal O}_5^{2-\text{loop}}= \{T2_1, T2_2, \dots
, T2_r\}$. We have identified 29 distinct topologies (see below) out
of which only a subset turns out to be relevant.  Once all topologies
have been identified, the next step is then that of specifying the
fermion and scalar internal lines ($F$ and $S$) as well as the
external lines ($L$ and $H$) of each topology, i.e. ``promoting''
topologies to diagrams. Here, {\it renormalizability} fixes possible
vertices to only dimension-four three and four point vertices (3-PVs
and 4-PVs). Due to the freedom when placing the two external $L$ and
$H$ lines, however, in general, a given two-loop order topology $T2_i$
can involve quite a few number of Feynman diagrams. At this point it
is possible to discard 11 topologies (see
fig. \ref{fig:non-renormalizable-topologies}), since these will always
lead to non-renormalizable diagrams, see appendix
\ref{sec:non-ren-topo-divergent-diag}.

From the remaining 18 topologies (see
figs. \ref{fig:box-based-topologies} and
\ref{fig:renormalizable-topologies-non-genuine}) only a subset
leads to genuine two-loop diagrams. In order to identify non-genuine
diagrams, one can assign arbitrary quantum numbers $q_i$ ($q_i$,
related with a new symmetry or in some cases with the gauge symmetry
itself, e.g. hypercharge) to the external and internal fields, and
then enforce conservation of these charges vertex by vertex. These
conservation rules define a set of conditions (denoted by $C^{2_i}$)
that whenever satisfied guarantee the presence of the corresponding
diagram.  These conditions should be confronted with those arising
from the 1-loop diagrams shown in fig. \ref{Fig:1lpdiags} (denoted by
$C^{1_i}$).  Thus, if $C^{2_i}\subset C^{1_i}$, the corresponding
2-loop diagram will be necessarily accompanied by a 1-loop diagram,
hence being non-genuine. Diagrams for which $C^{2_i}\subset C^{1_i}$
is not satisfied are potentially genuine, but their particle content
must satisfy further constraints which guarantee the absence of both
tree and 1-loop level diagrams (see sec. \ref{sec:genuine-models} for
more details). Once these constraints are assured, the full list of
truly genuine diagrams is fixed

Genuine diagrams define a set of renormalizable vertices, which will
lead to a 2-loop UV completion (Lagrangian) once the gauge quantum
numbers of the beyond SM fermion and scalar fields are specified. For
that purpose the lepton and Higgs gauge quantum numbers can be used to
constrain the possible quantum numbers of the internal fermion and
scalar fields.  This procedure, however, provides an unambiguous
determination only in the case of trilinear couplings involving two SM
fields.  Let us discuss this in more detail. Yukawa (or pure scalar
trilinear) couplings can involve two, one or none SM fields,
schematically: $\bar F L H$; $\bar F L S$ (which reduces to
$\overline{L^c}LS$ whenever $F=L^c$), $\bar F_1 F_2 H$; $\bar F_1F_2
S$. In the first case, clearly $F$ has to be---unambiguously---a
$SU(2)$ singlet or triplet (type-I or type-III seesaws) while in the
other cases $SU(2)\times U(1)_Y$ invariance requires:
\begin{align}
  \label{eq:SU2-U1-invariance}
  n_{\bar F}\otimes n_s&\supset 2\ ,\qquad Y_{\bar F}+Y_S+Y_L=0\ ,
  \nonumber\\
  n_{F_1}\otimes n_{F_2}&\supset 2\ ,\qquad Y_{\bar F_1}+Y_{F_2}+Y_{H}=0\
  ,
  \nonumber\\
  n_{F_1}\otimes n_{F_2}&\supset n_{\bar S}\ ,\qquad Y_{\bar
    F_1}+Y_{F_2}+Y_{S}=0\ ,
\end{align}
where $n_X$ corresponds to the $SU(2)$ representation under which the
$X$ field transforms.  From (\ref{eq:SU2-U1-invariance}), it can be
seen that---in principle---an infinite number of $SU(2)$
representations as well as hypercharge assignments are consistent with
the constraints implied by the lepton and Higgs quantum numbers. 4-PVs
allow even for more freedom. These vertices can involve three, two, one or
none SM fields (Higgses), schematically: $H H H S_1$, $H H S_1 S_2$, $H S_1 S_2
S_3$, $S_1S_2S_3S_4$. So, in this case gauge invariance implies:
\begin{align}
  \label{eq:quartic-couplings-GI}
  n_{S_1}&= \bar 4\ ,
  \qquad
  3Y_H+Y_{S_1}=0\ ,
  \nonumber\\
  n_{S_1}\otimes n_{S_2}&\supset \bar 3\ ,
  \qquad
  2Y_H+Y_{S_1}+Y_{S_2}=0\ ,
  \nonumber\\
  n_{S_1}\otimes n_{S_2}\otimes n_{S_3}&\supset 2\ ,
  \qquad
  Y_H+\sum_i Y_{S_i}=0\ ,
  \nonumber\\
  n_{S_1}\otimes n_{S_2}\otimes n_{S_3}\otimes n_{S_4}&\supset 1\ ,
  \qquad
  \sum_i Y_{S_i}=0\ ,
\end{align}
which shows that apart from $HHHS_1$, there is no coupling which
allows for an unambiguous determination of the SM gauge quantum
numbers of the new fields.

Since trilinear couplings involving two SM fields are only possible in
tree level realizations, the discussion above implies that once going
beyond the tree level, a given genuine diagram leads to an infinite
number of UV completions. From the practical point of view
nevertheless one can impose an upper limit on the dimensions of the
representations used. In our tables we list all combinations with
singlets, doublets and triplets of $SU(2)_L$. Tables for larger
representations can be easily derived. We also mention that we do not
give explicitly color quantum numbers in our tables since, as pointed
out in \cite{Bonnet:2012kz}, the inclusion of color is
straightforward, see also the discussion in section
\ref{sec:gauge-quantum-numbers}.

Finally, once the UV completions are identified the only step which
remains to be done is the determination of the light neutrino mass
matrix, which requires calculating $2$-loop integrals.  Although the
list of genuine diagrams is large this does not means that the number
of $2$-loop integrals to be evaluated is large.  Different diagrams,
not necessarily arising from the same topology, can involve the same
$2$-loop integral, essentially because after electroweak symmetry
breaking the couplings to Higgs legs are just couplings to a
background field: if coupled to fermions (scalars) they imply
chirality flips (scalar mixing). This observation allows to reduce the
number of integrals to be evaluated to just combinations of a few
basic integrals, which we list in appendix \ref{sec:formulas-2-loop}.

% -----------------
% Section
% -----------------
\subsection{Two-loop 1PI topologies}
\label{sec:2-loop-order-realizations}

\begin{figure}
  \centering
  \includegraphics[scale=0.88]{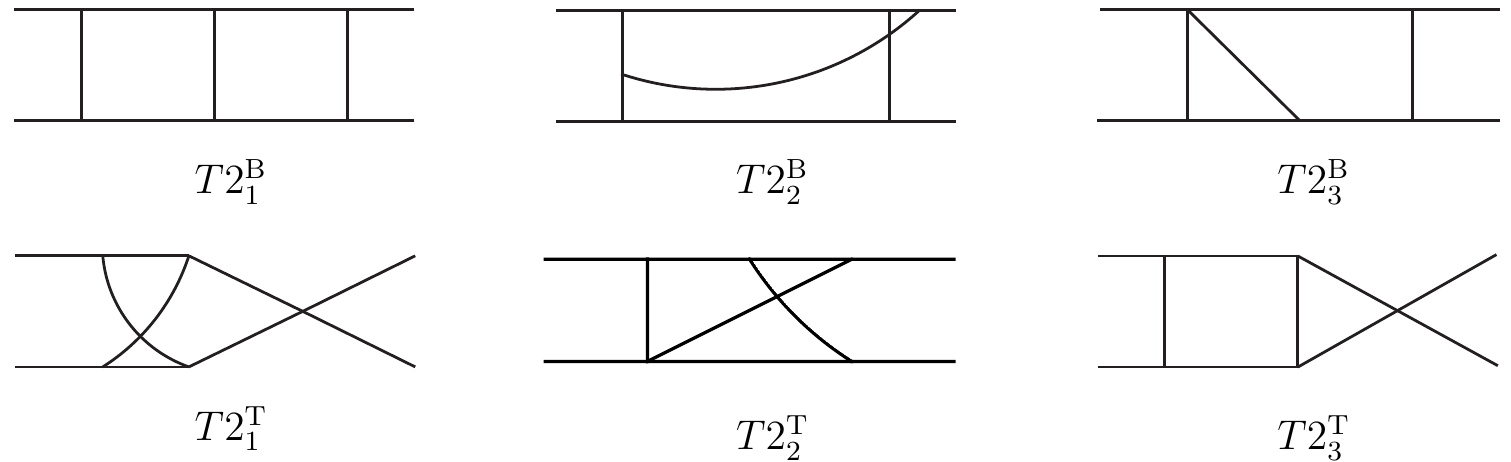}
  \caption{\it 1PI box-based (upper row) and triangle-based (lower
    row) two-loop topologies, which can lead to genuine 2-loop
    models. All other topologies can lead only to class-II models,
    infinite renormlizations of 1-loop diagrams or contain
    non-renormalizable vertices. Those topologies are given in the
    appendix \ref{sec:non-ren-topo-divergent-diag}.}
  \label{fig:box-based-topologies}
\end{figure}

Following the strategy described in sec. \ref{sec:generalities}, our
starting point consist in determining the complete set of two-loop
one-particle irreducible (1PI) inequivalent topologies. At the
two-loop order this set is expected to be large, so in order to
generate an exhaustive list we proceed with {\tt FeynArts}
\cite{Hahn:2000kx}. To simplify the output we suppress from the
calculation topologies involving external legs self-energies, tadpoles
and one-particle-reducible 2-loop topologies. The complete set of
topologies is displayed in fig. \ref{fig:box-based-topologies} and
figs.~\ref{fig:renormalizable-topologies-non-genuine} and
\ref{fig:non-renormalizable-topologies} in the appendix, respectively.
In total we have identified 29 topologies, but only topologies listed
in fig. \ref{fig:box-based-topologies} lead to genuine diagrams.
Topologies shown in
fig.~\ref{fig:renormalizable-topologies-non-genuine} will lead to
class-II models, while fig.\ref{fig:non-renormalizable-topologies}
shows for completeness the non-renormalizable topologies.

% --------------------------------------
% Box-based two-loop topologies
% --------------------------------------

Denoting by $(\# \text{3-PVs},\# \text{4-PVs})$ the number of 3-PVs
and 4-PVs entering in each diagram, all topologies can be placed in
four non-overlapping sets: (2,2), (4,1), (6,0) and (0,3). Since 4PVs
are only possible for scalars, topologies not satisfying the
renormalizability criterion should arise from the sets (2,2) and (0,3)
(those involving the largest number of 4PVs). Indeed, the set of
topologies not satisfying this criterion consist of the full (0,3)
subset and eight (2,2) topologies (see
fig.~\ref{fig:non-renormalizable-topologies}). 

\begin{figure}
  \centering
  \includegraphics[scale=1.1]{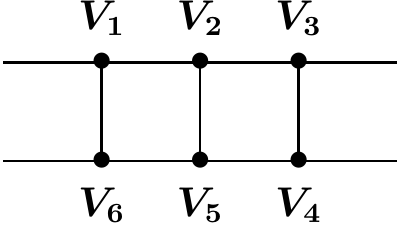}
  \qquad
  \qquad
  \includegraphics[scale=1]{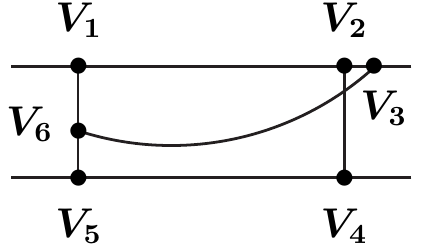}
  \caption{\it Vertex assignments for the two-loop (6,0) topologies
    $T2^\text{B}_1$ and $T2^\text{B}_2$.}
  \label{fig:vertices}
\end{figure}

\subsection{Constructing diagrams}
\label{sec:diag-const}
No matter the topology, any diagram can be constructed from the
following schematic Lagrangian:
\begin{align}
  \label{eq:schematic-Lagrangian}
  {\cal L}= &Y_L\bar L F_i S_k + Y_H \bar F_i F_j H + \mu_H S_i S_j H
  + \lambda_H S_i S_j S_k H
  \nonumber\\
  & + \lambda_{HH} S_i S_j H H
   + Y \bar F_i F_j S_k + \mu_s S_i S_j S_k + \mbox{H.c.}\ .
\end{align}
In order to illustrate the method we have used for constructing
diagrams we focus on the first two box-based topologies in
fig~\ref{fig:box-based-topologies} ($T2_1^B$ and $T2_2^B$), and base 
our discussion on the diagrams sketched in fig.~\ref{fig:vertices}.

% ---------------------------------
% Double box-based topology table
% ---------------------------------
\begin{table}[t!]
  \centering
  \renewcommand{\arraystretch}{1.5}
  \begin{tabular}{|c|c|c|c|c|c|c|}
    \hline
    % 0th row
    $\boldsymbol{V_1}$ & $\boldsymbol{V_2}$ & $\boldsymbol{V_3}$ &
    $\boldsymbol{V_4}$ & $\boldsymbol{V_5}$ & $\boldsymbol{V_6}$ & {\bf Diagram}
    \\\hline
    % 1st row    
    \multirow{7}{1.2cm}{$LSF$} & \multirow{3}{1.2cm}{$FSF$} 
    & $FSL$ & $HSS$ & $SSS$ & $SSH$ & 
    CLBZ-3\\\cline{3-7}
    % 2nd row
    &  & \multirow{2}{1.2cm}{$FFH$} & $LFS$ & $SSS$ 
    & $SSH$ & CLBZ-2\\\cline{4-7}
    % 3rd row
    &  &  & $HFF$ & $FSF$ & $FSL$ & 
    {RB-1}\\\cline{2-7}
    % 4th row
    & \multirow{4}{1.2cm}{$FFS$} & \multirow{2}{1.2cm}{$SFL$} 
    & $HFF$ & $FFS$ & $SSH$ & PTBM-2\\\cline{4-7}
    % 5th row
    &  &  & $LFS$ & $SFF$ & $FSL$ & {\large \ding{55}}\\\cline{3-7}
    % 6th row
    &  & \multirow{2}{1.2cm}{$SSH$} & $HSS$ & $SFF$ 
    & $FSL$ &  {RB-2}\\\cline{4-7}
    % 7th row
    &  &  & $LSF$ & $FFS$ & $SSH$ & PTBM-3\\
    \hline
    % 8th row    
    \multirow{7}{1.2cm}{$LFS$} & \multirow{4}{1.2cm}{$SFF$} 
    & \multirow{2}{1.2cm}{$FFH$} & $LFS$ & $SFF$ & $FFH$ & 
    PTBM-1\\\cline{4-7}
    % 9th row
    & & & $HFF$ & $FFS$ & $SFL$ & DIV% {\large \ding{55}}
    \\\cline{3-7}
    % 10th row
    &  & \multirow{2}{1.2cm}{$FSL$} & $LSF$ & $FFS$ 
    & $SFL$ & {\large \ding{55}}\\\cline{4-7}
    % 11th row
    &  &  & $HSS$ & $SFF$ & $FFH$ & 
    PTBM-2\\\cline{2-7}
    % 12th row
    & \multirow{4}{1.2cm}{$SSS$} & \multirow{2}{1.2cm}{$SFL$} 
    & $LFS$ & $SSS$ & $SFL$ & {\large \ding{55}}\\\cline{4-7}
    % 13th row
    &  &  & $HFF$ & $FSF$ & $FFH$ & CLBZ-1\\\cline{3-7}
    % 14th row
    &  & \multirow{2}{1.2cm}{$SSH$} & $LSF$ & $FSF$ 
    & $FFH$ & CLBZ-2\\\cline{4-7}
    % 15th row
    &  &  & $HSS$ & $SSS$ & $SFL$ &  ISC-i-3% {\large \ding{55}}
    \\
    \hline
  \end{tabular}
  \caption{\it Sequential vertex insertions leading to the 
    full set of diagrams for topology $T2_1^\text{B}$. 
    For $\boldsymbol{V_1}, \boldsymbol{V_2},\boldsymbol{V_3}$ 
    the field sequence goes from left to right while for 
    $\boldsymbol{V_4}, \boldsymbol{V_5},\boldsymbol{V_6}$ 
    from right to left. Crosses indicate diagrams that do 
    not correspond to ${\cal O}^{2-\text{loop}}_5$,
    while DIV a diagram involving a 2-loop divergent integral,
    hence of no interest.}
  \label{tab:box-based-diagrams-table}
\end{table}
% ---------------------------------------
In the (6,0) case, external vertices always involve $Y_L$, $Y_H$ or
$\mu_H$ couplings. So, in order to find an exhaustive list of possible
Feynman diagrams one can start by fixing any of these couplings at
$\boldsymbol{V_1}$ (see fig \ref{fig:vertices}) and then inserting
sequentially in clockwise direction all possible vertices
combinations. Table \ref{tab:box-based-diagrams-table} illustrates the
procedure for topology $T2_1^\text{B}$, where we have fixed at
$\boldsymbol{V_1}$ the $Y_L$ coupling. It can be seen that out of the
15 diagrams, 3 are possible only for four fermion external legs and so
have nothing to do with ${\cal O}_5^{2-\text{loop}}$. In addition two
diagrams appear twice (CLBZ-2 and PTBM-2), so at the end the 2-loop
box-based topology $T2^\text{B}_1$ involves 10 diagrams. For topology
$T2^\text{B}_2$, determining the complete list requires considering at
$\boldsymbol{V_1}$ not only $Y_L$ but also $Y_H$ and $\mu_H$, due to
its non-symmetric character.  By doing so, the resulting list involves
repeated diagrams (redundant diagrams) whose identification turns out
to be complex.  For that aim it is therefore useful to introduce the
following sextuplet
\begin{equation}
  \label{eq:six-tuple}
  (n_L, n_3, n_H, n_Y, n_S, n_4)\ ,
\end{equation}
where the different entries label the number of $Y_L$, $\mu_H$, $Y_H$,
$Y$, $\mu_s$ and $\lambda_{HH}$ vertices defining a given diagram, and
are such that depending on the topology obey certain constraints. For
(6,0)-based diagrams these constraints read:
\begin{align}
  \label{eq:constraints-six-tuple-6-0}
  &n_L + n_3 + n_H + n_Y + n_S + n_4=6\ ,\\
  &n_L=2\ ,\qquad n_3 + n_H=2\ , \qquad n_4=0\ ,\qquad n_Y +n_S=2\ ,
\end{align}
thus implying that the sextuplet structure of any diagram arising from
(6,0) topologies will necessarily belong to one of the following nine
sextuplets:
\begin{align}
  \label{eq:six-tuples-6-0}
  (2,2,0,2,0,0)\ ,\qquad &(2,2,0,0,2,0)\ ,\qquad (2,2,0,1,1,0)\ ,\nonumber\\
  (2,0,2,2,0,0)\ ,\qquad &(2,0,2,0,2,0)\ , \qquad (2,0,2,1,1,0)\ ,\nonumber\\
  (2,1,1,2,0,0)\ ,\qquad&(2,1,1,0,2,0)\ , \qquad(2,1,1,1,1,0)\ .
\end{align}

For $T2^\text{B}_2$, the procedure outlined above yields 22 diagrams
which can be grouped in five sets: one (2,2,0,0,2,0), six
(2,2,0,1,1,0), five (2,0,2,2,0,0), six (2,1,1,2,0,0) and four
(2,1,1,1,1,0). Possible redundant diagrams should belong to a specific
set, however since not all diagrams belonging to a given set are
redundant, the identification of superfluous diagrams requires
labelling the fermion and scalar lines of all diagrams within each set
and comparing the different couplings. If the couplings of a couple of
diagrams match, those diagrams count as one. So, proceeding in that
way we have found that the number of diagrams arising from the
$T2^\text{B}_2$ is ten.

Following this procedure for the remaining 16 topologies we have found
the full set of diagrams for 1PI 2-loop topologies.

\subsection{Genuine diagrams}

\label{sec:genuine-models}
\begin{figure}[t!]
  \centering
  \includegraphics[scale=1]{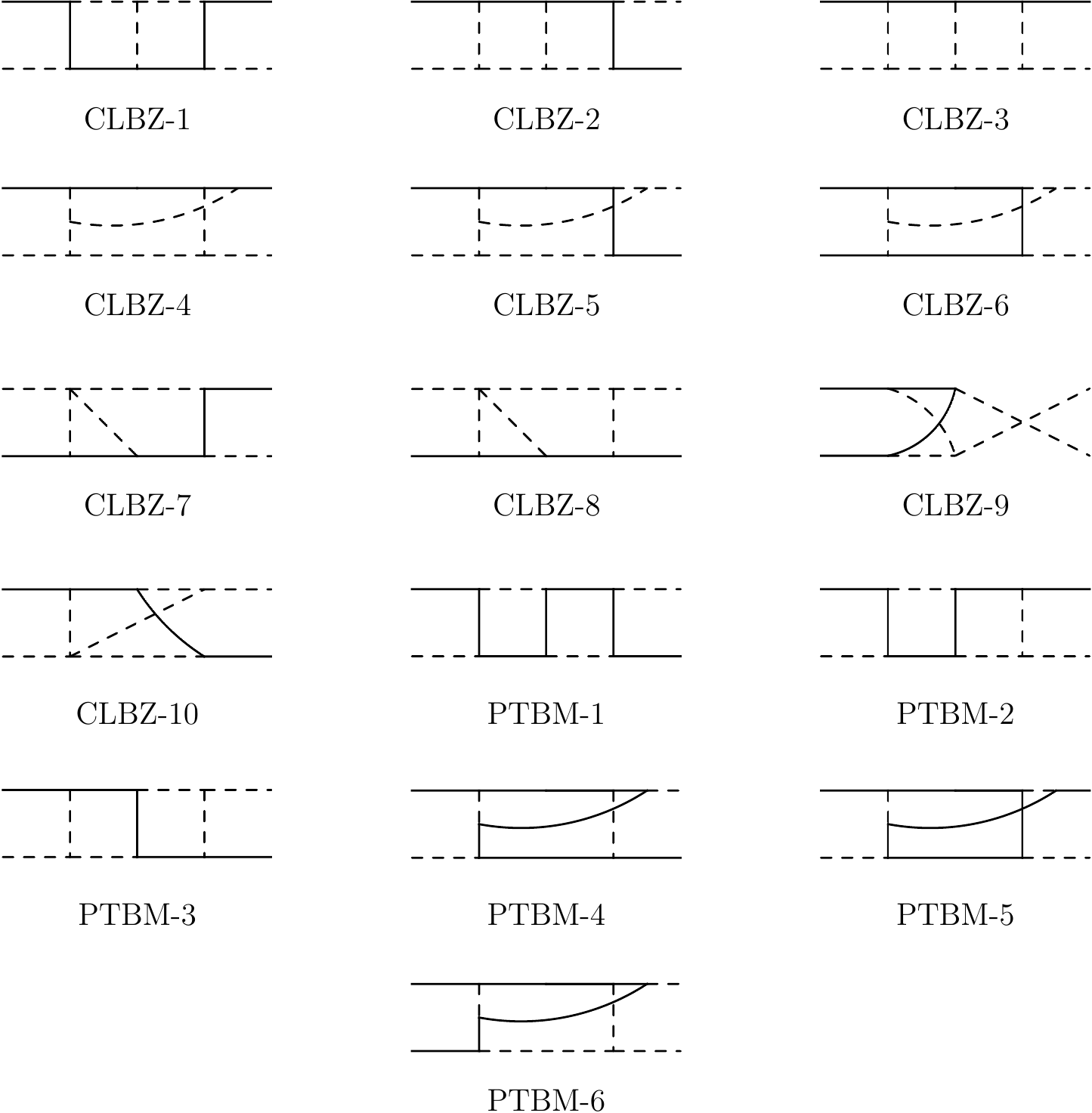}
  \caption{\it Genuine two-loop CLBZ (Cheng-Li-Babu-Zee) and PTBM
    (Petcov-Toshev-Babu-Ma) diagrams arising from the topologies in
    fig.~\ref{fig:box-based-topologies}. See the text for further
    details.}
  \label{fig:diagrams-1}
\end{figure}

\begin{figure}[t!]
  \centering
  \includegraphics[scale=1]{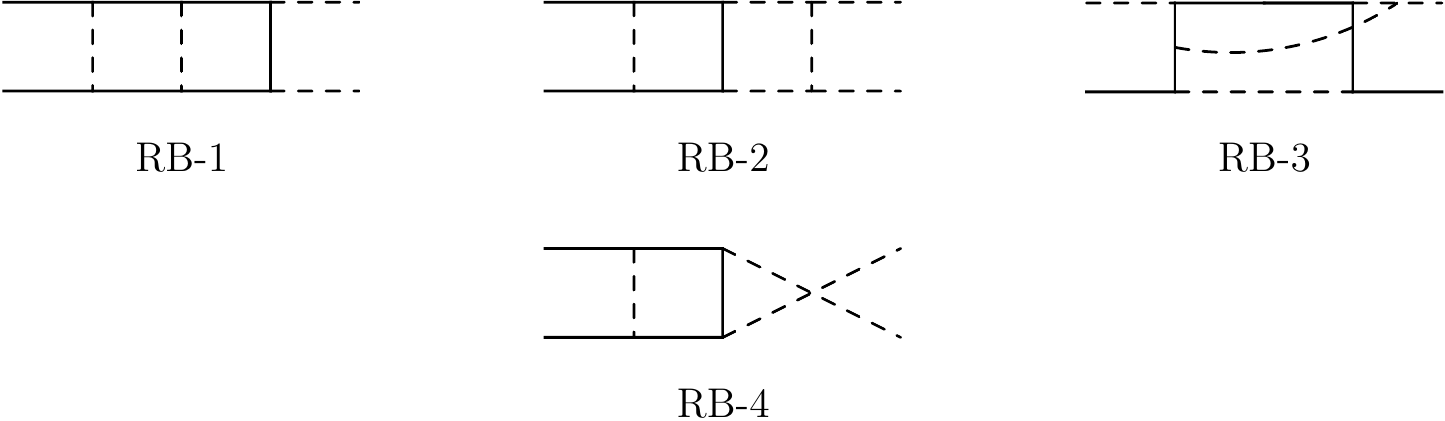}
  \caption{\it Genuine two-loop RB diagrams.  See the text for further
    details.}
  \label{fig:diagrams-2}
\end{figure}

Having identified all possible diagrams, we are now in a position to
build the full list of genuine diagrams. The procedure to be followed
involves two steps. First, we assume that the lepton and Higgs $SU(2)$
doublets as well as the heavy fields flowing in the loops carry
arbitrary charges $q_i$, and impose $q_i$ charge conservation vertex
by vertex, as outlined in sec. \ref{sec:generalities} (and exemplified
below). By doing so, we identify the non-genuine diagrams in our
list. We are then left with diagrams which potentially lead to genuine
models, see fig. \ref{fig:diagrams-1} and \ref{fig:diagrams-2}. Their
genuineness can then be guaranteed provided their particle content
obeys the additional constraints discussed near the end of this
section.

\begin{figure}[t]
  \centering
  \includegraphics[scale=0.92]{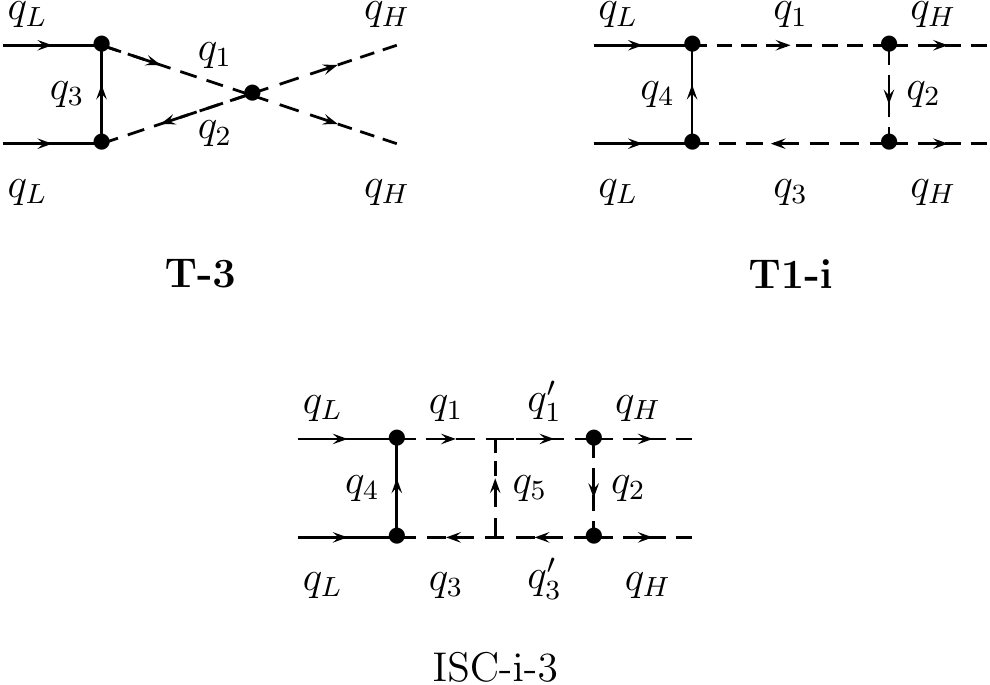}
  \caption{\it Charge $q_i$ flow for 1PI one-loop box and triangular
    diagrams ({\bf T1-i} and {\bf T-3}) as well as for the two-loop
    box-based diagram ICS-i-3.}
  \label{fig:charge-flow}
\end{figure}

Let us first illustrate the $q_i$ charge procedure we have employed to
identify non-genuine diagrams. The example we discuss is based on
the 1- and 2-loop diagrams displayed in fig.  \ref{fig:charge-flow}.

For diagram {\bf T1-i}, the equations for $q_i$ conservation
can be written as 
\begin{alignat}{2}
  \label{eq:one-loop-box-conservation-Eqs}
  q_L + q_3&=q_4\ , &\qquad\qquad 
  q_L + q_4&=q_1\ ,\nonumber\\
  q_H + q_2&=q_1\ , &
  q_H + q_3&=q_2\ .
\end{alignat} 
For diagram {\bf T-3} one has:
\begin{equation}
  \label{eq:one-loop-triangle-conservation-Eqs}
  q_L + q_3  = q_1\ ,\qquad
  2q_H + q_2 = q_1\ ,\qquad
  q_L + q_2  = q_3\ .
\end{equation}
The solution of these system of equations then leads to the following
charge constraints:
\begin{alignat}{4}
  \label{eq:one-loop-second-box-SOL}
  C^\text{T1-i}:&\qquad q_4&=&q_3 + q_L\ ,&\qquad
  q_1&=q_3 + 2 q_L\ ,
  &\qquad q_2&=q_4\ . \\
  \label{eq:one-loop-box-triangle-Eqs-SOL-2}
  C^\text{T-3}:&\qquad q_1&=&q_2 + 2 q_L\ ,&\qquad
  q_3&=q_2 + q_L\ , &&
\end{alignat}
where $C^\text{T1-i}$ is the solution for the
diagram {\bf T1-i} in fig. (\ref{fig:charge-flow}), 
whereas $C^\text{T-3}$ the solution for the one in eq.
(\ref{eq:one-loop-triangle-conservation-Eqs}).

The constraints in (\ref{eq:one-loop-second-box-SOL}) and
(\ref{eq:one-loop-box-triangle-Eqs-SOL-2}) are to be used to know
whether the 2-loop box-based diagram ISC-i-3 in fig.
\ref{fig:charge-flow} is non-genuine or not. For that aim the $q_i$
charge conservation equations for this diagram has to be written.
Conservation of charges implies:
\begin{alignat}{2}
  \label{eq:two-loop-box-second-case-conservation-Eqs}
  q_L + q_4 &=q_1\ , &\qquad\qquad q_1 + q_5&=q_1^\prime\ ,\nonumber\\
  q_H + q_2&=q_1^\prime\ , & q_H + q_3^\prime&=q_2\ ,\nonumber\\
  q_5 + q_3&=q_3^\prime\ , &q_L + q_3&=q_4\ ,
\end{alignat}
and their solution is given by
\begin{alignat}{4}
  \label{eq:two-loop-box-based-case-2-Eqs-SOL}
  C^\text{ISC-i-3}:&\qquad &q_1=&q_3 + 2q_L\ ,&\qquad
  q_4&=q_3 + q_L\ ,
  &\qquad q_2&=q_4 + q_5\ .
\end{alignat}
Comparing this solution with $C^\text{T1-i}_R$ in
(\ref{eq:one-loop-second-box-SOL}), one can see that $q_5\neq0$
forbids the 1-loop box-based diagram {\bf T1-i} (right-hand side in
fig.~\ref{fig:charge-flow}). However, when comparing with
$C^\text{T-3}$ (trading $q_2\to q_3$ and $q_3\to q_4$) in
~(\ref{eq:one-loop-box-triangle-Eqs-SOL-2}) it is clear that
constraints $C^\text{ISC-i-3}$ allow the 1-loop triangle-based diagram
{\bf T-3}, {\em independent of the choice of charges}. One can then
conclude that ISC-i-3 is not a genuine diagram.

Following this procedure, we have identified all non-genuine
diagrams. These emerge from the topologies in
fig. \ref{fig:renormalizable-topologies-non-genuine} in appendix
\ref{sec:non-ren-topo-divergent-diag}. Moreover, we have found that
the non-genuine but finite diagrams all belong to one of the following
five different types, namely: $(a)$ non-genuine CLBZ (NG-CLBZ), $(b)$
non-genuine PTBM (NG-PTBM), $(c)$ non-genuine RB (NG-RB), $(d)$ ISC-i,
$(e)$ ISC-ii.
Figs. \ref{fig:non-genuine-diagrams-i}-\ref{fig:non-genuine-diagrams-3}
in appendix \ref{sec:non-ren-topo-divergent-diag} show this complete
list of non-genuine but finite diagrams.

Let us now turn to the remaining (potentially) genuine diagrams that
can not be eliminated after this procedure has been applied to the
full list of diagrams. These are given in fig. \ref{fig:diagrams-1}
and fig. \ref{fig:diagrams-2}.  All of these fall, as already stressed
above, into the three classes: CLBZ, PTBM and RB. There is one
subtlety involved in these RB diagrams, which we want to discuss
briefly: In all non-genuine diagrams, see appendix
\ref{sec:non-ren-topo-divergent-diag}, it is possible to make a cut in
the diagram, such that the remaining sub-diagram is equivalent to a
vertex correction.  Looking superficially to the RB diagrams in
fig. \ref{fig:diagrams-2}, it seems that such a cut is possible too,
with the remaining sub-diagram being a correction to a fermion
propagator. However, in the RB case shrinking the remaining
sub-diagram to a point would generate a {\em non-renormalizable}
vertex F-F-H-H. These diagrams therefore present potentially genuine
models.

So far we have worked from the full set of (topologies and) diagrams,
excluding one after the other the non-interesting cases.  However, for
those remaining 20 diagrams, there is one more subtlety to be
discussed: One can write down Lagrangians, which produce, say, only
one neutrino mass at tree-level (or 1-loop) level, while the other
neutrino mass \footnote{Recall, that oscillation data require only two
  non-zero neutrino masses.}  (or masses) are generated
radiatively. In this case, restrictions on the particle content of the
model are determined by the requirements at that lower order. For
example, a model with one right-handed neutrino will produce one
non-zero neutrino mass at tree-level, while the other neutrino masses
are then automatically generated by the genuine 2-loop diagram PTBM-1
(with SM $W^+$ gauge bosons).

The following additional (but rather trivial) conditions, which
finally guarantee that Lagrangians producing the diagrams in
fig. \ref{fig:diagrams-1} are genuine 2-loop Lagrangians---in our
sense---should therefore be understood as constraints {\em per
  neutrino generation} for which one wants to generate genuine 2-loop
masses. Genuiness in this sense requires:

\begin{enumerate}[i)]
\item \label{gen-condition-1} Absence of hypercharge zero fermion
  electroweak singlets or triplets, or hypercharge $2$ scalar $SU(2)$
  triplets is required, otherwise the neutrino mass will be 
  determined by tree level type-(I,II,III) seesaw diagrams.
\item \label{gen-condition-2} Absence of hypercharge zero scalar
  $SU(2)$ singlets or triplets. The presence of these fields allow
  constructing 1-loop diagrams by (a) making a simple cut in the
  2-loop diagram, or (b) (only in case of triplets) allow to construct
  the 1-loop diagram {\bf T-3}.
\item \label{gen-condition-3} Internal scalars should not have the
  quantum numbers of the Higgs, otherwise for diagrams CLBZ-1 and 7,
  PTBM-1,4,5 and RB-1 a 1-loop diagram exists, no matter the position
  or flow of the Higgs quantum numbers. For the remaining diagrams,
  internal scalars with quantum numbers as the Higgs are allowed only
  if they ``flow out'' of the vertices connecting two fermions, i.e if
  calculating for $\overline{\nu_L^c}m_\nu\nu_L$ the Yukawa coupling
  has the structure: $\overline{F_a}F_bH^{\dagger}$ or
  $\overline{F_c}LH^\dagger$ \footnote{If instead one calculates for
    $\overline{\nu_R}m_\nu^* \nu_R^c$, the ``flow out'' will be
    defined by the Hermitian conjugate of these couplings.}.
  Otherwise also for those remaining diagrams a 1-loop diagram will be
  possible drawing as well.
\item \label{gen-condition-4} And, lastly, in order to guarantee
  absence of 1-loop contributions from {\bf T-3}, if not already
  excluded by the previous three conditions, one needs to check the
  $SU(2)$ quantum numbers of internal scalars. The quartic vertex in
  {\bf T-3} can be generated \cite{Bonnet:2012kz} by attaching a pair
  of Higgses to $S_1 S_2$ with $S_1=S_D$ and $S_2=S_D$ or $S_1=S_S$
  and $S_2=S_T$ or $S_1=S_T$ and $S_2=S_T$, where $S$, $D$ and $T$
  indicate singlet, doublet and triplet under $SU(2)$. If any of these
  combinations appear, the difference in hypercharge of these states
  must be different from $2 Y_H$ in order to forbid {\bf T-3}. For RB
  diagrams, this rule applies for $S_1$ being a scalar in the inner
  loop and $S_2$ a scalar in the outer loop.  Different from all
  previous conditions, this rule has (exactly) one exception, see
  table \ref{tab:ptbm-3-exception}.
\end{enumerate}

\begin{table}[t!]
  \renewcommand{\arraystretch}{1.2}
  \setlength{\tabcolsep}{4pt}
  \centering
  \begin{tabular}{|c||c|c|c|c|c|c|c|}
    \hline
    \multicolumn{8}{|c|}{\bf PTBM-3 model}
    \\\hline
    {\sc Fields}& $F_a$ & $F_b$ & $F_c$ & $S_1$ & $S_2$ & $S_3$ & $S_4$
    \\
    \hline
    $SU(2)_L$ & 2 & (1,3) & 2 & 1 & 2 & 2 & 1
    \\
    $U(1)_Y$ & $q+3$ & 2 & $q+1$ & $q+2$ & $q+3$ & $q+1$ & $q$ 
    \\\hline
  \end{tabular}
  \caption{\it Quantum number assignment for the new particles
    appearing in the PTBM-3 model, which is the one possible exception
    to rule (\ref{gen-condition-4}). Naming conventions for particles 
    as in fig. (\ref{fig:genuine-example-plot}). Strict application of
    rule (\ref{gen-condition-4}), would forbid this model. However,
    for any $q$ different from zero this model has no lower order
    neutrino mass diagram.}
  \label{tab:ptbm-3-exception}
\end{table}

% --------------------------
% section: quantum numbers
% --------------------------
\subsection{SM gauge quantum numbers }
\label{sec:gauge-quantum-numbers}
Due to the large number of diagrams involved, it is desirable to
apply a strategy where the quantum number assignments are done mostly at
the topology level rather than at the diagrammatic level. Since both
the leptons and the Higgs are $SU(2)$ doublets, for these quantum
number this turns out to be trivially possible. However, for
hypercharge ($Y=2(Q-T_3)$) the procedure is more subtle due to the
different hypercharges the lepton and Higgs doublets have.  This
implies that different external lepton-Higgs attachments lead to
different hypercharges for the internal fields.  So, when discussing
hypercharge assignments, in order to avoid a diagrammatic approach we
group the different diagrams according to the different external
lepton-Higgs structures, which once fixed lead to a unique set of
hypercharges for the internal fields. 

\begin{figure}
  \centering
  \includegraphics[scale=0.9]{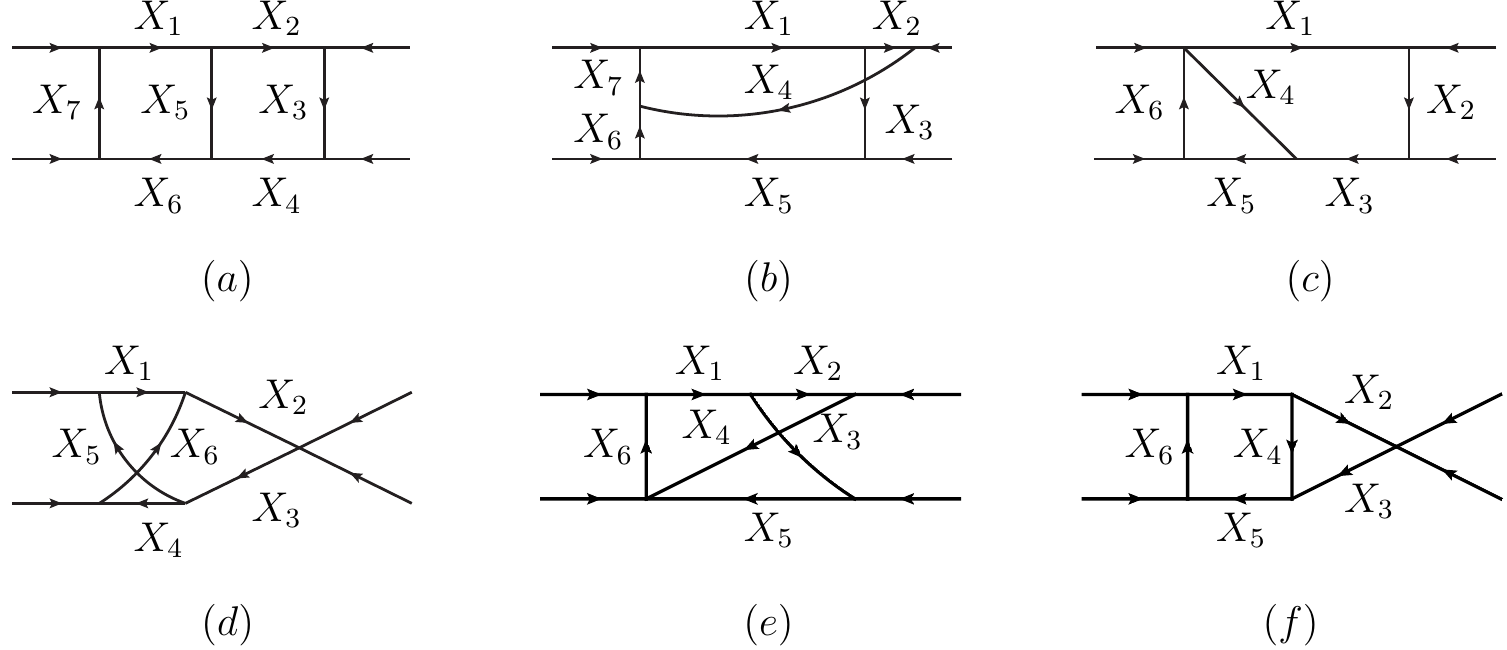}
  \caption{\it Symbolic internal field assignments for the different
    renormalizable and genuine two-loop diagrams in
    figs. \ref{fig:diagrams-1} and \ref{fig:diagrams-2}. $X_i$ holds
    either for fermion or bosons, the specific choice is determined by
    the diagrams in figs.~\ref{fig:diagrams-1} and
    \ref{fig:diagrams-2}.}
  \label{fig:hypercharge-field-assignments}
\end{figure}

For all the relevant topologies we will label the internal fields as
$X_i$ (see fig. \ref{fig:hypercharge-field-assignments}), where $X_i$
can be either a scalar or a fermion depending on the specific diagram
(fig. \ref{fig:diagrams-1} and \ref{fig:diagrams-2}). For the field
$X_i$, no matter whether it is a fermion or a scalar, we will use the
notation $r$ for the $SU(2)$ quantum numbers (with $r$ labelling the
$SU(2)$ representation $r=1,2,3$: singlet, doublet and
triplet). Hypercharge of a given field $X_i$ will be denoted by
$Y_i$. In what follows we discuss quantum number assignments for the
double-box diagrams CLBZ-i and PTBM-i (i=1,2,3) in
fig. \ref{fig:diagrams-1} and RB-j (j=1,2) in
fig.~\ref{fig:diagrams-2}. Results for the remaining diagrams are
summarized in appendix \ref{sec:non-ren-topo-divergent-diag}. For all
possible genuine diagrams we display the possible quantum number
assignments in tables.

\noindent
{\it Quantum numbers for diagrams of type $(a)$
    fig~\ref{fig:hypercharge-field-assignments}:}\\
  We start with $T2_1^\text{B}$-based diagrams, as shown in
  fig~\ref{fig:hypercharge-field-assignments}-$(a)$. $SU(2)$
  invariance of the different vertices imply the following
  constraints:
\begin{alignat}{2}
  \label{eq:external-LR-constraints1}
  X_1\otimes X_2&\supset \bar X_5\ , \qquad&
  X_1\otimes X_7&\supset 2\ ,
  \\
  \label{eq:external-LR-constraints2}
  X_7\otimes X_6&\supset 2\ , \qquad&
  X_2\otimes X_3&\supset 2\ ,
  \\
  \label{eq:external-LR-constraints3}
  X_3\otimes X_4&\supset 2 \ .
\end{alignat}
This means that fixing the representation of $X_{1,2}$ fixes
$X_5$. With $X_1$ fixed $X_7$ is determined too, and this in turn
allows settling $X_6$. With $X_2$ specified, $X_3$ can be determined
as well and this finally fixes $X_4$. The $SU(2)$ assignment ``chain''
is then given by: $X_{1,2}\to X_5$; $X_1\to X_7\to X_6$; $X_2\to
X_3\to X_4$.

The setup of constraints in (\ref{eq:external-LR-constraints1}),
(\ref{eq:external-LR-constraints2}) and
(\ref{eq:external-LR-constraints3}) are summarized in
tab. \ref{tab:double-box-QN}, where in addition to the $SU(2)$
possible quantum number assignments (upper table) we have as well
added a table with the different set of possible hypercharge
assignments (lower table).

% ---------------------------------------------------
% Double-box-based diagrams table (quantum numbers)
% ---------------------------------------------------
\begin{table}[t!]
  \renewcommand{\arraystretch}{1.2}
  \setlength{\tabcolsep}{4pt}
  \centering
  \begin{tabular}{ |c||c|c|c|c|c||c|c|c|c|c||c|c|c|c|c| }
    \hline
    \multicolumn{16}{|c|}{$\boldsymbol{SU(2)}$ {\bf quantum numbers}}\\\hline
    % First row
    \diaghead{\theadfont column wid}
    {$X_{2}$}{$X_{1}$}&\multicolumn{5}{c||}{$1$}
    &\multicolumn{5}{c||}{$2$} &\multicolumn{5}{c|}{$3$}\\
    \hline\hline
    & $X_{5}$&$X_{7}$&$X_{6}$ &$X_{3}$ & $X_{4}$& $X_{5}$&$X_{7}$&$X_{6}$ &$X_{3}$ 
    & $X_{4}$& $X_{5}$&$X_{7}$&$X_{6}$ &$X_{3}$ & $X_{4}$
    \\
    \hline\hline
    % Third row
    \multirow{2}{0.3cm}{$1$}&
    \multirow{2}{0.3cm}{$1$}&
    \multirow{2}{0.3cm}{$2$}&
    1&
    \multirow{2}{0.3cm}{$2$}&
    $1$ & 
    \multirow{2}{0.3cm}{2}&
    1& 
    \multirow{2}{0.3cm}{$2$}&
    \multirow{2}{0.3cm}{$2$}&
    1&
    \multirow{2}{0.3cm}{$3$}&
    \multirow{2}{0.3cm}{$2$}&
    1&
    \multirow{2}{0.3cm}{$2$}&
    1\\\cline{4-4}\cline{6-6}\cline{14-14}\cline{16-16}
    & &  & $3$  & & $3$ & & 3 & & & 3 & & & 3 && 3\\
    \hline
    % Fourth row
    \multirow{2}{0.3cm}{$2$}&
    \multirow{2}{0.3cm}{$2$}&
    \multirow{2}{0.3cm}{$2$}&
    1&
    1&
    \multirow{2}{0.3cm}{$2$}& 
    1&
    1& 
    \multirow{2}{0.3cm}{$2$}&
    1&
    \multirow{2}{0.3cm}{$2$}&
    \multirow{2}{0.3cm}{$2$}&
    \multirow{2}{0.3cm}{$2$}&
    1&
    1&
    \multirow{2}{0.3cm}{$2$}\\
    & &  & 3  & 3  & & 3 & 3 & & 3 & & & & 3 & 3 & \\
    \hline
     % Fifth row
    \multirow{2}{0.3cm}{$3$}&
    \multirow{2}{0.3cm}{$3$}&
    \multirow{2}{0.3cm}{$2$}&
    1&
    \multirow{2}{0.3cm}{$2$}&
    $1$ & 
    \multirow{2}{0.3cm}{2}&
    1& 
    \multirow{2}{0.3cm}{$2$}&
    \multirow{2}{0.3cm}{$2$}&
    1&
    1&
    \multirow{2}{0.3cm}{$2$}&
    1&
    \multirow{2}{0.3cm}{$2$}&
    1\\\cline{4-4}\cline{6-6}\cline{14-14}\cline{16-16}
    & &  & $3$  & & $3$ & & 3 & & & 3 & 3 & & 3 && 3\\
    \hline
  \end{tabular}
  
  \vspace{0.1cm}
  \begin{tabular}{|c|c|c|c|c|c|c|c|}\hline
    \multicolumn{8}{|c|}{\bf Hypercharge}\\\hline
    $S_i$ & $Y_1$ & $Y_2$ & $Y_3$ & $Y_4$ & $Y_5$ & $Y_6$ & $Y_7$
    \\\hline
    % $S_1$ & $-1+\alpha$ & $-1+\beta$ & $\beta$ & $1+\beta$ 
    % & $\alpha-\beta$ & $1+\alpha$ & $\alpha$
    % \\\hline
    $S_1$ & $-1+\alpha$ & $1+\beta$& $\beta$ & $1+\beta$ 
    & $-2+\alpha-\beta$ & $-1+\alpha$ & $\alpha$
    \\\hline
    $S_2$ & $-1+\alpha$ & $-1+\beta$ & $\beta$ & $-1+\beta$
    & $\alpha-\beta$ & $-1+\alpha$ & $\alpha$
    \\\hline
    $S_3$ & $-1+\alpha$ & $-1+\beta$ & $\beta$ & $1+\beta$
    & $\alpha-\beta$ & $1+\alpha$ & $\alpha$
    \\\hline
  \end{tabular}
  \caption{\it Electroweak quantum numbers for diagrams CLBZ-i, PTBM-i
    (i=1,2,3) and RB-j (j=1,2) in fig. \ref{fig:diagrams-1} and
    fig. \ref{fig:diagrams-2}.  Upper table: $SU(2)$ representations.
    Lower table: hypercharge assignments. Fields $X_i$ refer to
    internal fields in the symbolic diagram in fig
    \ref{fig:hypercharge-field-assignments}-$(a)$.  Symbols $S_i$
    refer to allowed external lepton-Higgs structures according to fig
    \ref{fig:bubbles}. Hypercharge of field $X_i$ is denoted by $Y_i$
    (see the text for further details). Since the lepton and Higgs
    doublets are color singlets, color charges can be trivially
    included.}
  \label{tab:double-box-QN}
\end{table}
\begin{figure}
  \centering
  \includegraphics[scale=0.85]{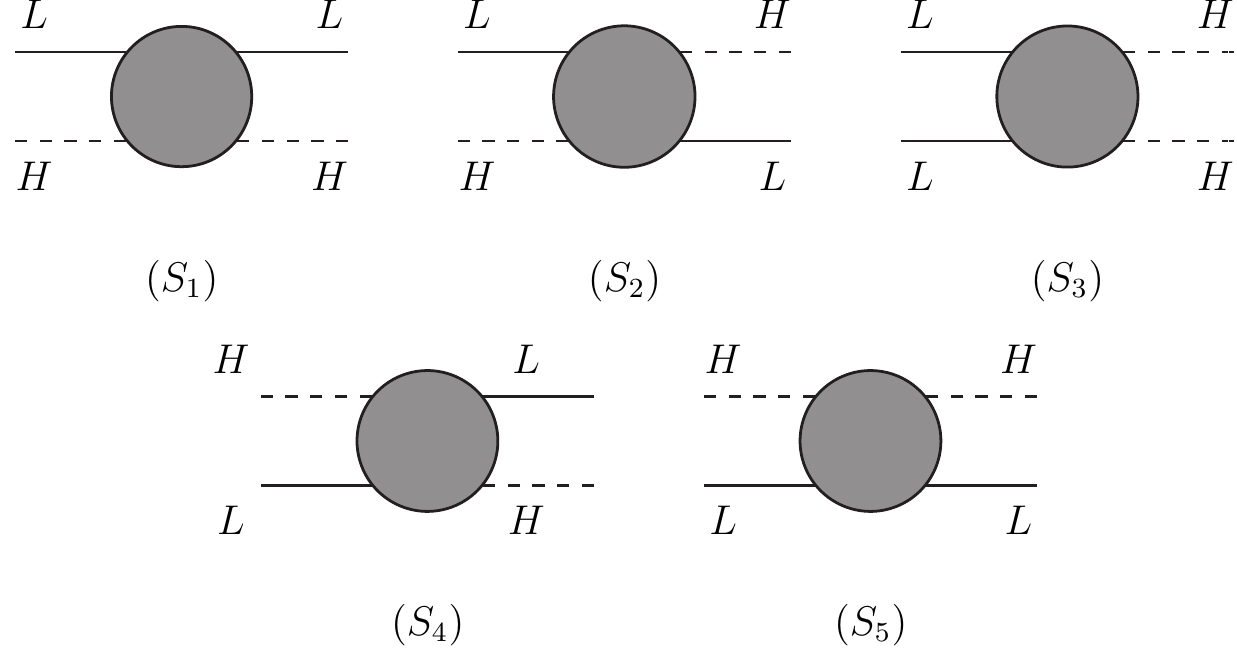}
  \caption{\it Possible external $LH$ structures used to determine the
    internal fields hypercharges.}
  \label{fig:bubbles}
\end{figure}

Some words are in order regarding tab. \ref{tab:double-box-QN}. The
upper table is divided in three subtables delimited by the double
vertical lines. The subtable in the left hand side shows the possible
$SU(2)$ charges of the internal fields for $X_1$ fixed to be a singlet
and for any $X_2$ (singlet, doublet and triplet). The following
subtables give the $SU(2)$ charges for $X_1$ transforming as a doublet
(middle subtable) or as a triplet (right hand side subtable) for any
$X_2$. Note that there is no relation between the choices for $X_1$
and $X_2$, e.g. while $X_1$ can transform as a triplet $X_2$ can do so
as a singlet. For fields which admit two $SU(2)$ charge assignments
within a certain row (see e.g. $X_6$ and $X_4$ for $X_{1,2}\sim 1$ or
$X_6$ and $X_3$ for $X_1\sim 1$ and $X_2\sim 2$ in
table~\ref{tab:double-box-QN}), a horizontal internal line indicates
that no crossed assignments are possible. For example, when
$X_{1,2}\sim 1$, $X_6$ can transform as either a singlet or a
triplet. If fixed to be a singlet (triplet), $X_4$ is fixed univocally
to be a singlet (triplet) too. If instead the horizontal internal line
is absent, crossing is possible. This is indeed the case for $X_6$ and
$X_3$ when $X_1\sim 1$ and $X_2\sim 2$. Fixing $X_6$ to be a singlet
(triplet) allows $X_3$ to be either a singlet or a triplet.

Finally, the lower table shows the different hypercharge assignments
derived by taking hypercharge flow according to fig
\ref{fig:hypercharge-field-assignments}, and the possible lepton-Higgs
external structures $S_1$, $S_2$ and $S_3$, schematically represented
in fig. \ref{fig:bubbles}. Note that since the number of internal
fields exceed the number of hypercharge conservation constraints (one
per each vertex), hypercharge is not univocally fixed. The
arbitrariness is encoded in the parameters $\alpha$ and $\beta$.

We do not give explicitly color quantum numbers in our tables. As
mentioned above, the inclusion of color is straightforward, since
Higgs and lepton doublets are color singlets. This implies that (pairs
of) internal particles coupled to either $L$ or $H$ can come only in
combinations of $1 \otimes 1$, $3\otimes\bar 3$, $6\otimes\bar 6$
etc. Moreover, once the colour quantum numbers for internal particles
coupled to $L$ or $H$ are chosen, the color quantum numbers of the
remaining inner particles are fixed by consistency conditions, derived
from $SU(3)$ rules such as $\bar{3} \otimes \bar{3} = 3_a \oplus
\bar{6}_s$ and $3 \otimes \bar{3} = 1 \oplus 8$.

\subsubsection{Assigning quantum numbers: some examples}
\label{assigning-QN-examples}
We now exemplify the use of these results by constructing a couple of
models. For that purpose we take diagrams CLBZ-1 and
PTBM-1 (see fig~\ref{fig:diagrams-1}):
\begin{itemize}
\item {\it A CLBZ-1-based model}:\\
  Starting with CLBZ-1, and comparing with the symbolic
  diagram in fig~\ref{fig:hypercharge-field-assignments} it can be
  seen that: $X_{1,2,5}\to S_{1,2,5}$ and $X_{3,4,6,7}\to
  F_{3,4,6,7}$. Whether the resulting model involves three (four)
  different scalar (fermion) fields should be determined by their
  transformation properties, for which
  tab. \ref{tab:non-planar-box-based-QN} should be used.

  Sticking to the case $X_{1,2}\sim 1$, one is left with $X_5\sim 1$
  and $X_7\sim 2$. For $X_6$ there are two possible choices, taking
  $X_6\sim 1$ one then has $X_3\sim 2$ and univocally $X_4\sim
  1$. Diagram CLBZ-1 follows a $S_1$ lepton-Higgs structure (see
  fig \ref{fig:bubbles}), so for hypercharge assignments one has to
  focus on the $S_1$ row in
  tab.~\ref{tab:non-planar-box-based-QN}. Fixing $\alpha=-1$ and
  $\beta=1$, one gets $Y_1=-2$, $Y_2=2$, $Y_4=2$, $Y_5=-4$,
  $Y_6=-2$. So, the resulting UV completion consist of: one
  hypercharge $+2$ scalar singlet and its complex conjugate
  ($S_2=S_1^*$), one hypercharge $+4$ scalar singlet ($S_5$), one
  hypercharge $-1$ fermion doublet and its conjugate ($F_3=\bar F_7$),
  and one hypercharge $-2$ fermion singlet and its conjugate
  ($F_4=\bar F_6$). Thus, the fermions can be identified with SM
  lepton doublets and singlets, and so the UV completion constructed
  in this way is nothing else but the CLBZ model
  \cite{Cheng:1980qt,Zee:1985id,Babu:1988ki}. Other quantum number
  choices, as dictated by tab. \ref{tab:double-box-QN}, will of course
  produce variants of the CLBZ model.
% ----------------
% Second example
% ----------------
\item {\it A PTBM-1-based model}:\\
  In this case comparing diagram PTBM-1 with that in
  fig. \ref{fig:hypercharge-field-assignments}-$(a)$ allows the
  identification: $X_{1,4}\to S_{1,4}$ and $X_{2,3,5,6,7}\to
  F_{2,3,5,6,7}$. For the $SU(2)$ charges we fix them as in the
  previous example. For hypercharge one has to bear in mind the
  lepton-Higgs structure, which for this diagram follows $S_2$ (see
  fig~\ref{fig:bubbles}). Thus, fixing $\alpha=\beta=1/3$ one gets
  the following UV completion: a hypercharge $+1/3$ fermion doublet
  and its copy ($F_3=F_7=F$), one hypercharge $-2/3$ fermion singlet
  and its copy ($F_2=F_6=F^\prime$), one vanishing hypercharge fermion
  singlet ($F_5=f$) and one hypercharge $-2/3$ scalar singlet and its
  copy ($S_1=S_4$). Assigning non-trivial color charges to these
  fields: $f\sim 8_c$, $F\sim 3_c$, $F\sim 3_c$ and $S_1=S_4\sim \bar 3$,
  one can then identify $F$ with quark $SU(2)$ doublets while
  $F^\prime$ with quark $SU(2)$ singlets. The resulting model in that
  case then matches the model of Angel et. al
  \cite{Angel:2013hla}. Using tab. \ref{tab:double-box-QN}, further
  variants can be constructed.
\end{itemize}

% ---------------------
% Example section
% ---------------------
\section{Constructing two-loop models}
\label{sec:constructing-model-guide}

\begin{figure}
  \centering
  \includegraphics[scale=0.9]{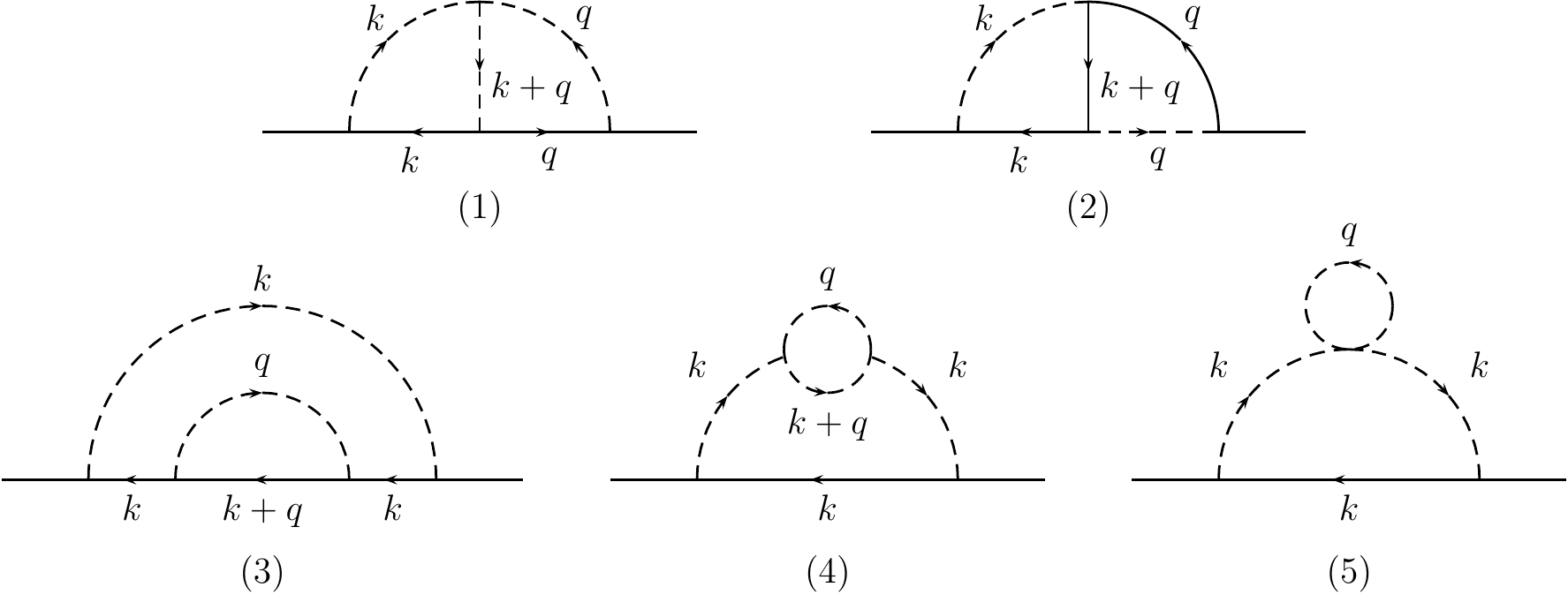}
  \caption{\it Diagrams determining the type of integrals one
    encounters in 2-loop models. Diagrams $(1)$ and $(2)$ in the upper
    row correspond to CLBZ and PTBM like models, whereas diagrams in
    the lower row correspond to RB $(3)$ and ISC ($(4)$ and
    $(5)$) models.  Note that the latter leads only to non-genuine
    2-loop models.}
  \label{fig:integrals}
\end{figure}

Fig. \ref{fig:integrals} shows diagramatically the different classes
of integrals that one encounters in the calculation of two-loop
models. Diagrams $(1)$ to $(3)$ show the case that can correspond to
``genuine'' or ``true 2-loop'' models, while diagrams corresponding to
ISC-i and ISC-ii diagrams (diagrams $(4)$ and $(5)$ respectively)
always correspond to non-genuine models.  We will discuss in the
following two examples, one for genuine models (PTBM-3) and one
non-geniune model (based on NG-RB-1).

\subsection{Genuine 2-loop models}

In genuine 2-loop models of either CLBZ or PTBM type one encounters
two types of integrals:\footnote{In the appendix we
    give also formulas for RB type diagrams.}
\begin{align}
  \label{eq:BZ}
  {\cal I}_{ab,\alpha\beta, X} &= \frac{1}{(2 \pi)^8}
  \int d^4k \int d^4q 
  \frac{1}{(k^2-m_a^2)(k^2-m_{\alpha}^2)(q^2-m_b^2)
    (q^2-m_{\beta}^2)[(q+k)^2-m_X^2]}\ ,
  \\
  \label{eq:PTBM-int}
  {\cal I}_{ab,\alpha\beta, X}^{\{k^2,q^2,(q+k)^2\}} &= \frac{1}{(2 \pi)^8}
  \int d^4k \int d^4q 
  \frac{\{k^2,q^2,(q+k)^2\}}{(k^2-m_a^2)(k^2-m_{\alpha}^2)(q^2-m_b^2)
    (q^2-m_{\beta}^2)[(q+k)^2-m_X^2]}\ .
\end{align}
Here, $\{ k^2,q^2,(q+k)^2\}$ implies that the numerator could be
any of $k^2$, $q^2$ or $(q+k)^2$, depending on the helicity structure
of the underlying Lagrangian, see discussion below. We choose the
convention of labelling the fermion masses as $a, b$ and the scalar
masses as $\alpha,\beta$. $X$ is the inner particle that can be either
a scalar (CLBZ-type) or fermion (PTBM-type).  Note that integral in
(\ref{eq:BZ}) is finite {\it per se}, while a finite result for 
integrals (\ref{eq:PTBM-int}) requires summation over internal mass
eigenstates.

\begin{figure}
  \centering
  \includegraphics[scale=0.9]{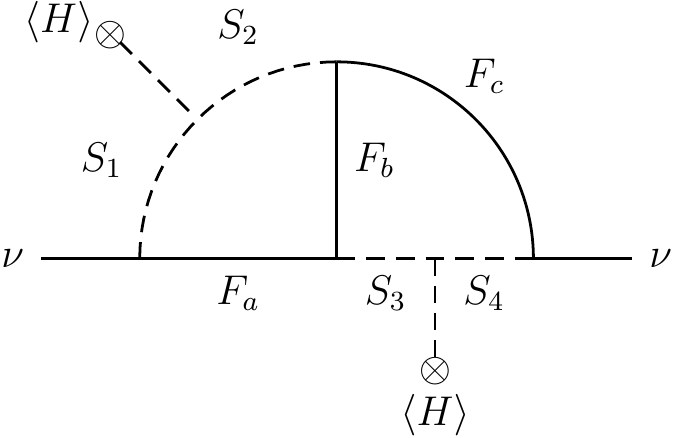}
  \caption{\it Example for a genuine 2-loop model. The diagram
    corresponds to PTBM-3 in fig. \ref{fig:diagrams-1} in
    sec. \ref{sec:genuine-models}.}
  \label{fig:genuine-example-plot}
\end{figure}

Integrals in (\ref{eq:BZ}) and (\ref{eq:PTBM-int}) can be evaluated by
rewriting them in terms of a ``master integral'' (see
eq. (\ref{eq:master-int}) in appendix \ref{sec:formulas-2-loop}).
In order to illustrate the way in which this is done, 
we write down a specific PTBM-3-based model which arises from the
diagram shown in fig.~\ref{fig:genuine-example-plot}. For all 
other possible genuine 2-loop models, the procedure follows 
very closely the one outlined for this particlar example. 

The diagram in fig.~\ref{fig:genuine-example-plot}  is
generated from the following Lagrangian
\begin{equation}
  \label{eq:lag-ptbm-3}
  {\cal L}_\text{int} =
  Y_{ia}\,\left(\overline{L^c_i} P_L S_1 \right)\cdot F^c_a
  +
  Y_{cj}\,\left(\overline{F_c} P_L L_j\right) \cdot S_4
  +
  h_{ab}\,\overline{F^c_a} \cdot \left(F^c_b S_3^\dagger\right)
  +
  h_{bc}\,\left(\overline{F^c_b} F_c\right) \cdot S_2^\dagger
  +
  \mbox{H.c.}\ ,
\end{equation}
and scalar potential terms
\begin{equation}
  \label{eq:scalar-pot-ptbm-3}
  V \supset
  \mu_{34}\,S_4^\dagger \cdot \left(S_3 H \right)
  +
  \mu_{12}\,S_2\cdot \left(S_1^\dagger H\right)
  +
  \mbox{H.c.}
  +
  \sum_{x=1}^4 m_{S_x}^2 \left|S_x\right|^2\ ,
\end{equation}

\begin{table}[t!]
  \renewcommand{\arraystretch}{1.2}
  \setlength{\tabcolsep}{4pt}
  \centering
  \begin{tabular}{|c||c|c|c|c|c|c|c|}
    \hline
    \multicolumn{8}{|c|}{\bf PTBM-3 model}
    \\\hline
    {\sc Fields}& $F_a$ & $F_b$ & $F_c$ & $S_1$ & $S_2$ & $S_3$ & $S_4$
    \\
    \hline
    $SU(2)_L$ & 1 & 2 & 2 & 2 & 1 & 2 & 1
    \\
    $U(1)_Y$ & 1 & 5 & $-4$ & 2 & 1 & $-4$ & $-3$ 
    \\\hline
  \end{tabular}
  \caption{\it Quantum number assignment for the new 
    particles appearing in the diagram shown in 
    fig.~\ref{fig:genuine-example-plot}. For simplicity, 
    all states are assumed to be color singlets.}
  \label{tab:QN-ptbm-3-exmaple}
\end{table}

\noindent
where the parenthesis indicate $SU(2)$ index contractions. The
particle content of the resulting model and its SM transformation
properties are displayed in tab.~\ref{tab:QN-ptbm-3-exmaple}. In 
addition, the fermions can have vectorlike mass terms, namely: 
\begin{equation}
  \label{eq:vectorlike-mass-terms-ptbm-3-exam}
  {\cal L}_M = \sum_{A=a,b,c} m_{F_A}\overline{F_A}F_A\ .
\end{equation}

Coupling $\mu_{34}$ in (\ref{eq:scalar-pot-ptbm-3}) induces mixing
between the $Q=3/2$ scalars, while $\mu_{12}$ mixing between the
$Q=1/2$ states. The mass matrices for these states then reads
\begin{equation}
  \label{eq:scalar-mass-matrix-ptbm-3-EX}
  M_{S^{Q=3/2}}^2=
  \begin{pmatrix}
    m_{S_3}^2   & \mu_{34}v\\
    \mu_{34}v & m_{S_4}^2
  \end{pmatrix}\ ,
  \qquad
  M_{S^{Q=1/2}}^2=
  \begin{pmatrix}
    m_{S_1}^2   & \mu_{12}v\\
    \mu_{12}v   & m_{S_2}^2
  \end{pmatrix}\ .
\end{equation}
Assuming the mixing parameters to be real, these matrices are
diagonalized by $2\times 2$ rotation matrices:
\begin{equation}
  \label{eq:rot-mat-ptbm-3}
  R_{Q}=
  \begin{pmatrix}
    \cos\theta_Q & \sin \theta_Q\\
    -\sin\theta_Q & \cos\theta_Q
  \end{pmatrix}\ ,
\end{equation}
with the rotation angles given by:
\begin{equation}
  \label{eq:roatation-angles-ptbm-3}
  \tan 2\theta_{Q=3/2}=\frac{2\mu_{34}v}{m_{S_3}^2 - m_{S_4}^2}\ ,
  \qquad
  \tan 2\theta_{Q=1/2}=\frac{2\mu_{12}v}{m_{S_1}^2 - m_{S_2}^2}\ .
\end{equation}
Rotating the interactions in (\ref{eq:lag-ptbm-3}) and
(\ref{eq:scalar-pot-ptbm-3}) to the scalar mass eigenstate basis, one
can then calculate the full neutrino mass matrix.

The chiral structures appearing in the diagram are 
determined by the different chiral projectors ($P_L$ or $P_R$)
entering in each of the Yukawa vertices involved. Since chirality of
external vertices (those where the SM neutrinos enter) is fixed by the
neutrino chirality, the number of possibilities is determined by the
different chiral structures of internal Yukawa vertices. For PTBM models
there are three chiral structures: internal vertices with $P_L-P_L$,
$P_R-P_L$ or $P_R-P_R$ stuctures. The (internal) combination $P_L-P_L$ 
leads to integrals of type eq. (\ref{eq:BZ}), while the other two 
possibilities project out integrals of type eq. (\ref{eq:PTBM-int}). 
The full final result reads:
\begin{align}
  \label{eq:neutrino-mm-ptbm-3}
  {\cal M}_\nu=\frac{1}{4(16\pi^2)^2}\left(Y_{ia}Y_{cj} + Y_{ja}Y_{ci}\right)
  h_{ab}h_{bc}\sin 2\theta_{Q=3/2}\sin 2\theta_{Q=1/2}
  \sum_{A=1}^4 \sum_{\alpha,\beta}(-1)^\alpha(-1)^\beta F_{ac,\alpha\beta,b}^{(A)}\ ,
\end{align}
with the different dimensionful functions $F_{ac,\alpha\beta,b}^{(A)}$,
determined by
\begin{align}
  \label{eq:functions-F-a-b-alpha-beta-PTBM-3-Eq1}
  F^{(1)}_{ab,\alpha\beta,b}&=\frac{m_{F_a}m_{F_c}}{m_{F_b}}\,
  \times\pi^{-4}\,\hat {\cal I}_{ac,\alpha\beta}\ ,
  \\
  \label{eq:functions-F-a-b-alpha-beta-PTBM-3-Eq2}
  F^{(2)}_{ab,\alpha\beta,b}&=\left(m_{F_a} + m_{F_b} + m_{F_c}\right)\,
  \times\pi^{-4}\,\hat {\cal I}_{ac,\alpha\beta}^{[(k+q)^2]}\ ,
  \\
  \label{eq:functions-F-a-b-alpha-beta-PTBM-3-Eq3}
  F^{(3)}_{ac,\alpha\beta,b}&=-(m_{F_a} + m_{F_b})\,
  \times\pi^{-4}\,\hat {\cal I}_{ac,\alpha\beta}^{(k^2)}\ ,
  \\
  \label{eq:functions-F-a-b-alpha-beta-PTBM-3-Eq4}
  F^{(4)}_{ac,\alpha\beta,b}&=-(m_{F_b} + m_{F_c})\,
  \times\pi^{-4}\,\hat {\cal I}_{ac,\alpha\beta}^{(q^2)}\ ,
\end{align}
With the aid of
eqs. (\ref{eq:BZVolkas})-(\ref{eq:int-prop-to-(q+k)Sq}) in appendix
\ref{sec:formulas-2-loop}, one can then express these functions in
terms of the ``master'' function $\hat g(s,t)$ (see
eq. (\ref{eq:ghat})).

\begin{figure}[t]
  \centering
  \includegraphics[scale=0.65]{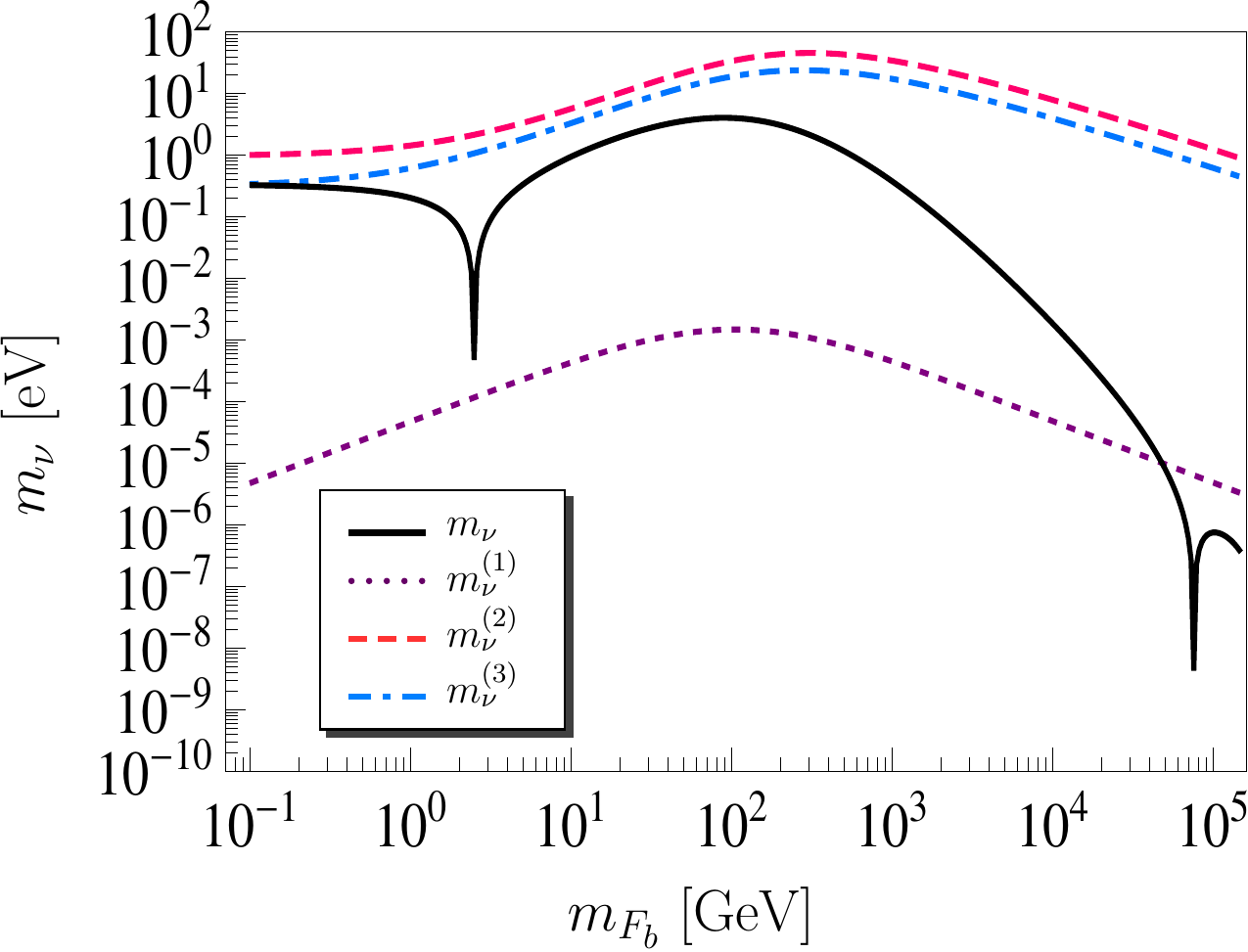}
  \hfill
  \includegraphics[scale=0.65]{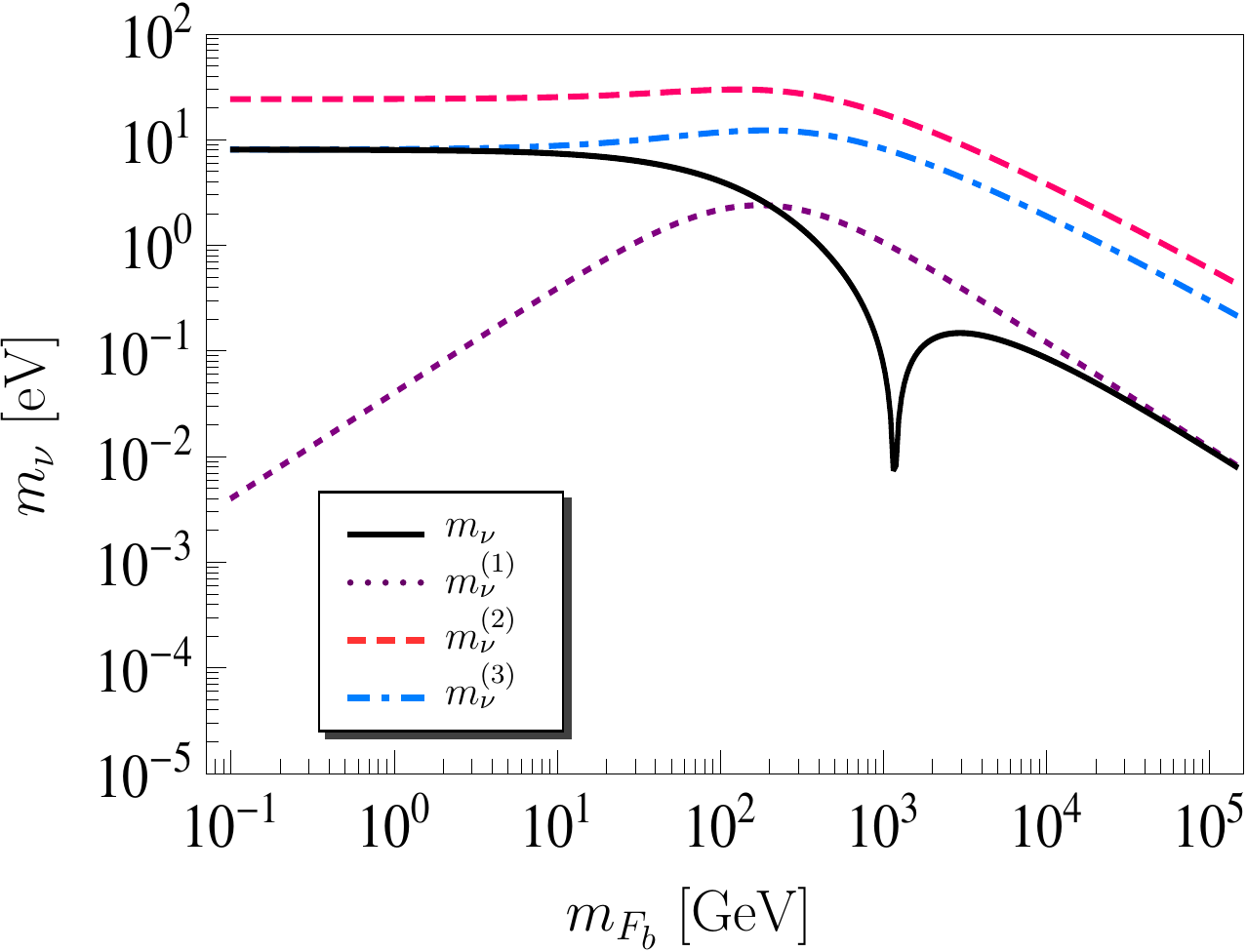}
  \includegraphics[scale=0.65]{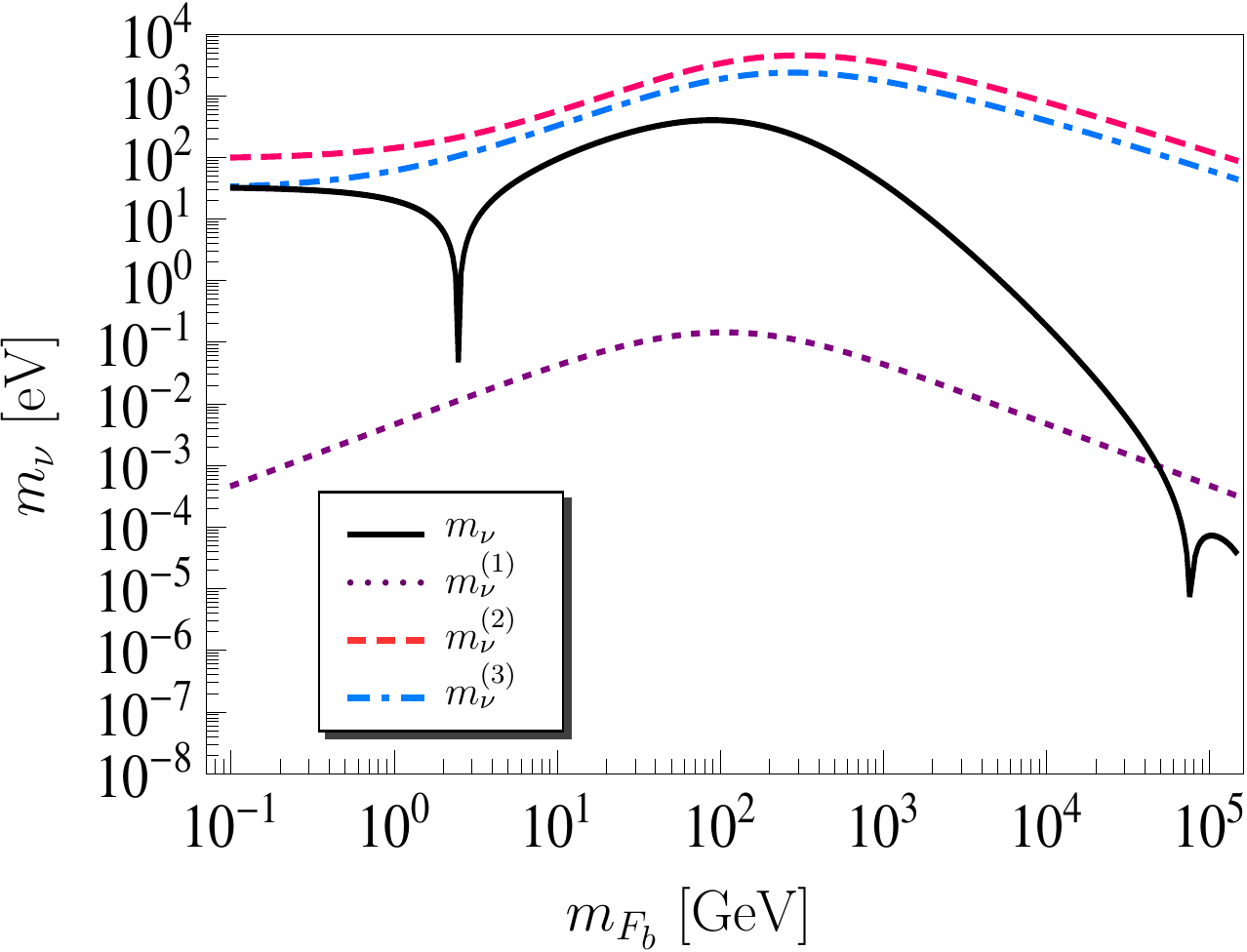}
  \hfill
  \includegraphics[scale=0.65]{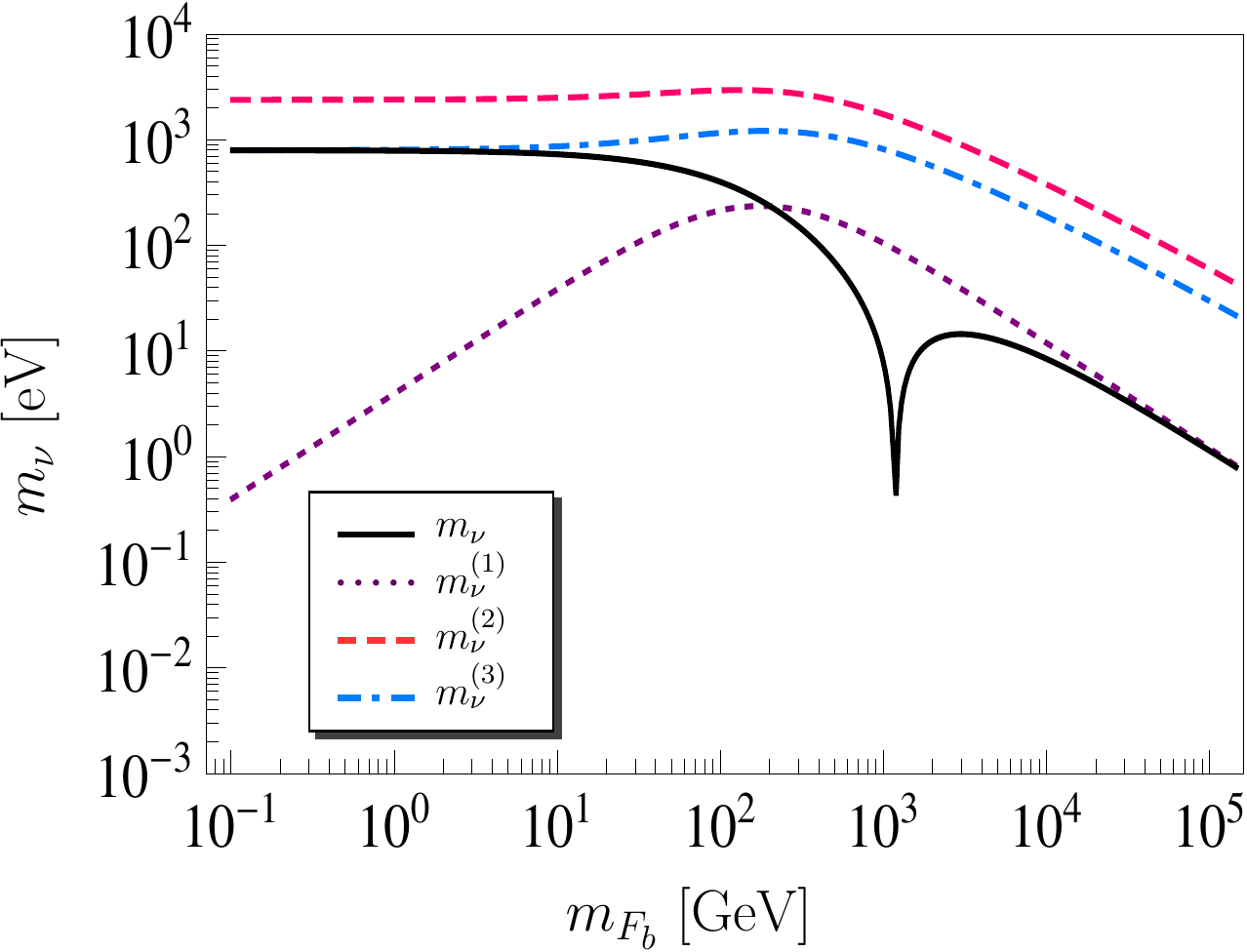}
  \caption{\it Examples of calculated neutrino mass as function of
    $m_{F_b}$ for four different sets of input parameters, see
    text. The full line shows the total $m_{\nu}$, the other lines the
    individual contributions $m_\nu^{(i)}$, determined by the
    functions $F^{(i)}_{ab,\alpha\beta,b}$ (see eqs
    (\ref{eq:functions-F-a-b-alpha-beta-PTBM-3-Eq1})-(\ref{eq:functions-F-a-b-alpha-beta-PTBM-3-Eq3}))
    and the common global factor in (\ref{eq:neutrino-mm-ptbm-3}). The
    plots are for Yukawa couplings equal to $1$ and thus show that
    neutrino masses of the correct order of magnitude can be obtained
    easily in this model.}
  \label{fig:integral-plot}
\end{figure}

Fig. \ref{fig:integral-plot} shows some examples of calculated
neutrino masses for different choices of input parameters. This
calculation does not take into account any flavour structure in the
indices of Yukawa couplings, i.e.  $Y_{ia}=Y_a$ etc, and puts the
values of all Yukawas $Y_a=Y_{c}=h_{ab}=h_{bc}=1$. The numerical
values of $m_{\nu}$ should therefore be understood as the typical
scale of neutrino mass and not as an exact prediction for the three
light neutrino mass eigenvalues, see the discussion on flavour fits
below. Also, note, that while the numerical values shown for $m_{\nu}$
are a bit too large compared to, say, the atmospheric neutrino scale,
$\sqrt{\Delta(m^2_{\rm Atm})} \simeq 0.05$ eV, this could be easily
adjusted for using smaller values for the Yukawas.

The plots then show $m_{\nu}$ as a function of $m_{F_b}$ for scalar
mass parameters $m_{S_1}^2 = 100^2$ GeV$^2$ and $m_{S_2}^2= m_{S_1}^2
+ \Delta m^2$ (with $\Delta m^2 =\mu\,v$), for two different, fixed
$\Delta m^2 =24.6$ GeV$^2$ (upper row) and 246 GeV$^2$ (lower row) and
two different values of $m_F=m_{F_a}=m_{F_c}$: to the left 1 GeV and
to the right 100 GeV. The black (full) line shows the total $m_{\nu}$,
the other lines show the different contributions $m^{(i)}_\nu$,
$i=1,2,3$ individually (determined by the functions
in~(\ref{eq:functions-F-a-b-alpha-beta-PTBM-3-Eq1})-(\ref{eq:functions-F-a-b-alpha-beta-PTBM-3-Eq3})
and the common global factor in (\ref{eq:neutrino-mm-ptbm-3})). Note,
that $m^{(4)}_\nu$ is numerically equal to $m^{(3)}_\nu$, while
$m^{(2)}_\nu < 0$ and we plot the absolute value. Usually the
contribution from $m^{(2)}_\nu-m^{(3)}_\nu$ dominates the neutrino
mass for small and moderate values of $m_{F_b}$, but at large values
of $m_{F_b}$, $m_\nu^{(2)}$ and $m_\nu^{(3)}+m_\nu^{(4)}$ tend to
cancel each other, such that the only remaining contribution comes
from $m_\nu^{(1)}$. In the plots there are some points for $m_{F_b}$,
for which the different contributions can actually exactly cancel each
other. Note also, that for $m_{F_b} \to \infty$, $m_{\nu}$ goes to
zero, as expected. Obviously, as these plots demonstrate, neutrino
masses of the correct order of magnitude can be achieved for a wide
range of input parameters.

We close this section with a brief discussion on neutrino flavour 
fits. Any model, aiming at explaining neutrino oscillation data, 
must of course not only reproduce the overall neutrino mass scale, 
but also have sufficient freedom to fit the two neutrino mass squared 
differences and the three neutrino angles. Our numerical examples 
have been done with only one non-zero neutrino mass, fits to all 
data can nevertheless easily be done. The actual form of the fit, 
however, depends on the number of copies of new fermions and scalars 
present in the model under consideration. For exactly one copy of new 
states, both fermions and scalar, eq. (\ref{eq:neutrino-mm-ptbm-3}) 
has rank-2. This implies that one can fit hierarchical neutrino 
spectra (both normal and inverted), but not degenerate neutrinos. 
With more copies of scalars or fermions, also degenerate neutrinos 
can be fitted. In this case, the simplest way to proceed is via 
a fit analogous to the Casas-Ibarra parametrization for the 
seesaw (type-I/III) \cite{Casas:2001sr}. The authors of 
\cite{Angel:2013hla} have spelled out this procedure for two copies 
of internal scalars and one vector of Yukawas, i.e. their case is 
also rank-2. One can devise in a completely analogous way the fit
for three-fermion or there-scalar models, simply adapting the 
formulas from  \cite{Casas:2001sr}, so we will not discuss this 
in further detail here.

\subsection{Non-genuine but finite 2-loop models}

As discussed at length in the previous sections, some CLBZ, PTBM or RB
diagrams will not correspond to genuine 2-loop models.  However,
models generating such kind of diagrams might still be interesting
constructions in the following sense: Consider, for example, a model
with some new fermions in which, invoking a non-Abelian flavor
symmetry, the direct coupling of the new fermions with the standard
model Higgs is forbidden. The flavour symmetry is then broken at some
large, unspecified scale and upon integrating out some heavy fields,
an effective fermion-fermion-Higgs vertex is generated at 1-loop
order. Such a construction would allow to understand, at least in
principle, why that particular coupling is small compared to all
others, simply due to the $1/(16\pi^2)$ suppression from the loop. An
example of this approach is the $d=7$ RB model of
\cite{Kajiyama:2013zla}, but the very same idea could, of course, be
applied to any of our non-genuine $d=5$ diagrams.

\begin{figure}[t]
  \centering
  \includegraphics[scale=0.8]{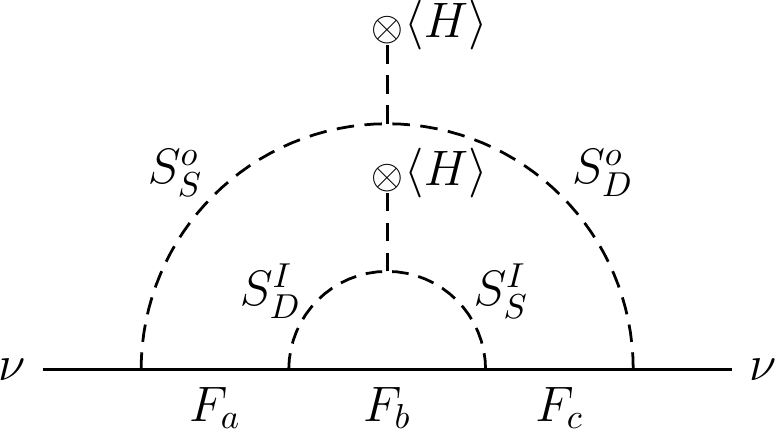}
  \caption{\it Example for a ``finite'' RB diagram. The diagram
    corresponds to NG-RB-1 in fig.~\ref{fig:non-genuine-diagrams-ii}
    in appendix \ref{sec:non-ren-topo-divergent-diag}.}
  \label{fig:RBfin}
\end{figure}

In all such cases one can carry out the calculation either by solving
the full 2-loop integral or by first calculating that particular
vertex at 1-loop order and then doing a 1-loop calculation for the
neutrino mass using this effective vertex in the second step. We will
call the former the ``full'' or ``2-loop'' calculation, while we call
the second approach ``2-step'' calculation in the following.  The two
calculations should, of course, lead to the same numerical result
(only) in the limit where there is a certain hierarchy of masses for
the particles in the loop. In this subsection, we will discuss one
particular example of such a model, based on the RB diagram
NG-RB-1, see fig. \ref{fig:RBfin}, in some detail. The treatment of
all other ``finite'' but non-genuine models is very similar.  Here, we
are mainly interested in demonstrating the numerical agreement between
full and 2-step calculation and therefore will not work out the
details of a suitable flavor symmetry model. We again refer to
\cite{Kajiyama:2013zla} for an example for the ($d=7$) RB diagram
based on the discrete symmetry $T_7$ and to \cite{Aoki:2013gzs} for
another example based on a variant of the ISC-i diagram using the
symmetry $\mathbb{Z}_2\times \mathbb{Z}_2'$.

The diagram in fig. \ref{fig:RBfin} can be generated from the
following interaction Yukawa Lagrangian:
\begin{eqnarray}
  \label{eq:LagRBi}
  {\cal L}_\text{int} = Y_{ia} ({\bar L_i^c}P_L F_a^c).S^{S}_O 
  + Y_{ci}  {\bar F_c}.(S^{D,\dagger}_OP_L L_i)
  + h_{ab} ({\bar F_a^c}S_I^{D,\dagger})P_R F_b 
  + h_{bc}  {\bar F_b^c}P_R F_c  S_I^{S}+ \mbox{H.c.}\ .
\end{eqnarray}
This fixes the SM quantum numbers as shown in table \ref{tab:QNEXa}.
The scalars appearing in the inner and outer loops, denoted by $S^I$
and $S^O$ respectively, have the same SM quantum numbers and thus
could be the same particles. For generality, however, and since in the
ultra-violet completion they could transform differently under the
flavour group, we will treat them as independent states.

\begin{table}[t!]
  \renewcommand{\arraystretch}{1.2}
  \setlength{\tabcolsep}{4pt}
  \centering
  \begin{tabular}{|c||c|c|c|c|c|c|c|}
    \hline
    \multicolumn{8}{|c|}{\bf NG-RB-1}
    \\\hline
    {\sc Fields}& $F_a$ & $F_b$ & $F_c$ & $ S^{O}_S$ 
    & $ S^{O}_D$ & $ S^{I}_S$ & $ S^{I}_D$
    \\
    \hline
    $SU(2)_L$ & 2 & 1 & 1 & 1 & 2 & 1 & 2
    \\
    $U(1)_Y$ & 1 & 0 & $-2$ & 2 & 1 & $-2$ & 1 
    \\\hline
  \end{tabular}
  \caption{\it Quantum number assignment for new particles appearing
    in the diagram shown in fig. \ref{fig:RBfin}. All states are
    assumed to be color singlets for simplicity.}
  \label{tab:QNEXa}
\end{table}

The scalar potential of the model contains the following terms:
\begin{eqnarray}
  \label{eq:LagRBm}
  V \supset 
  \mu_O (S^{O}_DH).S^{O,\dagger}_S
  + \mu_I (S^{I}_DH).S^{I}_S + \mbox{H.c.} 
  + \sum_{\substack{x=D,S\\y=O,I}} (m_{x}^{y})^2 |S_{x}^{y}|^2\ .
\end{eqnarray}
We add then the following three fermion 
mass terms:
\begin{eqnarray}\label{eq:LagRBFm}
{\cal L}_{M} =   m_{F_a} F_a{\bar F_a} +  m_{F_b} F_b{\bar F_b^c} 
            +  m_{F_c} F_c{\bar F_c}\ .
\end{eqnarray}
Only $F_b$ can have a Majorana mass, as indicated by the charge
conjugation ``$C$'' in eq.~(\ref{eq:LagRBFm}), but $F_a$ and $F_c$ can
have vector-like fermion mass terms \footnote{ We note that, the
  presence of the Majorana fermion, $F_b$, together with the scalar
  $S^D$ allows, in principle, to construct a 1-loop diagram for the
  neutrino mass, once the coefficient of $\lambda_5 (S^{D,\dagger}H)
  (S^{D,\dagger}H)$ is non-zero \cite{Ma:2006km}. This coupling must
  be forbidden by some symmetry in order to make the diagram NG-RB-1
  the leading contribution to the neutrino mass.}.

\noindent
The mass matrices for the inner and outer scalars can be written as
\begin{equation}
  \label{eq:ScMass}
  M_{S^k}^{2} = \left(
    \begin{array}{cc} 
      m_{D^k}^2  & \mu_k v \\
      \mu_k v & m_{S^k}^2
    \end{array}\right),
\end{equation}
with $k=I,O$. This matrix can be diagonalized, as in the example in
the previous section, by a simple rotation matrix with the angle given
as:
\begin{equation}\label{eq:tanth}
\tan(2\theta_k) = \frac{2\mu_k v}{m_{D^k}^2 - m_{S^k}^2}\ .
\end{equation}
The expression for the neutrino mass is given as
\begin{equation}
  \label{eq:RB2lp}
  (\Delta m_{\nu})_{ij} = \frac{1}{4(16 \pi^2)^2}
  (Y_{ia} Y_{jc}+Y_{ja} Y_{ic}) h_{ab}h_{bc}\sin(2\theta_O)\sin(2\theta_I)m_{F_b}
  \times \sum_{\alpha,\beta=1,2}(-1)^\alpha(-1)^\beta\pi^{-4}
  \hat{\cal I}^{(k^2)\text{RB}}_{ac,\alpha\beta, b}\ ,
\end{equation}
with $\hat{\cal I}_{ac,\alpha\beta}^{(k^2)\text{RB}}$ given by
eq. (\ref{eq:RB-int-num-kSq}) in appendix~\ref{sec:formulas-2-loop}.
% {\bf\color{red} The following two equations should be moved to the
%   appendix and only appear here cited:}

% with the two-loop integral defined as in (\ref{eq:PTBM-int}):
% \begin{eqnarray}\label{eq:IntRB}
%   {\cal I}^{RB,k^2}_{ac,\alpha\beta, b} = 
%   \int d^4k \int d^4q 
%   \frac{k^2}{(k^2-m_{F_a}^2)(k^2-m_{S_\alpha}^2)(k^2-m_{F_c}^2)
%     (q^2-m_{S_\beta}^2)[(q+k)^2-m_{F_b}^2]}\ . 
% \end{eqnarray}
% Rescaling the momenta we rewrite this integral in terms of
% dimensionless mass squared ratios. Using the formulas in appendix
% \ref{sec:formulas-2-loop} one finds:
% \begin{align}
%   \label{eq:SolRBk2}
%   {\hat {\cal I}}^{RB,k^2}_{ac,xy} &=
%   \frac{1}{r_a-t_y}\times
%   \left\{
%     [{\hat g}(r_a,t_x) - {\hat g}(t_y,t_x)] 
%   \right.
%   \\ \nonumber
%   &\left.
%     \quad
%     + \frac{r_c}{r_a-r_c}[{\hat g}(r_a,t_x)-{\hat g}(r_c,t_x)]
%     - \frac{r_c}{t_y-r_c}[{\hat g}(t_y,t_x)-{\hat g}(r_c,t_x)]
%   \right\}\ .
% \end{align}
% As before, $r_i = (m_{F_i}/m_{F_b})^2$ and $t_i=(m_{S_i}/m_{F_b})^2$. 

\noindent
Now consider the 2-step calculation. The inner loop can be evaluated 
as
\begin{equation}\label{eq:DelM}
\Delta_m = \frac{1}{2}h_{ab}h_{bc}m_{F_b}\sin(2\theta_I){\cal I}_{t_{x_1},t_{x_2}}\ ,
\end{equation}
with\footnote{${\cal I}_{t_{x_1},t_{x_2}}$ is essentially the difference of two 
Passarino-Veltman $B_0(0,s,t)$ functions, see appendix. We prefer to 
write it this way to make the contact with the notation in \cite{Ma:2006km}.}
\begin{equation}\label{eq:Iscot}
{\cal I}_{t_{x_1},t_{x_2}} = \frac{1}{(2\pi)^4} \int d^4q 
        \frac{1}{(q^2-t_{x_1})(q^2-t_{x_2})(q^2-1)} .
\end{equation}
The solution to eq. (\ref{eq:Iscot}) gives the well-known function
\begin{equation}\label{eq:Iscotsol}
{\cal I}_{t_{x_1},t_{x_2}} = \frac{i}{16\pi^2}
     \times \Big\{ \frac{t_{x_2}}{ t_{x_2}-1} \ln(t_{x_2}) -
            \frac{t_{x_1}}{ t_{x_1}-1} \ln(t_{x_1}) \Big\}.
\end{equation}
$\Delta_m$ gives an entry to the mass matrix of the fermions $F_a$ and 
$F_c$:
\begin{equation}\label{eq:mF}
{\cal M}_{F_aF_c} =  
         \left(
               \begin{array}{cc}
                 m_{F_a} & \Delta_m \\
                \Delta_m & m_{F_c} \\
               \end{array}
         \right).
\end{equation}
Diagonalizing this mass matrix gives two eigenvalues $m_{F_1}$ 
and $m_{F_2}$, which can be used in the calculation of the outer 
loop, which has the same form than the inner loop just calculated. 
This results in: 
\begin{equation}\label{eq:DelM2step}
(\Delta m_{\nu})_{ij}^{\rm 2-step} =  \frac{1}{2} 
(Y_{ia} Y_{jc}+Y_{ja} Y_{ic}) h_{ab}h_{bc}
\sum_{\alpha=1,2} m_{F_\alpha}V_{\alpha 1}V_{\alpha 2} 
\sin(2\theta_O){\cal I}_{t_{y_1},t_{y_2}},
\end{equation}
where $V_{ij}$ is the matrix which diagonalizes eq. (\ref{eq:mF}). 

\begin{figure}
  \centering
  \includegraphics[scale=0.75]{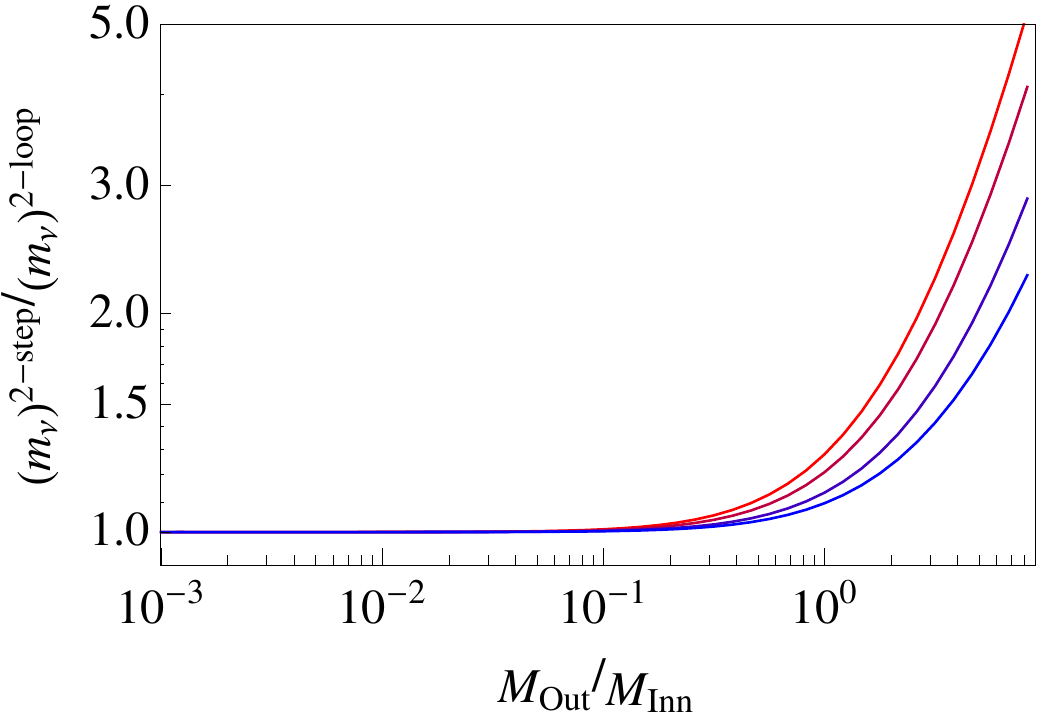}
  \includegraphics[scale=0.75]{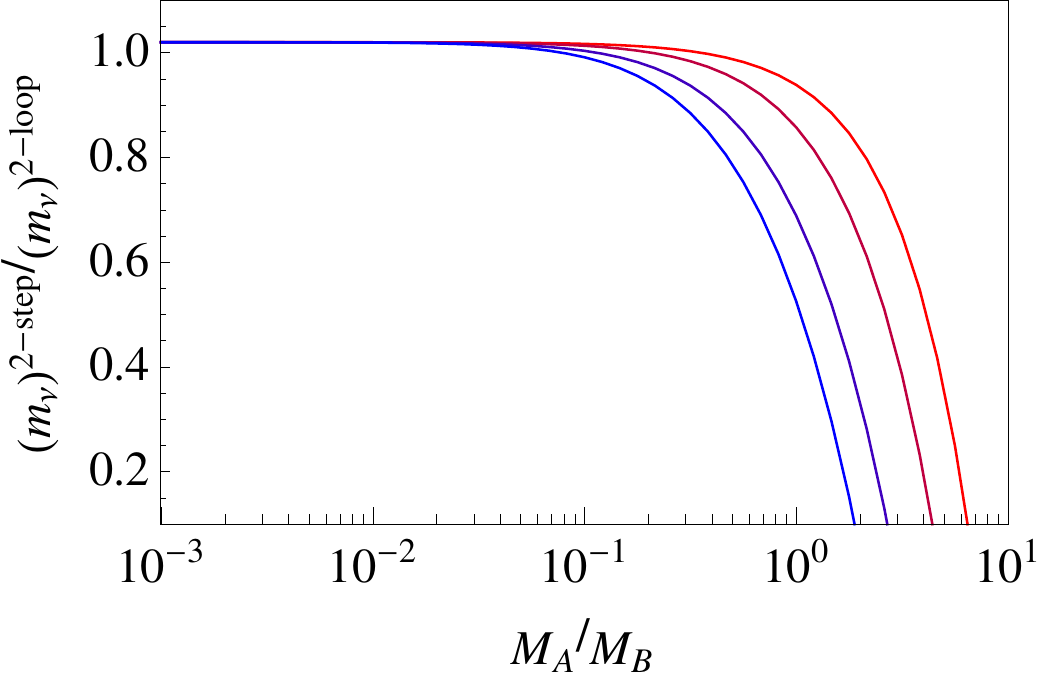}
  \caption{\it Comparison of the full 2-loop calculation to the
    approximate ``2-step'' calculation. The plots show the ratio of
    the approximate calculation to the full calculation. To the left:
    Outer fermions $F_a$ and $F_c$ have negligibly small masses,
    $m_{S^O_D}=m_{S^O_S}=M_{Out}$ and $m_{F_b}=m_{S^I_S}=M_{Inn}$. The
    four different lines are (from top to bottom)
    $m_{S^I_D}/M_{Inn}=1,2,5$ and $10$. To the right
    $m_{S^O_D}=m_{S^O_S}=M_{Out}$ and $m_{S^I_S}=m_{S^I_D}=10
    M_{Out}$, as a function of $m_{F_a}/m_{F_b}$ for four different
    values (from top to bottom) of $m_{F_c}/m_{F_a}=0.1, 0.2,0.5$ and
    $1$. }
  \label{fig:RBNum}
\end{figure}

Fig. \ref{fig:RBNum} shows a comparison of the neutrino mass
calculated with eq. (\ref{eq:RB2lp}) (full 2-loop result) and
eq. (\ref{eq:DelM2step}) (2-step result) for different combinations of
internal masses. We show the ratio of the two calculations, thus all
coupling constants cancel and need not to be specified. The
calculation is for one neutrino mass only and not meant to be a
complete fit to all neutrino data. The plot on the left shows the
result for negligibly small outer fermion masses, varying the (common)
mass of the inner-loop particles, keeping the masses of the outer
scalars constant. The plot on the right shows the result for fixed
values of the scalar masses, but varying the ratio of inner to outer
fermion mass. In both cases, it is clear that if there is a hierarchy
in the masses of the particles in the inner loop with respect to the
masses in the outer loop, then the two calculations agree very
well. Comparison of the plot on the right to the plot on the left
demonstrates that especially the value of the ratio of the fermion
masses is important: Fermion $F_b$ should be heavier than the outer
fermions, otherwise the 2-step calculation starts to fail.

We close of this discussion with one more
comment. Eq. (\ref{eq:LagRBi}) specifies that the vertex connecting
$F_a$ and $F_b$ has a projector $P_R$. However, we have given a
vector-like mass term to these fermions and vector-like fermions can
couple, in prinicple, with both chiralities. A model, in which the
other projector $P_L$ also appears, however, will produce terms
proportional to $(m_{F_a}+m_{F_c})\,q.k$ and, different from the case
discussed here, can not be calculated in the ``2-step'' method
described here, which relies on picking the internal fermion mass
$m_{F_b}$.

% -------------
% Conclusions
% -------------
\section{Conclusions}
\label{sec:conclusions}
Using a diagrammatic approach we have systematically studied the $d=5$
Weinberg operator at the 2-loop order. Out of the large number of
possible diagrams the majority are just corrections to lower order
diagrams. We have shown that the relevant 2-loop models can be
classified as follows: (A) Class-I models, which only involve genuine
diagrams, i.e. diagrams for which the absence of lower order diagrams
is assured. Interestingly, we have found that class-I models implicate
only variants of the CLBZ (Cheng-Li-Babu-Zee) models
\cite{Cheng:1980qt,Zee:1985id,Babu:1988ki}, PTBM
(Petcov-Toshev-Babu-Ma) models \cite{Petcov:1984nz,Babu:1988ig} and RB
(rainbow) models. (B) Class-II models, which involve non-genuine but
finite 2-loop diagrams. Diagrams belonging to this class correspond to
1-loop diagrams that contain a 1-loop generated vertex, and are
variations of just five different diagrams which we have dubbed:
NG-CLBZ (non-genuine CLBZ), NG-PTBM (non-genuine PTBM), NG-RB
(non-genuine RB), ISC-i and ISC-ii (internal scalar correction type i
and ii).

We provided the full list of class-I diagrams in
fig.~\ref{fig:diagrams-1} and \ref{fig:diagrams-2}. This list combined
with our results for the internal fields SM quantum number assignments
(summarized in tabs.~\ref{tab:double-box-QN},
\ref{tab:non-planar-box-based-QN}-\ref{tab:RB-4-table}) , allows the
construction of novel 2-loop neutrino mass models, something that we
have exemplified in sec.~\ref{assigning-QN-examples} and, in more
details in section~\ref{sec:constructing-model-guide}. We have given
as well the full list of non-genuine but finite 2-loop diagrams in
figs.~\ref{fig:non-genuine-diagrams-i}-\ref{fig:non-genuine-diagrams-3}.
This list enables the construction of novel 2-loop models where the
smallness of certain coupling can be, in principle, understood as due
to its 1-loop radiative origin. Also, the ``tools'' needed for
numerical calculations have been collected in
appendix~\ref{sec:formulas-2-loop}.

In summary, we have identified the possible 2-loop neutrino mass
models arising from the $d=5$ Weinberg operator. Our findings can be
understood as a guide for the construction of 2-loop neutrino mass
models, which arguably might serve for several purposes, e.g:
systematic study of neutrino mass model signals at the LHC (testing
the origin of neutrino masses at the LHC, as has been pointed out at
the 1-loop level e.g. in
\cite{AristizabalSierra:2007nf,Sierra:2008wj,FileviezPerez:2009ud},
and at the 2-loop level in \cite{Cai:2014kra}) or systematic
construction of common frameworks for neutrino masses and dark matter
(in the same vein it has been done for the 1-loop case
\cite{Restrepo:2013aga}).

%\newpage
\section*{Acknowledgments}
\addcontentsline{toc}{section}{Acknowledgements} We thank Renato
Fonseca for discussions and for pointing out an inconsistency in the
first version of the draft.  DAS wants to thank Yasaman Farzan for
conversations and useful comments. DAS is supported by a “Charg\`e de
Recherches” contract funded by the Belgian FNRS agency.  MH
acknowldeges support by Spanish MICINN grants FPA2011-22975, MULTIDARK
CSD2009-00064, the Generalitat Valenciana grant PrometeoII/2014/084
and FCT grant EXPL/FIS-NUC/0460/2013.

\clearpage
% ----------------
% Appendices
% ----------------
\appendix                       %
\section{Non-renormalizable topologies and finite non-genuine
  diagrams}
\label{sec:non-ren-topo-divergent-diag}
\begin{figure}[h!]
  \centering
  \includegraphics[scale=1]{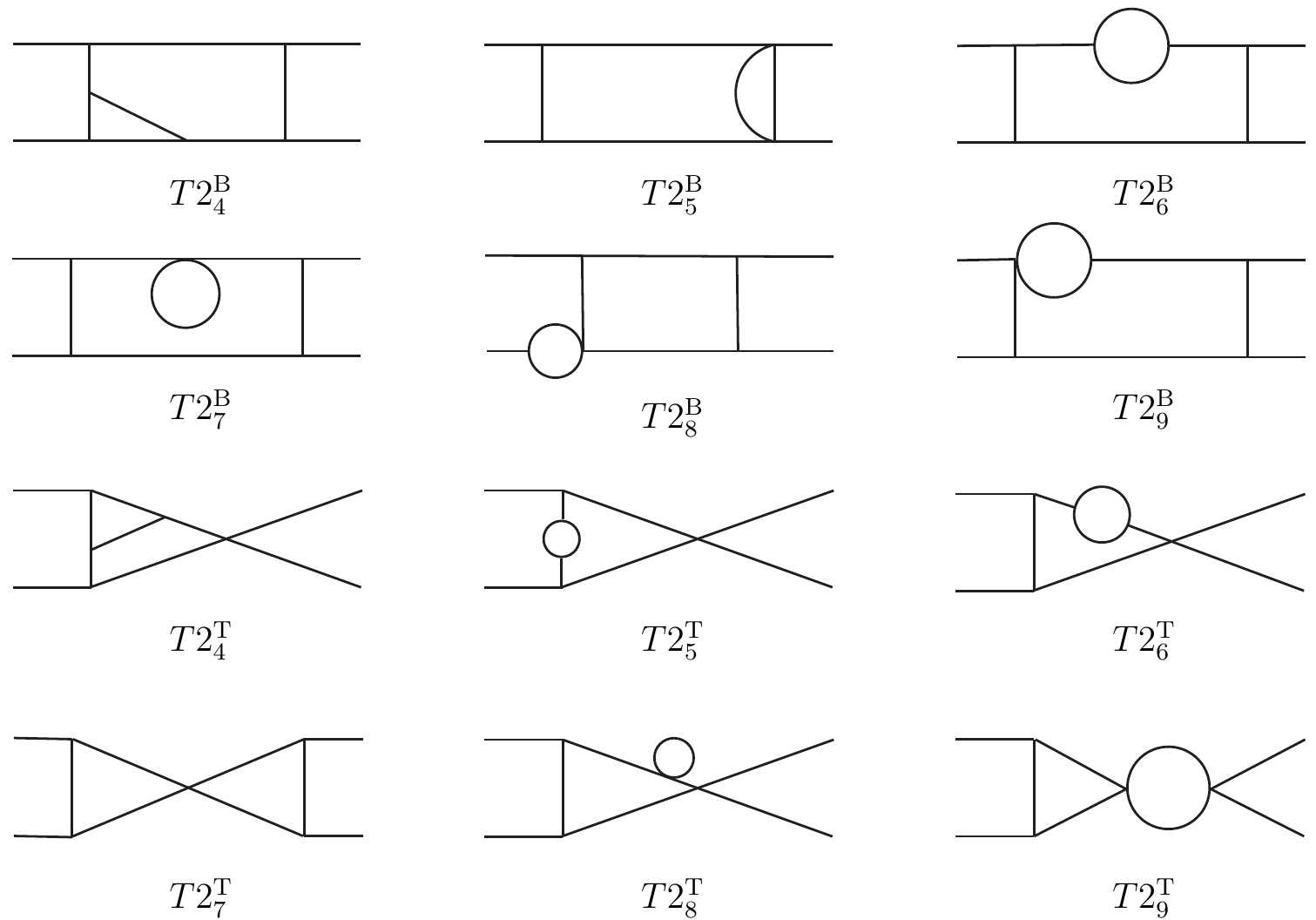}
  \caption{\it 1PI two-loop topologies leading to non-genuine finite
    or infinite diagrams.  Topologies $T2_{4,6}^B$ belong to set
    (6,0), topologies $T_5^B$ and $T_{8,9}^T$ to (2,2), while the
    remaining to the (4,1) set. Further details can be found in
    sec.~\ref{sec:2-loop-order-realizations}.}
  \label{fig:renormalizable-topologies-non-genuine}
\end{figure}

\begin{figure}[h!]
  \centering
  \includegraphics[scale=1]{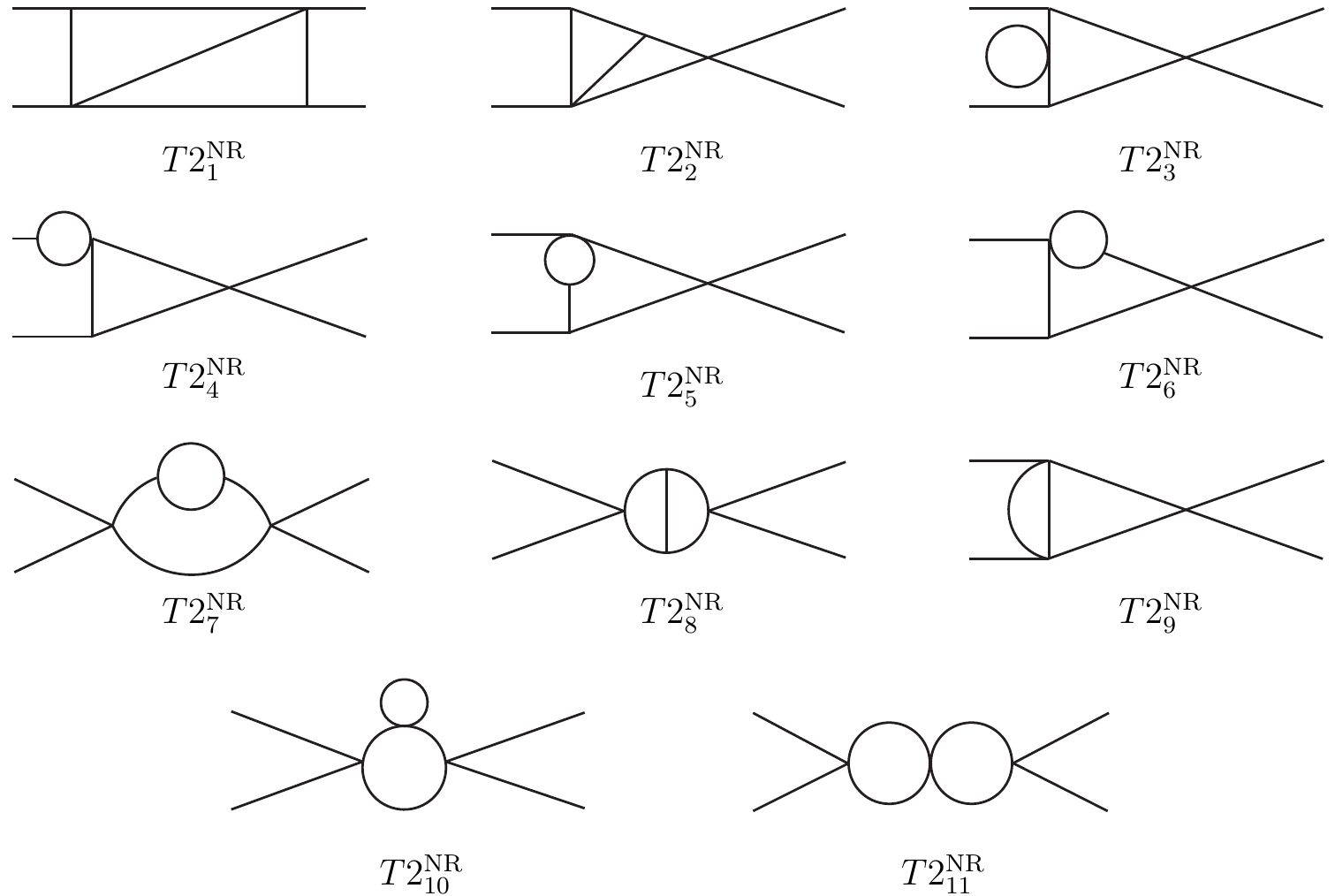}
  \caption{\it 1PI two-loop topologies not satisfying the
    renormalizability condition. The first eight topologies belong to
    (2,2) set while the last three to the set (0,3) set. Further
    details can be found in sec.~\ref{sec:2-loop-order-realizations}.}
  \label{fig:non-renormalizable-topologies}
\end{figure}

\begin{figure}%[h!]
  \centering
  \includegraphics[scale=1]{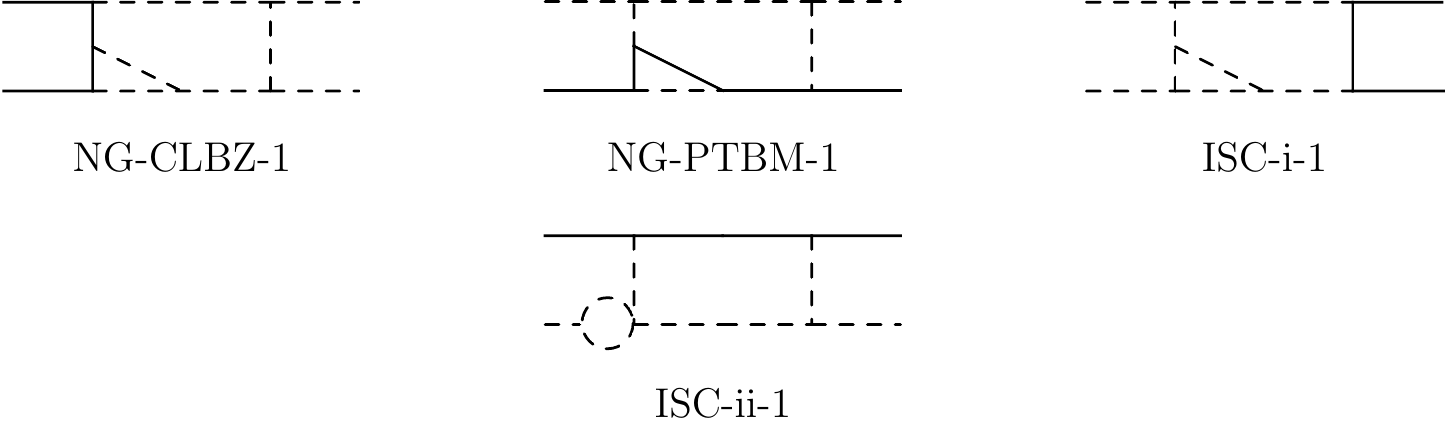}
  \caption{\it Non-genuine and finite two-loop diagrams which
    correspond to the one-loop generation of one of the couplings
    entering in the one-loop diagram T1-i (see
    fig \ref{Fig:1lpdiags}).}
  \label{fig:non-genuine-diagrams-i}
\end{figure}

\begin{figure}[h!]
  \centering
  \includegraphics[scale=1]{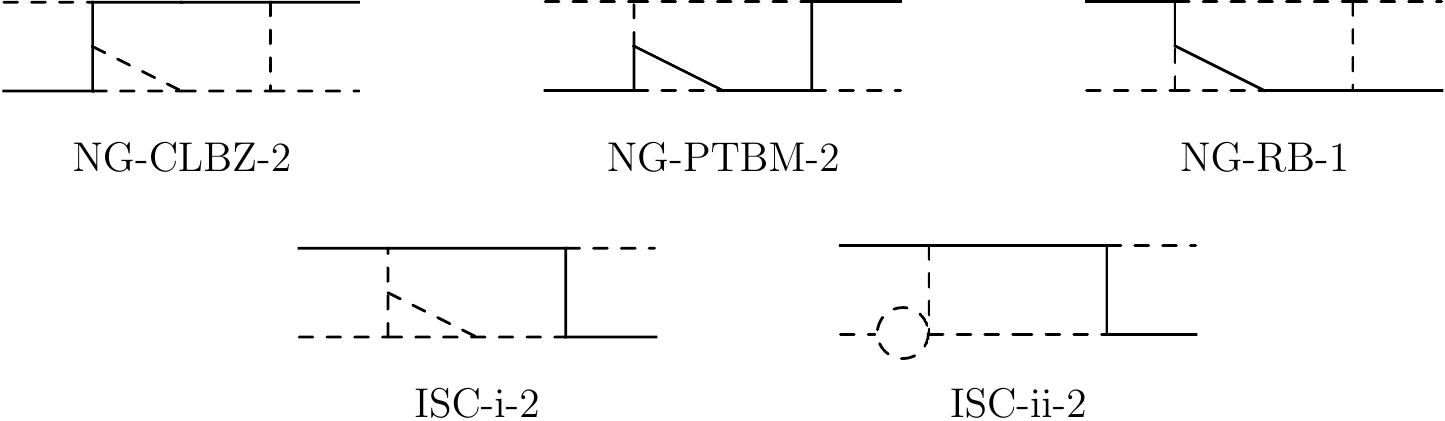}
  \caption{\it Non-genuine and finite two-loop diagrams which
    correspond to the one-loop generation of one of the couplings
    entering in the one-loop diagram T1-ii (see
    fig \ref{Fig:1lpdiags}).}
  \label{fig:non-genuine-diagrams-ii}
\end{figure}

\begin{figure}[h!]
  \centering
  \includegraphics[scale=1]{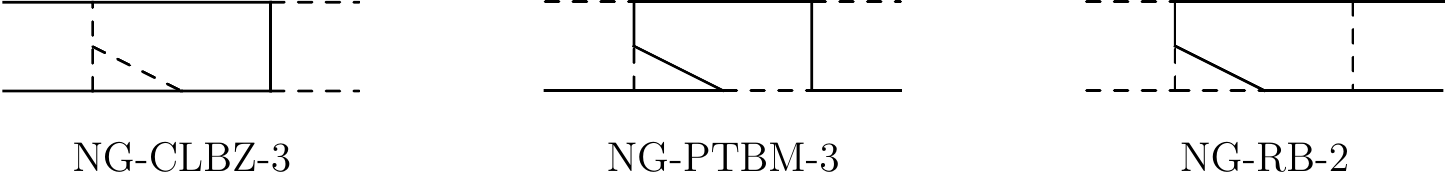}
  \caption{\it Non-genuine and finite two-loop diagrams which
    correspond to the one-loop generation of one of the couplings
    entering in the one-loop diagram T1-iii (see
    fig \ref{Fig:1lpdiags}).}
  \label{fig:non-genuine-diagrams-iii}
\end{figure}

\begin{figure}[h!]
  \centering
  \includegraphics[scale=1]{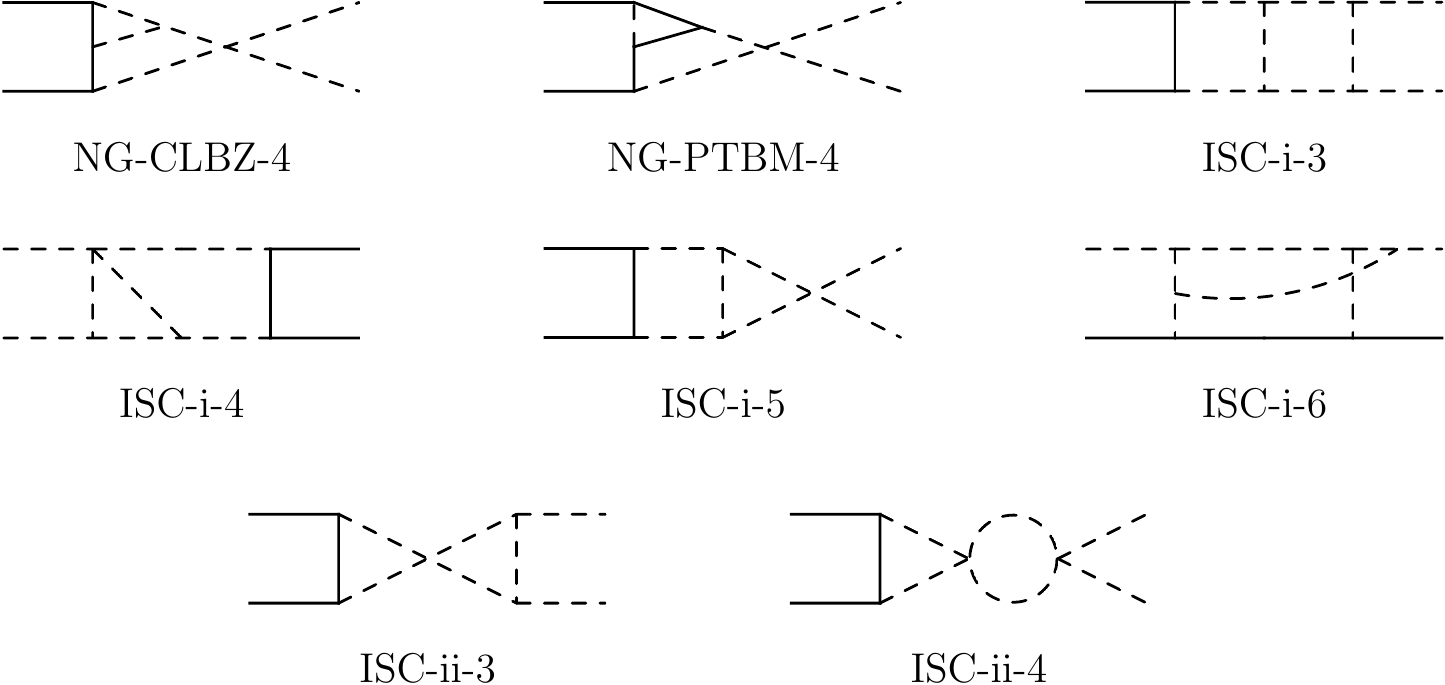}
  \caption{\it Non-genuine and finite two-loop diagrams which
    correspond to the one-loop generation of one of the couplings
    entering in the one-loop diagram T-3 (see
    fig \ref{Fig:1lpdiags}).}
  \label{fig:non-genuine-diagrams-3}
\end{figure}

In this appendix, we present in
fig.~\ref{fig:renormalizable-topologies-non-genuine} the list of
renormalizable topologies involving non-genuine but finite 2-loop
diagrams. For completeness in
fig~\ref{fig:non-renormalizable-topologies} we display as well the
full set of non-renormalizable topologies we have found. As we have
already mentioned, 2-loop non-genuine but finite diagrams arise from
1-loop diagrams where one of the internal vertices is generated at
the 1-loop order. 2-loop finite non-genuine diagrams can therefore be
classified according to the 1-loop diagram from which they originate.
Figs.~\ref{fig:non-genuine-diagrams-i}-\ref{fig:non-genuine-diagrams-3}
show the different finite non-genuine diagrams classified according to
this scheme.
% -----------------------------------------------------------
% Renormalizable topologies leading to non-genuine diagrams
% -----------------------------------------------------------
% ------------------------------
% non-renormalizable topologies
% ------------------------------
% ----------------------------------------------
% Non-genuine and finite diagrams (type T1-i) 
% ----------------------------------------------
% ----------------------------------------------
% Non-genuine and finite diagrams (type T1-ii) 
% ----------------------------------------------
% ----------------------------------------------
% Non-genuine and finite diagrams (type T1-iii) 
% ----------------------------------------------
% ----------------------------------------------
% Non-genuine and finite diagrams (type T1-3) 
% ----------------------------------------------

\clearpage
\section{Quantum numbers}
\label{sec:quantum-numbers-appendix}
In this appendix, we give tables from which the SM quantum numbers of
genuine diagrams CLBZ-i and PTBM-i (i=4,5,6,$\dots$) as well as RB-j
(j=3,4) can be determined. The tables obey the same conventions as
tab.~\ref{tab:double-box-QN}, i.e. symbols $S_i$ refer to allowed
external lepton-Higgs structures according to fig. \ref{fig:bubbles},
and hypercharge of field $X_i$ is denoted by $Y_i$. Their utilization
requires as well using fig. \ref{fig:hypercharge-field-assignments},
as already discussed and exemplified in
sec.~\ref{sec:gauge-quantum-numbers}.

We start with tab.~\ref{tab:non-planar-box-based-QN}, which provides
the possible quantum number assignments for genuine diagrams CLBZ-i
and PTBM-i (with i=4,5,6) in fig.~\ref{fig:diagrams-1} and RB-3 in
fig.~\ref{fig:diagrams-2}. Tab. \ref{tab:diagonal-box-based-QN},
instead, gives the possible assignments for genuine diagrams CLBZ-7
and CLBZ-8, whereas tabs. \ref{tab:diagram-d-triangle-based-QN} and
\ref{tab:diagram-e-triangle-based-QN} for genuine diagrams CLBZ-9 and
CLBZ-10, respectively. Finally, tab.~\ref{tab:RB-4-table} gives the
assignments for diagram RB-4 in fig. \ref{fig:diagrams-2}.  We note
again that due to the lepton and Higgs doublets being color singlets,
color charges for internal fields can be straightforwardly included,
and so we do not list them.

We point out that in order to construct compact tables, we have
written in some cases two possibilities for $SU(2)$ assignments of
particles. Usually this would lead to 8 possible combinations, for
which, however, not all are allowed. Exceptions are those where
vertices are (obviously) forbidden by $SU(2)$ invariance or do not
yield the Weinberg operator. Therefore, when using tables
\ref{tab:non-planar-box-based-QN}-\ref{tab:RB-4-table}, one should
bear in mind that neither triple vertices with combinations of
representations 1-1-3 (or any of its permutations) nor combinations
which lead to (an effective) quartic vertex $HH-1-1$ are allowed.

% ----------------------------
% Table for diagrams type-(b)
% ----------------------------
\begin{table}[h!]
  \renewcommand{\arraystretch}{1.2}
  \setlength{\tabcolsep}{4pt}
  \centering
  \begin{tabular}{ |c||c|c|c|c|c||c|c|c|c|c||c|c|c|c|c| }
    \hline
    % First row
    \multicolumn{16}{|c|}{$\boldsymbol{SU(2)}$ {\bf quantum numbers}}\\\hline
    \diaghead{\theadfont column wid}
    {$X_{5}$}{$X_{1}$}&\multicolumn{5}{c||}{$1$}
    &\multicolumn{5}{c||}{$2$} 
    &\multicolumn{5}{c|}{$3$}\\
    \hline\hline
    % Second row
    & $X_{3}$ & $X_{2}$ & $X_{4}$ & $X_{6}$ & $X_{7}$ & $X_{3}$
    & $X_{2}$ & $X_{4}$ & $X_{6}$ & $X_{7}$ & $X_{3}$ & $X_{2}$
    & $X_{4}$ & $X_{6}$ & $X_{7}$  \\
    \hline\hline
    % Third row
    \multirow{2}{0.3cm}{$1$}&
    \multirow{2}{0.3cm}{$2$}&
    \multirow{2}{0.3cm}{$2$}&
    1&
    \multirow{2}{0.3cm}{$2$}&
    \multirow{2}{0.3cm}{$2$} &
    \multirow{2}{0.3cm}{$2$}& 
    1&
    \multirow{2}{0.3cm}{$2$}&
    \multirow{2}{0.3cm}{$2$}&
    1&
    \multirow{2}{0.3cm}{$2$}&
    \multirow{2}{0.3cm}{$2$}&
    1&
    \multirow{2}{0.3cm}{$2$}&
    \multirow{2}{0.3cm}{$2$}\\
     & &  & 3 & & & & 3 & & & 3 & & & 3 & & \\
\hline
  % Fourth row
    \multirow{2}{0.3cm}{$2$}&
    3&
    1&
    \multirow{2}{0.3cm}{$2$}&
    1&
    \multirow{2}{0.3cm}{$2$}&
    1& 
    \multirow{2}{0.3cm}{$2$}&
    1& 
    1&
    1&
    1&
    3&
    \multirow{2}{0.3cm}{$2$}&
    1&
    \multirow{2}{0.3cm}{$2$}\\\cline{2-2}\cline{3-3}\cline{12-12}\cline{13-13}
     & 1 & 3 &  & 3  & & 3 &  & 3 & 3 & 3 & 3 & 1 & & 3 & \\
     \hline
     % Fifth row
    \multirow{2}{0.3cm}{$3$}&
    \multirow{2}{0.3cm}{$2$}&
    \multirow{2}{0.3cm}{$2$}&
    1&
    \multirow{2}{0.3cm}{$2$}&
    \multirow{2}{0.3cm}{$2$}&
    \multirow{2}{0.3cm}{2}&
    1& 
    \multirow{2}{0.3cm}{$2$}&
    \multirow{2}{0.3cm}{$2$}&
    1&
    \multirow{2}{0.3cm}{$2$}&
    \multirow{2}{0.3cm}{$2$}&
    1&
    \multirow{2}{0.3cm}{$2$}&
    \multirow{2}{0.3cm}{$2$}\\
     & &  & $3$  & & & & 3 & & & 3 & & & 3 &&\\
\hline
\end{tabular}

\vspace{0.1cm}
\begin{tabular}{|c|c|c|c|c|c|c|c|}\hline
  \multicolumn{8}{|c|}{\bf Hypercharge}\\\hline
  $S_i$ & $Y_1$ & $Y_2$ & $Y_3$ & $Y_4$ & $Y_5$ & $Y_6$ & $Y_7$
  \\\hline
  $S_1$ & $-1+\alpha$ & $-1+\alpha-\beta$& $\beta$ & $-2+\alpha-\beta$ 
  & $1+\beta$ & $2+\beta$ & $\alpha$
  \\\hline
  $S_2$ & $-1+\alpha$ & $-1+\alpha-\beta$ & $\beta$ & $\alpha-\beta$
  & $-1+\beta$ & $\beta$ & $\alpha$
  \\\hline
  $S_3$ & $-1+\alpha$ & $-1+\alpha-\beta$ & $\beta$ & $\alpha-\beta$ 
  & $1+\beta$ & $\beta$ & $\alpha$
    \\\hline
  $S_5$ & $1+\alpha$ & $1+\alpha-\beta$ & $\beta$ & $2+\alpha-\beta$ 
  & $-1+\beta$ & $-2+\beta$ & $\alpha$
  \\\hline
\end{tabular}
\caption{\it Electroweak quantum numbers for diagrams CLBZ-i and
  PTBM-i (i=4,5,6) in fig. \ref{fig:diagrams-1} and RB-3 
  in fig.~\ref{fig:diagrams-2}. Upper table: $SU(2)$
  representations.  Lower table: hypercharge assignments. Fields $X_i$
  refer to internal fields in the symbolic diagram in fig
  \ref{fig:hypercharge-field-assignments}-$(b)$.}
\label{tab:non-planar-box-based-QN}
\end{table}
% ----------------------------
% Table for diagrams type-(c)
% ----------------------------
\begin{table}[h!]
  \renewcommand{\arraystretch}{1.2}
  \setlength{\tabcolsep}{4pt}
  \centering
  \begin{tabular}{ |c||c|c|c|c||c|c|c|c||c|c|c|c| }
    \hline
    \multicolumn{13}{|c|}{$\boldsymbol{SU(2)}$ {\bf quantum numbers}}\\\hline
    % First row
    \diaghead{\theadfont column wid}
    {$X_{6}$}{$X_{1}$}&\multicolumn{4}{c||}{$1$}
    &\multicolumn{4}{c||}{$2$} &\multicolumn{4}{c|}{$3$}\\
    \hline\hline
    % Second row
    & $X_{2}$ & $X_{3}$ & $X_{4}$ & $X_{5}$ &
    $X_{2}$ & $X_{3}$ & $X_{4}$ & $X_{5}$ & 
    $X_{2}$ & $X_{3}$ & $X_{4}$ & $X_{5}$   \\
    \hline\hline
    % Third row
    \multirow{2}{0.3cm}{$1$}&
    \multirow{2}{0.3cm}{$2$}&
    1&
    \multirow{2}{0.3cm}{$2$}&
    \multirow{2}{0.3cm}{$2$}&
    1&
    \multirow{2}{0.3cm}{$2$} & 
   1& 
    \multirow{2}{0.3cm}{$2$}&
    \multirow{2}{0.3cm}{$2$}&
    1&
    \multirow{2}{0.3cm}{$2$}&
    \multirow{2}{0.3cm}{$2$}\\
    & & 3  & &  & 3 & & 3 & & & 3 & & \\
    \hline
    % Fourth row
    \multirow{2}{0.3cm}{$2$}&
    \multirow{2}{0.3cm}{$2$}&
    1&
    \multirow{2}{0.3cm}{$2$}&
    \multirow{2}{0.3cm}{$2$}& 
    1&
    \multirow{2}{0.3cm}{$2$}&
    \multirow{2}{0.3cm}{$2$}&
    1&
    \multirow{2}{0.3cm}{$2$}&
    1&
    1&
    1\\
    & & 3 &   &   & 3 &  & & 3 &  & 3 & 3 & 3 \\
    \hline
    % Fifth row
    \multirow{2}{0.3cm}{$3$}&
    \multirow{2}{0.3cm}{$2$}&
    1&
    \multirow{2}{0.3cm}{$2$}&
    \multirow{2}{0.3cm}{2}&
    1& 
    \multirow{2}{0.3cm}{$2$}&
    1&
    \multirow{2}{0.3cm}{$2$}&
    \multirow{2}{0.3cm}{$2$}&
    1&
    \multirow{2}{0.3cm}{$2$}&
    \multirow{2}{0.3cm}{$2$}
    \\
    & & 3 & & & $3$ & & 3 & & & 3 &  & \\
    \hline
  \end{tabular}
  
  \vspace{0.1cm}
  \begin{tabular}{|c|c|c|c|c|c|c|}\hline
    \multicolumn{7}{|c|}{\bf Hypercharge}\\\hline
    $S_i$ & $Y_1$ & $Y_2$ & $Y_3$ & $Y_4$ & $Y_5$ & $Y_6$ 
    \\\hline
    $S_4$ & $1+\beta$ & $\beta$ &$1 + \beta$ & $\alpha-\beta$ 
    & $1+\alpha$ & $\alpha$  
    \\\hline
    $S_5$ & $-1+\beta$ & $\beta$ &$-1 + \beta$ & $2+\alpha-\beta$ 
    & $1+\alpha$ & $\alpha$ 
    \\\hline
  \end{tabular}
  \caption{\it Electroweak quantum numbers for diagrams 
    CLBZ-7 and CLBZ-8 in fig. \ref{fig:diagrams-1}. Upper table: 
    $SU(2)$ representations. Lower table: hypercharge assignments. 
    Fields $X_i$ refer to internal fields in the symbolic diagram in fig
    \ref{fig:hypercharge-field-assignments}-$(c)$.}
  \label{tab:diagonal-box-based-QN}
\end{table}
% ----------------------------
% Table for diagrams type-(d)
% ----------------------------
\begin{table}[h!]
  \renewcommand{\arraystretch}{1.2}
  \setlength{\tabcolsep}{4pt}
  \centering
  \begin{tabular}{ |c||c|c|c|c||c|c|c|c||c|c|c|c| }
    \hline
    \multicolumn{13}{|c|}{$\boldsymbol{SU(2)}$ {\bf quantum numbers}}\\\hline
    % First row
    \diaghead{\theadfont column wid}
    {$X_{4}$}{$X_{1}$}&\multicolumn{4}{c||}{$1$}
    &\multicolumn{4}{c||}{$2$} &\multicolumn{4}{c|}{$3$}\\
    \hline\hline
    % Second row
    & $X_{5}$ & $X_{6}$ & $X_{2}$ & $X_{3}$ &
    $X_{5}$ & $X_{6}$ & $X_{2}$ & $X_{3}$ & 
    $X_{5}$ & $X_{6}$ & $X_{2}$ & $X_{3}$   \\
    \hline\hline
  % Third row
    \multirow{2}{0.3cm}{$1$}&
    \multirow{2}{0.3cm}{$2$}&
    \multirow{2}{0.3cm}{$2$}&
    \multirow{2}{0.3cm}{$2$}&
    \multirow{2}{0.3cm}{$2$}&
    1&
    \multirow{2}{0.3cm}{$2$} & 
   1& 
    1&
    \multirow{2}{0.3cm}{$2$}&
    \multirow{2}{0.3cm}{$2$}&
    \multirow{2}{0.3cm}{$2$}&
    \multirow{2}{0.3cm}{$2$}\\\cline{6-6}\cline{9-9}
     & & & &  & 3 & & 3 & 3 & &  & & \\
\hline
  % Fourth row
    \multirow{2}{0.3cm}{$2$}&
    \multirow{2}{0.3cm}{$2$}&
    1&
    1&
    1& 
    1&
    1&
    \multirow{2}{0.3cm}{$2$}&
    \multirow{2}{0.3cm}{$2$}&
    \multirow{2}{0.3cm}{$2$}&
    1&
    3&
    1\\\cline{3-3}\cline{4-4}\cline{11-11}\cline{12-12}
     & & 3 & 3 & 3 & 3 & 3 &  &  &  & 3 & 1 & 3 \\
\hline
     % Fifth row
    \multirow{2}{0.3cm}{$3$}&
    \multirow{2}{0.3cm}{$2$}&
    \multirow{2}{0.3cm}{$2$}&
    \multirow{2}{0.3cm}{$2$}&
    \multirow{2}{0.3cm}{2}&
   1& 
    \multirow{2}{0.3cm}{$2$}&
    1&
    3&
    \multirow{2}{0.3cm}{$2$}&
    \multirow{2}{0.3cm}{$2$}&
    \multirow{2}{0.3cm}{$2$}&
    \multirow{2}{0.3cm}{$2$}
    \\\cline{6-6}\cline{9-9}
     & & & & & $3$ & & 3 & 1 & &  &  & \\
\hline
\end{tabular}

  \vspace{0.1cm}
  \begin{tabular}{|c|c|c|c|c|c|c|}\hline
    \multicolumn{7}{|c|}{\bf Hypercharge}\\\hline
    $S_i$ & $Y_1$ & $Y_2$ & $Y_3$ & $Y_4$ & $Y_5$ & $Y_6$ 
    \\\hline
    $S_3$ & $-2-\alpha-\beta$ & $-2+\beta$ &$ \beta$ & $1+\alpha$ 
    & $\alpha$ & $-1-\alpha+\beta$ 
    \\\hline
  \end{tabular}
  \caption{\it Electroweak quantum numbers for diagrams CLBZ-9 
    in fig. \ref{fig:diagrams-1}. Upper table: $SU(2)$ representations.
    Lower table: hypercharge assignments. Fields $X_i$ refer to internal 
    fields in the symbolic diagram in fig
    \ref{fig:hypercharge-field-assignments}-$(d)$.}
  \label{tab:diagram-d-triangle-based-QN}
\end{table}
% ----------------------------
% Table for diagrams type-(e)
% ----------------------------
\begin{table}[h!]
  \renewcommand{\arraystretch}{1.2}
  \setlength{\tabcolsep}{4pt}
  \centering
  \begin{tabular}{ |c||c|c|c|c||c|c|c|c||c|c|c|c| }
    \hline
    \multicolumn{13}{|c|}{$\boldsymbol{SU(2)}$ {\bf quantum numbers}}\\\hline
    % First row
    \diaghead{\theadfont column wid}
    {$X_{3}$}{$X_{1}$}&\multicolumn{4}{c||}{$1$}&
    \multicolumn{4}{c||}{$2$} &\multicolumn{4}{c|}{$3$}\\
    \hline\hline
    % Second row
    & $X_{2}$ & $X_{4}$ & $X_{5}$ & $X_{6}$ &
    $X_{2}$ & $X_{4}$ & $X_{5}$ & $X_{6}$  & 
    $X_{2}$ & $X_{4}$ & $X_{5}$ & $X_{6}$    \\
    \hline\hline
    % Third row
    \multirow{2}{0.3cm}{$1$}&
    \multirow{2}{0.3cm}{$2$}&
    \multirow{2}{0.3cm}{$2$}&
    1&
    \multirow{2}{0.3cm}{$2$}&
    \multirow{2}{0.3cm}{$2$}&
    1& 
    \multirow{2}{0.3cm}{$2$}& 
    1&
    \multirow{2}{0.3cm}{$2$}&
    \multirow{2}{0.3cm}{$2$}&
    1&
    \multirow{2}{0.3cm}{$2$}\\
    & & & 3 &  & & 3 & & 3 & &  & 3 & \\
    \hline
    % Fourth row
    \multirow{2}{0.3cm}{$2$}&
    1&
    1&
    \multirow{2}{0.3cm}{$2$}&
    \multirow{2}{0.3cm}{$2$}&
    1&
    \multirow{2}{0.3cm}{$2$}&
    1&
    1&
    1&
    3&
    \multirow{2}{0.3cm}{$2$}&
    \multirow{2}{0.3cm}{$2$}
    \\\cline{2-2}\cline{3-3}\cline{10-10}\cline{11-11}
    & 3 & 3 & & & 3 & & 3 &3  &3  & 1 & & \\
    \hline
    % Fifth row
    \multirow{2}{0.3cm}{$3$}&
    \multirow{2}{0.3cm}{$2$}&
    \multirow{2}{0.3cm}{$2$}&
    1&
    \multirow{2}{0.3cm}{2}&
    \multirow{2}{0.3cm}{$2$}&
    1&
    \multirow{2}{0.3cm}{$2$}&
    1&
    \multirow{2}{0.3cm}{$2$}&
    \multirow{2}{0.3cm}{$2$}&
    1&
    \multirow{2}{0.3cm}{$2$}
    \\
    & & & 3 & & & 3 & & 3 & &  & 3 & \\
    \hline
  \end{tabular}
  
  \vspace{0.1cm}
  \begin{tabular}{|c|c|c|c|c|c|c|}\hline
    \multicolumn{7}{|c|}{\bf Hypercharge}\\\hline
    $S_i$ & $Y_1$ & $Y_2$ & $Y_3$ & $Y_4$ & $Y_5$ & $Y_6$ 
    \\\hline
    $S_1$ & $-1+\alpha$ & $1+\beta$ &$ \beta$ & $-2+\alpha-\beta$ 
    & $-1+\alpha-\beta$ & $\alpha$ 
    \\\hline
  \end{tabular}
  \caption{\it Electroweak quantum numbers for diagram CLBZ-10
    in fig. \ref{fig:diagrams-1}. Upper table: $SU(2)$ representations.
    Lower table: hypercharge assignments. Fields $X_i$ refer to internal 
    fields in the symbolic diagram in fig.
    \ref{fig:hypercharge-field-assignments}-$(e)$.}
  \label{tab:diagram-e-triangle-based-QN}
\end{table}

\begin{table}
  \renewcommand{\arraystretch}{1.2}
  \setlength{\tabcolsep}{4pt}
  \centering
  \begin{tabular}{ |c||c|c|c|c||c|c|c|c||c|c|c|c| }
    \hline
    \multicolumn{13}{|c|}{$\boldsymbol{SU(2)}$ {\bf quantum numbers}}\\\hline
% First row
 \diaghead{\theadfont column wid}
    {$X_{3}$}{$X_{1}$}&\multicolumn{4}{c||}{$1$}
    &\multicolumn{4}{c||}{$2$} &\multicolumn{4}{c|}{$3$}\\
    \hline\hline
    % Second row
    & $X_{2}$ & $X_{4}$ & $X_{5}$ & $X_{6}$ &
    $X_{2}$ & $X_{4}$ & $X_{5}$ & $X_{6}$  & 
    $X_{2}$ & $X_{4}$ & $X_{5}$ & $X_{6}$    \\
    \hline\hline
    % Third row
    \multirow{2}{0.3cm}{$1$}&
    \multirow{2}{0.3cm}{$3$}&
    \multirow{2}{0.3cm}{$3$}&
    \multirow{2}{0.3cm}{$3$}&
    \multirow{2}{0.3cm}{$2$}&
    \multirow{2}{0.3cm}{$3$}&
    \multirow{2}{0.3cm}{$2$}&
    \multirow{2}{0.3cm}{$2$}& 
    1&
    \multirow{2}{0.3cm}{$3$}&
    1&
    1&
    \multirow{2}{0.3cm}{$2$}
    \\\cline{12-12}\cline{11-11}
    & & &  &  & &  & & 3 & & 3 & 3 & \\
    \hline
    % Fourth row
    \multirow{2}{0.3cm}{$2$}&
    \multirow{2}{0.3cm}{$2$}&
    \multirow{2}{0.3cm}{$2$}&
    1&
    \multirow{2}{0.3cm}{$2$}&
    \multirow{2}{0.3cm}{$2$}&
    1&
    \multirow{2}{0.3cm}{$2$}&
    1&
    \multirow{2}{0.3cm}{$2$}&
    \multirow{2}{0.3cm}{$2$}&
    1&
    \multirow{2}{0.3cm}{$2$}\\
    &  &  &3 & &  & 3 &  &3  &  &  & 3 & \\
    \hline
    % Fifth row
    \multirow{2}{0.3cm}{$3$}&
    1&
    1&
    1&
    \multirow{2}{0.3cm}{2}&
    1&
    \multirow{2}{0.3cm}{$2$}&
    \multirow{2}{0.3cm}{$2$}&
    1&
    1&
    1&
    1&
    \multirow{2}{0.3cm}{$2$}
    \\\cline{2-2}\cline{3-3}\cline{2-2}\cline{3-3
    }
    &3 &3 & 3 & & 3&  & & 3 &3 &3  & 3 & \\
    \hline
  \end{tabular}
  
  \vspace{0.1cm}
  \begin{tabular}{|c|c|c|c|c|c|c|}\hline
    \multicolumn{7}{|c|}{\bf Hypercharge}\\\hline
    $S_i$ & $Y_1$ & $Y_2$ & $Y_3$ & $Y_4$ & $Y_5$ & $Y_6$ 
    \\\hline
    $S_3$ & $\alpha$ & $\alpha-\beta$ &$2+\alpha- \beta$ & $\beta$ 
    & $2+\alpha$ & $1+\alpha$ 
    \\\hline
  \end{tabular}
  \caption{\it Electroweak quantum numbers for diagram RB-4
    in fig. \ref{fig:diagrams-2}. Upper table: $SU(2)$ representations.
    Lower table: hypercharge assignments. Fields $X_i$ refer to internal 
    fields in the symbolic diagram in fig.
    \ref{fig:hypercharge-field-assignments}-$(f)$.}
  \label{tab:RB-4-table}
\end{table}

\clearpage
\section{Useful formulas for 2-loop calculations}
\label{sec:formulas-2-loop}
The integrals appearing in the 2-loop diagrams have been evaluated
several times in the literature. We follow
\cite{Angel:2013hla,McDonald:2003zj}, both of which are based on
\cite{vanderBij:1983bw}. We repeat here only the basic definitions and
final results, for more details see
\cite{Angel:2013hla,McDonald:2003zj,vanderBij:1983bw}.

As a starting point, define \cite{vanderBij:1983bw}
\begin{eqnarray}\label{eq:IntGen}
(M_{11},\cdots,M_{1n_1}|M_{21},\cdots,M_{2n_2}|M_{31},\cdots,M_{3n_3})
\\ \nonumber
=\int d^np\int\ d^nq \Pi_{i=1}^{n_1}\Pi_{j=1}^{n_2}\Pi_{k=1}^{n_3}
\frac{1}{(p^2 + M_{1i}^2)}\frac{1}{(q^2 + M_{2j}^2)}
\frac{1}{[(p+q)^2 + M_{3k}^2]}\ .
\end{eqnarray}
Here, $n$ is the number of dimensions. In the case of infinite 
integrals one has to carefully evaluate all terms for $n=4+\epsilon$ 
before taking the limit $\epsilon \to 0$. Since we are interested 
only in models with finite integrals, we will not write out  
terms containing poles in $1/\epsilon$.

Using partial fractions, one can rewrite integrals of 
the form of eq. (\ref{eq:IntGen}) as sums over integrals 
with fewer denominators:
\begin{equation}\label{eq:partp1}
(m,m_0|m_1|m_2)=\frac{1}{m^2-m_0^2}
\Big\{(m_0|m_1|m_2)-(m|m_1|m_2)\Big\}\ .
\end{equation}
Similarly, integrals with three denominators can be recombined 
into less divergent ones, using \cite{vanderBij:1983bw}
\begin{equation}\label{eq:partp2}
(m_0|m_1|m_2) = \frac{1}{3-n} 
\Big\{ m_0^2(2 m_0| m_1 | m_2) + m_1^2(2 m_1 | m_0|m_2) +
m_2^2(2 m_2 | m_0 | m_1)\Big\}\ .
\end{equation}
Here, $(2 m|m_i|m_j)$ is a short-hand for $(m,m|m_i|m_j)$. The ``$p^2$
decomposition'' is another relation which proves to be useful for
calculating integrals with momentum-dependent numerators, namely
\begin{equation}
  \label{eq:kSq-decomposition}
  \frac{p^2}{(p^2-m_1^2)(p^2-m_2^2)} = 
  \frac{1}{(p^2-m_1^2)} + \frac{m_2^2}{(p^2-m_1^2)(p^2-m_2^2)}\ .
\end{equation}

Using only eq. (\ref{eq:partp1}) results in expressions
\cite{Angel:2013hla} which are more compact than those given in
\cite{McDonald:2003zj,vanderBij:1983bw}, which make repeated use of
both, eq. (\ref{eq:partp1}) and eq. (\ref{eq:partp2}). 

Also, for numerical evaluation, it is useful to define the final
expressions in terms of dimensionless quantities. By convention we
scale all masses appearing in the integrals with respect to the
``innermost'' scalar/fermion mass. This implies rescaling the momenta,
and for ${\cal I}_{ab,\alpha\beta,X}$ factoring out this overall mass
squared. We thus write
\begin{align}
  \label{eq:resc}
  {\cal I}_{ab,\alpha\beta, X}&=\frac{1}{(2\pi)^8}
  \frac{1}{m_X^2}{\hat{\cal I}}_{ab,\alpha\beta}\ ,
  \\
  {\cal I}_{ab,\alpha\beta}^{\{ k^2,q^2,(q+k)^2\}}&=\frac{1}{(2\pi)^8}
  {\hat{\cal I}}_{ab,\alpha\beta}^{\{ k^2,q^2,(q+k)^2\}}\ ,
\end{align}
with
\begin{align}
  \label{eq:BZgen}
  {\hat{\cal I}}_{ab,\alpha\beta} &=
  \int d^4k \int d^4q 
  \frac{1}{(k^2-r_a)(k^2-t_{\alpha})(q^2-r_b)(q^2-t_{\beta})([q+k]^2-1)}\ ,
  \\
  \label{eq:remaining-PTBM-int}
  {\hat{\cal I}}_{ab,\alpha\beta}^{\{k^2,q^2,(q+k)^2\}} &=
  \int d^4k \int d^4q 
  \frac{\{ k^2,q^2,(q+k)^2\}}
  {(k^2-r_a)(k^2-t_{\alpha})(q^2-r_b)(q^2-t_{\beta})([q+k]^2-1)}\ ,
\end{align}
for CLBZ and PTBM models, while
\begin{align}
  \label{eq:RB-1-numerator}
  {\hat{\cal I}}_{ab,\alpha\beta}^\text{RB} &=
  \int d^4k \int d^4q 
  \frac{1}{(k^2-r_a)(k^2-r_b)(k^2-t_\beta)(q^2-t_\alpha)([q+k]^2-1)}\ ,
  \\
  \label{eq:RB-non-trivial-numerator}
  {\hat{\cal I}}_{ab,\alpha\beta}^{\{k^2,q^2,(q+k)^2\}\text{RB}} &=
  \int d^4k \int d^4q 
  \frac{\{ k^2,q^2,(q+k)^2\}}
  {(k^2-r_a)(k^2-r_b)(k^2-t_\beta)(q^2-t_\alpha)([q+k]^2-1)}\ ,
\end{align}
for RB models. Here $\{k^2,q^2,(q+k)^2\}$ stands for $k^2$, $q^2$ or
$(k+q)^2$, $r_a =(m_{F_a}/m_X)^2$ and
$t_{\alpha}=(m_{S_\alpha}/m_X)^2$.  The ``strategy'' then for
calculating these integrals consists of reducing them to ``master
integral'' form:
\begin{equation}
  \label{eq:master-int}
  I(s,t)=\mu^\epsilon\int d^nk\;\int d^nq \frac{1}{(k^2-s)(q^2-t)[(k+q)^2-1]}\ .
\end{equation}
This integral, which involves an infinite and a finite piece, has been
calculated in \cite{Angel:2013hla} (see below).  Thus, with the aid of
eqs. (\ref{eq:partp1}) and (\ref{eq:kSq-decomposition}), the
calculation of integral in (\ref{eq:BZgen}) results in
\cite{Angel:2013hla}
\begin{equation}\label{eq:BZVolkas}
  \pi^{-4}\,{\hat{\cal I}}_{ab,\alpha\beta} = 
  \frac{1}{(t_{\alpha}-r_a)(t_{\beta}-r_b)}
  \left\{ -{\hat g}(t_{\alpha},t_{\beta})
    +{\hat g}(r_a,t_{\beta})
    +{\hat g}(t_{\alpha},r_b)
    -{\hat g}(r_a,r_b)\right\}\ ,
\end{equation}
while the result for integrals in (\ref{eq:remaining-PTBM-int}) reads
\begin{align}
    \label{eq:int-prop-to-kSq}
    % integral 1
    \pi^{-4}\,{\hat{\cal I}}_{ab,\alpha\beta}^{(k^2)} &=
    \left\{
      \frac{1}{t_\beta-r_b}
      \left[
        - \hat g(r_a,t_\beta) 
        + \hat g(r_a,r_b) 
      \right]\right.
    \nonumber\\
    &\left.
      +\frac{t_\alpha}{(t_\alpha - r_a)(t_\beta - r_b)}
      \left[
        - \hat g(t_\alpha,t_\beta) 
        + \hat g(t_\alpha,r_b)
        + \hat g(r_a,t_\beta) 
        - \hat g(r_a,r_b)
      \right]
    \right\}
    \ ,
    \\
    % integral 2
    \label{eq:int-prop-to-qSq}
    \pi^{-4}\,{\hat{\cal I}}_{ab,\alpha\beta}^{(q^2)} &=
        \left\{
      \frac{1}{t_\alpha-r_a}
      \left[
        - \hat g(t_\alpha,r_b) 
        + \hat g(r_a,r_b) 
      \right]\right.
    \nonumber\\
    &\left.
      + \frac{t_\beta}{(t_\alpha - r_a)(t_\beta - r_b)}
      \left[
        - \hat g(t_\alpha,t_\beta) 
        + \hat g(t_\alpha,r_b)
        + \hat g(r_a,t_\beta) 
        - \hat g(r_a,r_b)
      \right]
    \right\}\ ,
    \\
    % integral 3
    \label{eq:int-prop-to-(q+k)Sq}
    \pi^{-4}\,{\hat{\cal I}}_{ab,\alpha\beta}^{\{(k+q)^2\}} &=
    \left\{
      \widehat B_0^\prime(0,r_a,t_\alpha)\widehat B_0^\prime(0,r_b,t_\beta)
      + \frac{
        - \hat g(t_\alpha,t_\beta) 
        + \hat g(t_\alpha,r_b)
        + \hat g(r_a,t_\beta) 
        - \hat g(r_a,r_b)}
      {(t_\alpha - r_a)(t_\beta - r_b)}
    \right\}
    \ .
\end{align}
Calculation of the integrals in (\ref{eq:RB-1-numerator}) and
(\ref{eq:RB-non-trivial-numerator}) gives, instead, the following
results:
\begin{align}
  \label{eq:RB-int-num-1}
  % RB integral 1
  \pi^4\,{\hat{\cal I}}_{ab,\alpha\beta}^\text{RB}&=
  \frac{1}{t_\beta-r_a}
  \left\{
    \frac{1}{t_\beta-r_b}
    \left[
      -\hat g(t_\beta,t_\alpha) + \hat g(r_b,t_\alpha)
    \right]
    -
    \frac{1}{r_b-r_a}
    \left[
      -\hat g(r_b,t_\alpha) + \hat g(r_a,t_\alpha)
    \right]
  \right\}\ ,
  \\
  % RB integral 2
  \label{eq:RB-int-num-kSq}
  \pi^4\,{\hat{\cal I}}_{ab,\alpha\beta}^{(k^2)\text{RB}}&=
  \frac{1}{r_a-t_\beta}
  \left\{
    -\hat g(r_a,t_\alpha) + \hat g(t_\beta,t_\alpha)
  \right.
  \nonumber\\
  &\left.
    + \frac{r_b}{r_a-r_b}
    \left[
      -\hat g(r_a,t_\alpha) + \hat g(r_b,t_\alpha)
    \right]
    - \frac{r_b}{t_\beta-r_b}
    \left[
      -\hat g(t_\beta,t_\alpha) + \hat g(r_b,t_\alpha)
    \right]
  \right\}\ ,
  \\
  % RB integral 3
  \label{eq:RB-int-num-qSq}
  \pi^4\,{\hat{\cal I}}_{ab,\alpha\beta}^{(q^2)\text{RB}}&=
  \frac{t_\alpha}{r_a-r_b}
  \left\{
    \frac{1}{r_a-t_\beta}
    \left[
      -\hat g(r_a,t_\alpha) + \hat g(t_\beta,t_\alpha)
    \right]
   -\frac{1}{r_b-t_\beta}
   \left[
     -\hat g(r_b,t_\alpha) + \hat g(t_\beta,t_\alpha)
   \right]
  \right\}\ ,
  \\
  % RB integral 4
  \label{eq:RB-int-num-kplusqSq}
  \pi^4\,{\hat{\cal I}}_{ab,\alpha\beta}^{\{(k+q)^2\}\text{RB}}&=
  \frac{t_1-t_2}{r_a-t_\beta}
  \left[
    \hat B_0^\prime(0,r_a,r_b) - \hat B_0^\prime(0,t_\beta,r_b)
  \right]\hat B_0^\prime(0,t_1,t_2)
  \nonumber\\
  &+\frac{1}{t_\beta-r_a}
  \left\{
    \frac{1}{t_\beta-r_b}\left[-\hat g(t_\beta,t_\alpha) + \hat g(r_b,t_\alpha)\right]
    -
    \frac{1}{r_b-r_a}\left[-\hat g(r_b,t_\alpha) + \hat g(r_a,t_\alpha)\right]
  \right\}\ .
\end{align}
Here, ${\hat g}(s,t)$ is the solution to (\ref{eq:master-int}), while
$\widehat B_0^\prime(0,x,y)$ is given as follows. The well-known
one-loop scalar Passarino-Veltman function $B_0
$\cite{Passarino:1978jh} in the vanishing external momentum limit
($p\to 0$) reads:
\begin{equation}
  \label{eq:passarino-veltman-function}
  B_0(0,s,t)=\int \frac{d^4k}{(2\pi)^4}\frac{1}{(k^2-s)(k^2-t)}\ .
\end{equation}
Defining
\begin{equation}
  \label{eq:B0hat}
  B_0(0,s,t)=\frac{1}{(2\pi)^4}\widehat B_0(0,s,t)\ ,
\end{equation}
the finite part of the $\widehat B_0^\prime(0,s,t)$ function can be
written according to
\begin{equation}
  \label{eq:finite-B0}
  \widehat B_0(0,s,t)=-\pi^2 i\left(\frac{s \log s - t \log t}{s-t}\right)
  =\pi^2\,\widehat B_0^\prime(0,s,t)\ ,
\end{equation}
whereas the finite piece for $\hat g(s,t)$ reads:
\begin{eqnarray}\label{eq:ghat}
{\hat g}(s,t) & = & \frac{s}{2}\ln s \ln t +
\sum_{\pm} \pm \frac{s(1-s)+3st+2(1-t)x_{\pm}}{2\omega}
\\ \nonumber
& \times & \Big[\text{Li}_2(\frac{x_{\pm}}{x_{\pm}-s}) 
              -\text{Li}_2(\frac{x_{\pm}-s}{x_{\pm}}) 
              +\text{Li}_2(\frac{t-1}{x_{\pm}}) 
              -\text{Li}_2(\frac{t-1}{x_{\pm}-s})  \Big]\ ,
\end{eqnarray}
with the standard di-logarithm
\begin{equation}\label{eq:dilog}
\text{Li}_2(x) = - \int_0^x \frac{\ln(1-y)}{y}dy\ ,
\end{equation}
and
\begin{eqnarray}\label{eq:xpm}
x_{\pm}=\frac{1}{2}(-1+s+t \pm \omega) & & \omega=\sqrt{1+s^2+t^2-2(s+t+st)}\ .
\end{eqnarray}

In both cases, $B_0$ and $\hat g(s,t)$, we have giving expressions
only for their finite pieces. The reason is that for the CLBZ or PTBM
integral (\ref{eq:int-prop-to-(q+k)Sq}), we have found that the
divergent piece in the first term cancels upon summation over the
different contributions. Cancellation of the divergent piece in $\hat
g(s,t)$, in eqs. (\ref{eq:BZVolkas})-(\ref{eq:int-prop-to-(q+k)Sq}),
occurs as well when summing of the the different contributions. For RB
integrals, cancellation of divergences proceeds differently: the
divergent term from (\ref{eq:RB-int-num-kplusqSq}) cancels with the
divergent piece from (\ref{eq:RB-int-num-qSq}). Thus, always rendering
finite results for genuine diagrams, as of course it has to be.

Below we repeat the solution for ${\hat{\cal I}}_{ij,\alpha\beta}$ 
in the convention of  \cite{McDonald:2003zj}. This solution rewrites 
eq. (\ref{eq:master-int}) into a less divergent expression 
using eq. (\ref{eq:partp2}). Introducing dimensionless arguments 
as before, one finds
\begin{align}\label{eq:BZMcMc}
{\hat{\cal I}}_{ij,\alpha\beta} = 
\frac{\pi^4}{(t_{\alpha}-r_i)(t_{\beta}-r_j)}
&\Big\{r_i [f(t_{\beta}/r_i,1/r_i)-f(r_j/r_i,1/r_i)]
\\ \nonumber
&+ r_j [f(t_{\alpha}/r_j,1/r_j)-f(r_i/r_j,1/r_j)]
\\ \nonumber
&+ t_{\alpha} [f(r_j/t_{\alpha},1/t_{\alpha})-f(t_{\beta}/t_{\alpha},1/t_{\alpha})]
\\ \nonumber
&+ t_{\beta} [f(r_i/t_{\beta},1/t_{\beta})-f(t_{\alpha}/t_{\beta},1/t_{\beta})]
\\ \nonumber
&+[f(t_{\alpha},r_j)-f(r_i,r_j)-f(t_{\alpha},t_{\beta})+f(r_i,t_{\beta})]\Big\}\ .
\end{align}
Here,
\begin{align}\label{eq:fab}
f(a,b)=-\frac{1}{2}\ln a \ln b -\frac{1}{2}\Big(\frac{a+b-1}{\kappa}\Big)
       &\Big\{\text{Li}_2(\frac{-x_2}{y_1}) + \text{Li}_2(\frac{-y_2}{x_1}) 
           - \text{Li}_2(\frac{-x_1}{y_2}) - \text{Li}_2(\frac{-y_1}{x_2})
\\ \nonumber
         &+ \text{Li}_2(\frac{b-a}{x_2}) + \text{Li}_2(\frac{a-b}{y_2})
          - \text{Li}_2(\frac{b-a}{x_1}) - \text{Li}_2(\frac{a-b}{y_1})\Big\},
\end{align}
with
\begin{eqnarray}\label{eq:defxy}
x_{1,2} = \frac{1}{2}(1+b-a \pm \kappa)\ ,
\\ \nonumber
y_{1,2} = \frac{1}{2}(1+a-b \pm \kappa)\ ,
\end{eqnarray}
and
\begin{equation}\label{eq:defkap}
\kappa = \sqrt{1-2(a+b)+(a-b)^2}\ .
\end{equation}
Eq. (\ref{eq:BZMcMc}) is more complicated than eq. (\ref{eq:BZVolkas}), 
but leads to exactly the same numerical result. We have found it 
therefore a useful cross-check for our calculation.


\begin{thebibliography}{99}
\addcontentsline{toc}{section}{References}
% -------------
% References
% -------------

%\cite{Fukuda:1998mi}
\bibitem{Fukuda:1998mi} 
  Y.~Fukuda {\it et al.}  [Super-Kamiokande Collaboration],
  %``Evidence for oscillation of atmospheric neutrinos,''
  Phys.\ Rev.\ Lett.\  {\bf 81}, 1562 (1998)
  [hep-ex/9807003].
  %%CITATION = HEP-EX/9807003;%%

%\cite{Ahmad:2002jz}
\bibitem{Ahmad:2002jz} 
  Q.~R.~Ahmad {\it et al.}  [SNO Collaboration],
  %``Direct evidence for neutrino flavor transformation from neutral
  % current interactions in the Sudbury Neutrino Observatory,''
  Phys.\ Rev.\ Lett.\  {\bf 89}, 011301 (2002)
  [nucl-ex/0204008].
  %%CITATION = NUCL-EX/0204008;%%

%\cite{Eguchi:2002dm}
\bibitem{Eguchi:2002dm} 
  K.~Eguchi {\it et al.}  [KamLAND Collaboration],
  %``First results from KamLAND: Evidence for reactor anti-neutrino
  % disappearance,''
  Phys.\ Rev.\ Lett.\  {\bf 90}, 021802 (2003)
  [hep-ex/0212021].
  %%CITATION = HEP-EX/0212021;%%

%\cite{Forero:2014bxa}
\bibitem{Forero:2014bxa} 
  D.~V.~Forero, M.~Tortola and J.~W.~F.~Valle,
  %``Neutrino oscillations refitted,''
  arXiv:1405.7540 [hep-ph].
  %%CITATION = ARXIV:1405.7540;%%

%\cite{Capozzi:2013csa}
\bibitem{Capozzi:2013csa} 
  F.~Capozzi, G.~L.~Fogli, E.~Lisi, A.~Marrone, D.~Montanino and A.~Palazzo,
  %``Status of three-neutrino oscillation parameters, circa 2013,''
  Phys.\ Rev.\ D {\bf 89}, 093018 (2014)
  [arXiv:1312.2878 [hep-ph]].
  %%CITATION = ARXIV:1312.2878;%%

%\cite{GonzalezGarcia:2012sz}
\bibitem{GonzalezGarcia:2012sz} 
  M.~C.~Gonzalez-Garcia, M.~Maltoni, J.~Salvado and T.~Schwetz,
  %``Global fit to three neutrino mixing: critical look at present precision,''
  JHEP {\bf 1212}, 123 (2012)
  [arXiv:1209.3023 [hep-ph]].
  %%CITATION = ARXIV:1209.3023;%%
  %378 citations counted in INSPIRE as of 14 Aug 2014

%\cite{Weinberg:1980bf}
\bibitem{Weinberg:1980bf} 
  S.~Weinberg,
  %``Varieties of Baryon and Lepton Nonconservation,''
  Phys.\ Rev.\ D {\bf 22}, 1694 (1980).
  %%CITATION = PHRVA,D22,1694;%%
  %206 citations counted in INSPIRE as of 14 Aug 2014

%\cite{Ma:1998dn}
\bibitem{Ma:1998dn} 
  E.~Ma,
  %``Pathways to naturally small neutrino masses,''
  Phys.\ Rev.\ Lett.\  {\bf 81}, 1171 (1998)
  [hep-ph/9805219].
  %%CITATION = HEP-PH/9805219;%%
  %305 citations counted in INSPIRE as of 27 Sep 2013

%\cite{Minkowski:1977sc}
\bibitem{Minkowski:1977sc} 
  P.~Minkowski,
  %``mu --> e gamma at a Rate of One Out of 1-Billion Muon Decays?,''
  Phys.\ Lett.\ B {\bf 67}, 421 (1977).
  %%CITATION = PHLTA,B67,421;%%

%\cite{Yanagida:1979as}
\bibitem{Yanagida:1979as} 
  T.~Yanagida,
  %``Horizontal Symmetry And Masses Of Neutrinos,''
  Conf.\ Proc.\ C {\bf 7902131}, 95 (1979).
  %%CITATION = CONFP,C7902131,95;%%

%\cite{GellMann:1980vs}
\bibitem{GellMann:1980vs} 
  M.~Gell-Mann, P.~Ramond and R.~Slansky,
  %``Complex Spinors and Unified Theories,''
  Conf.\ Proc.\ C {\bf 790927}, 315 (1979)
  [arXiv:1306.4669 [hep-th]].
  %%CITATION = ARXIV:1306.4669;%%

%\cite{Mohapatra:1979ia}
\bibitem{Mohapatra:1979ia} 
  R.~N.~Mohapatra and G.~Senjanovic,
  %``Neutrino Mass and Spontaneous Parity Violation,''
  Phys.\ Rev.\ Lett.\  {\bf 44}, 912 (1980).
  %%CITATION = PRLTA,44,912;%%

%\cite{Schechter:1980gr}
\bibitem{Schechter:1980gr}
  J.~Schechter and J.~W.~F.~Valle,
  % ``Neutrino Masses In SU(2) X U(1) Theories,''
  Phys.\ Rev.\  D {\bf 22}, 2227 (1980);
  %% CITATION = PHRVA,D22,2227;%%

%\cite{Magg:1980ut}
\bibitem{Magg:1980ut} 
  M.~Magg and C.~Wetterich,
  %``Neutrino Mass Problem and Gauge Hierarchy,''
  Phys.\ Lett.\ B {\bf 94}, 61 (1980).
  %%CITATION = PHLTA,B94,61;%%

%\cite{Mohapatra:1980yp}
\bibitem{Mohapatra:1980yp} 
  R.~N.~Mohapatra and G.~Senjanovic,
  %``Neutrino Masses and Mixings in Gauge Models with Spontaneous
  % Parity Violation,''
  Phys.\ Rev.\ D {\bf 23}, 165 (1981).
  %%CITATION = PHRVA,D23,165;%%

%\cite{Cheng:1980qt}
\bibitem{Cheng:1980qt} 
  T.~P.~Cheng and L.~F.~Li,
  % ``Neutrino Masses, Mixings and Oscillations in SU(2) x U(1) Models
  % of Electroweak Interactions,''
  Phys.\ Rev.\ D {\bf 22}, 2860 (1980).
  %%CITATION = PHRVA,D22,2860;%%
  %543 citations counted in INSPIRE as of 07 Aug 2014

%\cite{Foot:1988aq}
\bibitem{Foot:1988aq}
  R.~Foot, H.~Lew, X.~G.~He and G.~C.~Joshi,
  %``SEESAW NEUTRINO MASSES INDUCED BY A TRIPLET OF LEPTONS,''
  Z.\ Phys.\  C {\bf 44}, 441 (1989).
  %%CITATION = ZEPYA,C44,441;%%

%\cite{Mohapatra:1986bd}
\bibitem{Mohapatra:1986bd} 
  R.~N.~Mohapatra and J.~W.~F.~Valle,
  %``Neutrino Mass and Baryon Number Nonconservation in Superstring Models,''
  Phys.\ Rev.\ D {\bf 34}, 1642 (1986).
  %%CITATION = PHRVA,D34,1642;%%
%\cite{Akhmedov:1995ip}
\bibitem{Akhmedov:1995ip} 
  E.~K.~Akhmedov, M.~Lindner, E.~Schnapka and J.~W.~F.~Valle,
  %``Left-right symmetry breaking in NJL approach,''
  Phys.\ Lett.\ B {\bf 368}, 270 (1996)
  [hep-ph/9507275].
  %%CITATION = HEP-PH/9507275;%%
%\cite{Akhmedov:1995vm}
\bibitem{Akhmedov:1995vm} 
  E.~K.~Akhmedov, M.~Lindner, E.~Schnapka and J.~W.~F.~Valle,
  %``Dynamical left-right symmetry breaking,''
  Phys.\ Rev.\ D {\bf 53}, 2752 (1996)
  [hep-ph/9509255].
  %%CITATION = HEP-PH/9509255;%%

%\cite{Zee:1980ai}
\bibitem{Zee:1980ai} 
  A.~Zee,
  % ``A Theory of Lepton Number Violation, Neutrino Majorana Mass, and
  % Oscillation,''
  Phys.\ Lett.\ B {\bf 93}, 389 (1980)
  [Erratum-ibid.\ B {\bf 95}, 461 (1980)].
  %%CITATION = PHLTA,B93,389;%%
  %756 citations counted in INSPIRE as of 14 Aug 2014

%\cite{Wolfenstein:1980sy}
\bibitem{Wolfenstein:1980sy}
  L.~Wolfenstein,
  %``A Theoretical Pattern for Neutrino Oscillations,''
  Nucl.\ Phys.\ B {\bf 175}, 93 (1980).
  %%CITATION = NUPHA,B175,93;%%
  %158 citations counted in INSPIRE as of 16 Aug 2014

%\cite{Balaji:2001ex}
\bibitem{Balaji:2001ex} 
  K.~R.~S.~Balaji, W.~Grimus and T.~Schwetz,
  %``The Solar LMA neutrino oscillation solution in the Zee model,''
  Phys.\ Lett.\ B {\bf 508}, 301 (2001)
  [hep-ph/0104035].
  %%CITATION = HEP-PH/0104035;%%
  %91 citations counted in INSPIRE as of 14 Aug 2014

%\cite{He:2003ih}
\bibitem{He:2003ih} 
  X.~G.~He,
  %``Is the Zee model neutrino mass matrix ruled out?,''
  Eur.\ Phys.\ J.\ C {\bf 34}, 371 (2004)
  [hep-ph/0307172].
  %%CITATION = HEP-PH/0307172;%%
  %32 citations counted in INSPIRE as of 16 Aug 2014

%\cite{AristizabalSierra:2006ri}
\bibitem{AristizabalSierra:2006ri} 
  D.~Aristizabal Sierra and D.~Restrepo,
  %``Leptonic Charged Higgs Decays in the Zee Model,''
  JHEP {\bf 0608}, 036 (2006)
  [hep-ph/0604012].
  %%CITATION = HEP-PH/0604012;%%
  %13 citations counted in INSPIRE as of 14 Aug 2014

%\cite{Zee:1985id}
\bibitem{Zee:1985id} 
  A.~Zee,
  %``Quantum Numbers of Majorana Neutrino Masses,''
  Nucl.\ Phys.\ B {\bf 264}, 99 (1986).
  %%CITATION = NUPHA,B264,99;%%
  %220 citations counted in INSPIRE as of 07 Aug 2014

%\cite{Babu:1988ki}
\bibitem{Babu:1988ki} 
  K.~S.~Babu,
  %``Model of 'Calculable' Majorana Neutrino Masses,''
  Phys.\ Lett.\ B {\bf 203}, 132 (1988).
  %%CITATION = PHLTA,B203,132;%%
  %384 citations counted in INSPIRE as of 07 Aug 2014

%\cite{Choudhury:1994vr}
\bibitem{Choudhury:1994vr} 
  D.~Choudhury, R.~Gandhi, J.~A.~Gracey and B.~Mukhopadhyaya,
  %``Two loop neutrino masses and the solar neutrino problem,''
  Phys.\ Rev.\ D {\bf 50}, 3468 (1994)
  [hep-ph/9401329].
  %%CITATION = HEP-PH/9401329;%%


%\cite{Kitabayashi:2000nf}
\bibitem{Kitabayashi:2000nf} 
  T.~Kitabayashi and M.~Yasue,
  %``Neutrino oscillations induced by two loop radiative mechanism,''
  Phys.\ Lett.\ B {\bf 490}, 236 (2000)
  [hep-ph/0006014].
  %%CITATION = HEP-PH/0006014;%%

 


%\cite{Babu:2005ia}
\bibitem{Babu:2005ia} 
  K.~S.~Babu and C.~Macesanu,
  %``Neutrino masses and mixings in a minimal SO(10) model,''
  Phys.\ Rev.\ D {\bf 72}, 115003 (2005)
  [hep-ph/0505200].
  %%CITATION = HEP-PH/0505200;%%
  %98 citations counted in INSPIRE as of 14 Aug 2014


%\cite{AristizabalSierra:2006gb}
\bibitem{AristizabalSierra:2006gb}
  D.~Aristizabal Sierra and M.~Hirsch,
  % ``Experimental tests for the Babu-Zee two-loop model of Majorana
  % neutrino masses,''
  JHEP {\bf 0612}, 052 (2006) [hep-ph/0609307].
  %% CITATION = HEP-PH/0609307;
  %%% 54 citations counted in INSPIRE as of 14 Aug 2014
 
%\cite{Nebot:2007bc}
\bibitem{Nebot:2007bc} 
  M.~Nebot, J.~F.~Oliver, D.~Palao and A.~Santamaria,
  % ``Prospects for the Zee-Babu Model at the CERN LHC and low energy
  % experiments,''
  Phys.\ Rev.\ D {\bf 77}, 093013 (2008)
  [arXiv:0711.0483 [hep-ph]].
  %% CITATION = ARXIV:0711.0483;%%
  % 51 citations counted in INSPIRE as of 14 Aug 2014

%\cite{Schmidt:2014zoa}
\bibitem{Schmidt:2014zoa}
  D.~Schmidt, T.~Schwetz and H.~Zhang, 
  % ``Status of the Zee-Babu model for neutrino mass and possible
  % tests at a like-sign linear collider,''
  arXiv:1402.2251 [hep-ph].
  %% CITATION = ARXIV:1402.2251;%% 
  % 4 citations counted in INSPIRE as of 14 Aug 2014
  
%\cite{Herrero-Garcia:2014hfa}
\bibitem{Herrero-Garcia:2014hfa}
  J.~Herrero-Garcia, M.~Nebot, N.~Rius and A.~Santamaria,
  % ``The Zee-Babu Model revisited in the light of new data,''
  arXiv:1402.4491 [hep-ph].
  %% CITATION = ARXIV:1402.4491;%%
  % 4 citations counted in INSPIRE as of 14 Aug 2014


%\cite{Babu:1999me}
\bibitem{Babu:1999me} 
  K.~S.~Babu and S.~Nandi,
  %``Natural fermion mass hierarchy and new signals for the Higgs boson,''
  Phys.\ Rev.\ D {\bf 62}, 033002 (2000)
  [hep-ph/9907213].
  %%CITATION = HEP-PH/9907213;%%

%\cite{Bonnet:2009ej}
\bibitem{Bonnet:2009ej} 
  F.~Bonnet, D.~Hernandez, T.~Ota and W.~Winter,
  %``Neutrino masses from higher than d=5 effective operators,''
  JHEP {\bf 0910}, 076 (2009)
  [arXiv:0907.3143 [hep-ph]].
  %%CITATION = ARXIV:0907.3143;%%
  %53 citations counted in INSPIRE as of 14 Aug 2014

%\cite{Babu:2009aq}
\bibitem{Babu:2009aq} 
  K.~S.~Babu, S.~Nandi and Z.~Tavartkiladze,
  % ``New Mechanism for Neutrino Mass Generation and Triply Charged
  % Higgs Bosons at the LHC,''
  Phys.\ Rev.\ D {\bf 80}, 071702 (2009)
  [arXiv:0905.2710 [hep-ph]].
  %%CITATION = ARXIV:0905.2710;%%
  %61 citations counted in INSPIRE as of 16 Feb 2015

%\cite{Bonnet:2012kz}
\bibitem{Bonnet:2012kz}
  F.~Bonnet, M.~Hirsch, T.~Ota and W.~Winter,
  % ``Systematic study of the d=5 Weinberg operator at one-loop
  % order,''
  JHEP {\bf 1207}, 153 (2012)
  [arXiv:1204.5862 [hep-ph]].
  %%CITATION = ARXIV:1204.5862;%%
  %20 citations counted in INSPIRE as of 27 Sep 2013

%\cite{Ma:2006km}
\bibitem{Ma:2006km} 
  E.~Ma,
  % ``Verifiable radiative seesaw mechanism of neutrino mass and dark
  % matter,''
  Phys.\ Rev.\ D {\bf 73}, 077301 (2006)
  [hep-ph/0601225].
  %%CITATION = HEP-PH/0601225;%%

%\cite{Hahn:2000kx}
\bibitem{Hahn:2000kx} 
  T.~Hahn,
  %``Generating Feynman diagrams and amplitudes with FeynArts 3,''
  Comput.\ Phys.\ Commun.\  {\bf 140}, 418 (2001)
  [hep-ph/0012260].
  %%CITATION = HEP-PH/0012260;%%
  %716 citations counted in INSPIRE as of 01 Oct 2013

%\cite{Petcov:1984nz}
\bibitem{Petcov:1984nz} 
  S.~T.~Petcov and S.~T.~Toshev,
  %``Conservation of Lepton Charges, Massive Majorana and Massless Neutrinos,''
  Phys.\ Lett.\ B {\bf 143}, 175 (1984).
  %%CITATION = PHLTA,B143,175;%%
  %46 citations counted in INSPIRE as of 18 Aug 2014

%\cite{Babu:1988ig}
\bibitem{Babu:1988ig} 
  K.~S.~Babu and E.~Ma,
  % ``Natural Hierarchy of Radiatively Induced Majorana Neutrino
  % Masses,''
  Phys.\ Rev.\ Lett.\  {\bf 61}, 674 (1988).
  %%CITATION = PRLTA,61,674;%%
  %78 citations counted in INSPIRE as of 07 Aug 2014

%\cite{Ma:2012xj}
\bibitem{Ma:2012xj} 
  E.~Ma and J.~Wudka,
  %``Vector-Boson-Induced Neutrino Mass,''
  Phys.\ Lett.\ B {\bf 712}, 391 (2012)
  [arXiv:1202.3098 [hep-ph]].
  %%CITATION = ARXIV:1202.3098;%%

%\cite{Bamba:2008jq}
\bibitem{Bamba:2008jq} 
  K.~Bamba, C.~Q.~Geng and S.~H.~Ho,
  %``Radiative neutrino mass generation and dark energy,''
  JCAP {\bf 0809}, 001 (2008)
  [arXiv:0806.0952 [hep-ph]].
  %%CITATION = ARXIV:0806.0952;%%
  %4 citations counted in INSPIRE as of 15 Aug 2014

%\cite{Lindner:2011it}
\bibitem{Lindner:2011it} 
  M.~Lindner, D.~Schmidt and T.~Schwetz,
  % ``Dark Matter and Neutrino Masses from Global $U(1)_{B-L}$
  % Symmetry Breaking,''
  Phys.\ Lett.\ B {\bf 705}, 324 (2011)
  [arXiv:1105.4626 [hep-ph]].
  %%CITATION = ARXIV:1105.4626;%%

%\cite{Aoki:2010ib}
\bibitem{Aoki:2010ib} 
  M.~Aoki, S.~Kanemura, T.~Shindou and K.~Yagyu,
  % ``An R-parity conserving radiative neutrino mass model without
  % right-handed neutrinos,''
  JHEP {\bf 1007}, 084 (2010)
  [Erratum-ibid.\  {\bf 1011}, 049 (2010)]
  [arXiv:1005.5159 [hep-ph]].
  %%CITATION = ARXIV:1005.5159;%%
  %28 citations counted in INSPIRE as of 15 Aug 2014

%\cite{Kohda:2012sr}
\bibitem{Kohda:2012sr} 
  M.~Kohda, H.~Sugiyama and K.~Tsumura,
  %``Lepton number violation at the LHC with leptoquark and diquark,''
  Phys.\ Lett.\ B {\bf 718}, 1436 (2013)
  [arXiv:1210.5622 [hep-ph]].
  %%CITATION = ARXIV:1210.5622;%%

%\cite{Ma:2007gq}
\bibitem{Ma:2007gq} 
  E.~Ma,
  %``Z(3) Dark Matter and Two-Loop Neutrino Mass,''
  Phys.\ Lett.\ B {\bf 662}, 49 (2008)
  [arXiv:0708.3371 [hep-ph]].
  %%CITATION = ARXIV:0708.3371;%%

%\cite{delAguila:2011gr}
\bibitem{delAguila:2011gr} 
  F.~del Aguila, A.~Aparici, S.~Bhattacharya, A.~Santamaria and J.~Wudka,
  % ``A realistic model of neutrino masses with a large neutrinoless
  % double beta decay rate,''
  JHEP {\bf 1205}, 133 (2012)
  [arXiv:1111.6960 [hep-ph]].
  %%CITATION = ARXIV:1111.6960;%%

%\cite{Guo:2012ne}
\bibitem{Guo:2012ne} 
  G.~Guo, X.~G.~He and G.~N.~Li,
  %``Radiative Two Loop Inverse Seesaw and Dark Matter,''
  JHEP {\bf 1210}, 044 (2012)
  [arXiv:1207.6308 [hep-ph]].
  %%CITATION = ARXIV:1207.6308;%%

%\cite{Li:2012mu}
\bibitem{Li:2012mu} 
  G.~N.~Li, G.~Guo, B.~Ren, Y.~J.~Zheng and X.~G.~He,
  % ``Lepton number violation and $h\to \gamma\gamma$ in a radiative
  % inverse seesaw dark matter model,''
  JHEP {\bf 1304}, 026 (2013)
  [arXiv:1212.5528].
  %%CITATION = ARXIV:1212.5528;%%

%\cite{Chang:2006aa}
\bibitem{Chang:2006aa} 
  D.~Chang and H.~N.~Long,
  % ``Interesting radiative patterns of neutrino mass in an SU(3)(C) x
  % SU(3)(L) x U(1)(X) model with right-handed neutrinos,''
  Phys.\ Rev.\ D {\bf 73}, 053006 (2006)
  [hep-ph/0603098].
  %%CITATION = HEP-PH/0603098;%%

%\cite{Aoki:2014cja}
\bibitem{Aoki:2014cja} 
  M.~Aoki and T.~Toma,
  %``Impact of semi-annihilation of $\mathbb{Z}_3$ symmetric dark matter with radiative neutrino masses,''
  JCAP {\bf 1409}, 016 (2014)
  [arXiv:1405.5870 [hep-ph]].

%\cite{Okada:2014qsa}
\bibitem{Okada:2014qsa} 
  H.~Okada, T.~Toma and K.~Yagyu,
  %``Inert Extension of the Zee-Babu Model,''
  Phys.\ Rev.\ D {\bf 90}, no. 9, 095005 (2014)
  [arXiv:1408.0961 [hep-ph]].


%\cite{Chen:2006vn}
\bibitem{Chen:2006vn} 
  C.~S.~Chen, C.~Q.~Geng and J.~N.~Ng,
  % ``Unconventional Neutrino Mass Generation, Neutrinoless Double
  % Beta Decays, and Collider Phenomenology,''
  Phys.\ Rev.\ D {\bf 75}, 053004 (2007)
  [hep-ph/0610118].
  %%CITATION = HEP-PH/0610118;%%

%\cite{Chen:2007dc}
\bibitem{Chen:2007dc} 
  C.~S.~Chen, C.~Q.~Geng, J.~N.~Ng and J.~M.~S.~Wu,
  %``Testing radiative neutrino mass generation at the LHC,''
  JHEP {\bf 0708}, 022 (2007)
  [arXiv:0706.1964 [hep-ph]].
  %%CITATION = ARXIV:0706.1964;%%

%\cite{King:2014uha}
\bibitem{King:2014uha} 
  S.~F.~King, A.~Merle and L.~Panizzi,
  % ``Effective theory of a doubly charged singlet scalar:
  % complementarity of neutrino physics and the LHC,''
  arXiv:1406.4137 [hep-ph].
  %%CITATION = ARXIV:1406.4137;%%

%\cite{Gustafsson:2014vpa}
\bibitem{Gustafsson:2014vpa} 
  M.~Gustafsson, J.~M.~No and M.~A.~Rivera,
  % ``Radiative neutrino mass generation linked to neutrino mixing and
  % $0\nu\beta\beta$-decay predictions,''
  Phys.\ Rev.\ D {\bf 90}, no. 1, 013012 (2014)
  [arXiv:1402.0515 [hep-ph]].
  %%CITATION = ARXIV:1402.0515;%%
  %4 citations counted in INSPIRE as of 16 Feb 2015

%\cite{Gustafsson:2012vj}
\bibitem{Gustafsson:2012vj} 
  M.~Gustafsson, J.~M.~No and M.~A.~Rivera,
  % ``Predictive Model for Radiatively Induced Neutrino Masses and
  % Mixings with Dark Matter,''
  Phys.\ Rev.\ Lett.\  {\bf 110}, no. 21, 211802 (2013)
  [Erratum-ibid.\  {\bf 112}, no. 25, 259902 (2014)]
  [arXiv:1212.4806 [hep-ph]].
  %%CITATION = ARXIV:1212.4806;%%
  %42 citations counted in INSPIRE as of 16 Feb 2015

%\cite{Angel:2013hla}
\bibitem{Angel:2013hla} 
  P.~W.~Angel, Y.~Cai, N.~L.~Rodd, M.~A.~Schmidt and R.~R.~Volkas,
  % ``Testable two-loop radiative neutrino mass model based on an
  % $LLQd^cQd^c$ effective operator,''
  JHEP {\bf 1310}, 118 (2013)
  [arXiv:1308.0463 [hep-ph]].
  %%CITATION = ARXIV:1308.0463;%%

%\cite{Borzumati:2002bf}
\bibitem{Borzumati:2002bf} 
  F.~Borzumati and J.~S.~Lee,
  %``Novel constraints on Delta L = 1 interactions from neutrino masses,''
  Phys.\ Rev.\ D {\bf 66}, 115012 (2002)
  [hep-ph/0207184].
  %%CITATION = HEP-PH/0207184;%%

%\cite{Dey:2008ht}
\bibitem{Dey:2008ht} 
  P.~Dey, A.~Kundu, B.~Mukhopadhyaya and S.~Nandi,
  % ``Two-loop neutrino masses with large R-parity violating
  % interactions in supersymmetry,''
  JHEP {\bf 0812}, 100 (2008)
  [arXiv:0808.1523 [hep-ph]].
  %%CITATION = ARXIV:0808.1523;%%

%\cite{Babu:2010vp}
\bibitem{Babu:2010vp} 
  K.~S.~Babu and J.~Julio,
  %``Two-Loop Neutrino Mass Generation through Leptoquarks,''
  Nucl.\ Phys.\ B {\bf 841}, 130 (2010)
  [arXiv:1006.1092 [hep-ph]].
  %%CITATION = ARXIV:1006.1092;%%

%\cite{Babu:2011vb}
\bibitem{Babu:2011vb} 
  K.~S.~Babu and J.~Julio,
  %``Radiative Neutrino Mass Generation through Vector-like Quarks,''
  Phys.\ Rev.\ D {\bf 85}, 073005 (2012)
  [arXiv:1112.5452 [hep-ph]].
  %%CITATION = ARXIV:1112.5452;%%
  %9 citations counted in INSPIRE as of 15 Aug 2014

%\cite{Kajiyama:2013rla}
\bibitem{Kajiyama:2013rla} 
  Y.~Kajiyama, H.~Okada and T.~Toma,
  % ``Multicomponent dark matter particles in a two-loop neutrino
  % model,''
  Phys.\ Rev.\ D {\bf 88}, no. 1, 015029 (2013)
  [arXiv:1303.7356].
  %%CITATION = ARXIV:1303.7356;%%

%\cite{Baek:2013fsa}
\bibitem{Baek:2013fsa} 
  S.~Baek, H.~Okada and T.~Toma,
  % ``Two loop neutrino model and dark matter particles with global
  % B-L symmetry,''
  JCAP {\bf 1406}, 027 (2014)
  [arXiv:1312.3761 [hep-ph], arXiv:1312.3761].
  %%CITATION = ARXIV:1312.3761;%%

%\cite{Ma:2007yx}
\bibitem{Ma:2007yx} 
  E.~Ma and U.~Sarkar,
  %``Revelations of the E(6)/U(1)(N) Model: Two-Loop Neutrino Mass and Dark Matter,''
  Phys.\ Lett.\ B {\bf 653}, 288 (2007)
  [arXiv:0705.0074 [hep-ph]].
  %%CITATION = ARXIV:0705.0074;%%

%\cite{Aoki:2013gzs}
\bibitem{Aoki:2013gzs} 
  M.~Aoki, J.~Kubo and H.~Takano,
  % ``Two-loop radiative seesaw mechanism with multicomponent dark
  % matter explaining the possible γ excess in the Higgs boson decay
  % and at the Fermi LAT,''
  Phys.\ Rev.\ D {\bf 87}, no. 11, 116001 (2013)
  [arXiv:1302.3936 [hep-ph]].
  %%CITATION = ARXIV:1302.3936;%%

%\cite{Grimus:2000kv}
\bibitem{Grimus:2000kv} 
  W.~Grimus and L.~Lavoura,
  % ``A Neutrino mass matrix with seesaw mechanism and two loop mass
  % splitting,''
  Phys.\ Rev.\ D {\bf 62}, 093012 (2000)
  [hep-ph/0007011].
  %%CITATION = HEP-PH/0007011;%

%\cite{Davidson:2006tg}
\bibitem{Davidson:2006tg} 
  S.~Davidson, G.~Isidori and A.~Strumia,
  %``The Smallest neutrino mass,''
  Phys.\ Lett.\ B {\bf 646}, 100 (2007)
  [hep-ph/0611389].
  %%CITATION = HEP-PH/0611389;%%

%\cite{Joshipura:1999xe}
\bibitem{Joshipura:1999xe} 
  A.~S.~Joshipura and S.~D.~Rindani,
  %``Neutrino anomalies in an extended Zee model,''
  Phys.\ Lett.\ B {\bf 464}, 239 (1999)
  [hep-ph/9907390].
  %%CITATION = HEP-PH/9907390;%%
  %67 citations counted in INSPIRE as of 15 Aug 2014

%\cite{Chang:1999hga}
\bibitem{Chang:1999hga} 
  D.~Chang and A.~Zee,
  %``Radiatively induced neutrino Majorana masses and oscillation,''
  Phys.\ Rev.\ D {\bf 61}, 071303 (2000)
  [hep-ph/9912380].
  %%CITATION = HEP-PH/9912380;%%

%\cite{Kitabayashi:2000za}
\bibitem{Kitabayashi:2000za} 
  T.~Kitabayashi,
  % ``Two loop radiative corrections to neutrino masses in SU(3)(L) x
  % U(1)(N) gauge models,''
  hep-ph/0010341.
  %%CITATION = HEP-PH/0010341;%%

%\cite{Kitabayashi:2001jp}
\bibitem{Kitabayashi:2001jp} 
  T.~Kitabayashi,
  % ``Comment on neutrino masses and oscillations in an SU(3)(L) X
  % times U(1)(N) model with radiative mechanism,''
  Phys.\ Rev.\ D {\bf 64}, 057301 (2001)
  [hep-ph/0103195].
  %%CITATION = HEP-PH/0103195;%%
  %25 citations counted in INSPIRE as of 15 Aug 2014

%\cite{Kitabayashi:2001id}
\bibitem{Kitabayashi:2001id} 
  T.~Kitabayashi and M.~Yasue,
  % ``Large mixing angle MSW solution in an SU(3)-L x U(1)-N gauge
  % model with two loop radiative mechanism,''
  Phys.\ Lett.\ B {\bf 508}, 85 (2001)
  [hep-ph/0102228].
  %%CITATION = HEP-PH/0102228;%%
  %39 citations counted in INSPIRE as of 15 Aug 2014

%\cite{Kitabayashi:2000nq}
\bibitem{Kitabayashi:2000nq} 
  T.~Kitabayashi and M.~Yasue,
  % ``Radiatively induced neutrino masses and oscillations in an
  % SU(3)(L) x U(1)(N) gauge model,''
  Phys.\ Rev.\ D {\bf 63}, 095002 (2001)
  [hep-ph/0010087].
  %%CITATION = HEP-PH/0010087;%%

%\cite{Kajiyama:2013zla}
\bibitem{Kajiyama:2013zla} 
  Y.~Kajiyama, H.~Okada and K.~Yagyu,
  %``Two Loop Radiative Seesaw Model with Inert Triplet Scalar Field,''
  Nucl.\ Phys.\ B {\bf 874}, 198 (2013)
  [arXiv:1303.3463 [hep-ph]].
  %%CITATION = ARXIV:1303.3463;%%

%\cite{Okada:2014vla}
\bibitem{Okada:2014vla} 
  H.~Okada,
  % ``Two loop Induced Dirac Neutrino Model and Dark Matters with
  % Global $U(1)'$ Symmetry,''
  arXiv:1404.0280 [hep-ph].
  %%CITATION = ARXIV:1404.0280;%%

%\cite{Farzan:2012ev}
\bibitem{Farzan:2012ev} 
  Y.~Farzan, S.~Pascoli and M.~A.~Schmidt,
  %``Recipes and Ingredients for Neutrino Mass at Loop Level,''
  JHEP {\bf 1303}, 107 (2013)
  [arXiv:1208.2732 [hep-ph]].
  %%CITATION = ARXIV:1208.2732;%%


%\cite{Babu:2001ex}
\bibitem{Babu:2001ex} 
  K.~S.~Babu and C.~N.~Leung,
  %``Classification of effective neutrino mass operators,''
  Nucl.\ Phys.\ B {\bf 619}, 667 (2001)
  [hep-ph/0106054].
  %%CITATION = HEP-PH/0106054;%%
  %74 citations counted in INSPIRE as of 14 Aug 2014

%\cite{Choi:2002bb}
\bibitem{Choi:2002bb} 
  K.~w.~Choi, K.~S.~Jeong and W.~Y.~Song,
  %``Operator analysis of neutrinoless double beta decay,''
  Phys.\ Rev.\ D {\bf 66}, 093007 (2002)
  [hep-ph/0207180].
  %%CITATION = HEP-PH/0207180;%%
  %15 citations counted in INSPIRE as of 15 Aug 2014

%\cite{degouvea:2007xp}
\bibitem{degouvea:2007xp} 
  A.~de Gouvea and J.~Jenkins,
  % ``A Survey of Lepton Number Violation Via Effective Operators,''
  Phys.\ Rev.\ D {\bf 77}, 013008 (2008)
  [arXiv:0708.1344 [hep-ph]].
  %%CITATION = ARXIV:0708.1344;%%
  %65 citations counted in INSPIRE as of 14 Aug 2014

%\cite{Angel:2012ug}
\bibitem{Angel:2012ug} 
  P.~W.~Angel, N.~L.~Rodd and R.~R.~Volkas,
  % ``Origin of neutrino masses at the LHC: $\Delta L = 2$ effective
  % operators and their ultraviolet completions,''
  Phys.\ Rev.\ D {\bf 87}, no. 7, 073007 (2013)
  [arXiv:1212.6111 [hep-ph]].
  %%CITATION = ARXIV:1212.6111;%%
  %18 citations counted in INSPIRE as of 14 Aug 2014


%\cite{delAguila:2012nu}
\bibitem{delAguila:2012nu} 
  F.~del Aguila, A.~Aparici, S.~Bhattacharya, A.~Santamaria and J.~Wudka,
  %``Effective Lagrangian approach to neutrinoless double beta decay and neutrino masses,''
  JHEP {\bf 1206}, 146 (2012)
  [arXiv:1204.5986 [hep-ph]].
  %%CITATION = ARXIV:1204.5986;%%
  %15 citations counted in INSPIRE as of 16 Feb 2015


% The following two are also never cited??

%\cite{Okada:2014qsa}
%\bibitem{Okada:2014qsa} 
%  H.~Okada, T.~Toma and K.~Yagyu,
%  %``Inert Extension of the Zee-Babu Model,''
%  arXiv:1408.0961 [hep-ph].
%  %%CITATION = ARXIV:1408.0961;%%
%\cite{Aparici:2011nu}
%\bibitem{Aparici:2011nu} 
%  A.~Aparici, J.~Herrero-Garcia, N.~Rius and A.~Santamaria,
%  %``Neutrino masses from new generations,''
%  JHEP {\bf 1107}, 122 (2011)
%  [arXiv:1104.4068 [hep-ph]].
%  %%CITATION = ARXIV:1104.4068;%%
%  %10 citations counted in INSPIRE as of 18 Aug 2014


%\cite{Casas:2001sr}
\bibitem{Casas:2001sr} 
  J.~A.~Casas and A.~Ibarra,
  %``Oscillating neutrinos and muon ---> e, gamma,''
  Nucl.\ Phys.\ B {\bf 618}, 171 (2001)
  [hep-ph/0103065].


%\cite{Liao:2009nq}
%\bibitem{Liao:2009nq} 
%  Y.~Liao, J.~Y.~Liu and G.~Z.~Ning,
%  %``Radiative Neutrino Mass in Type III Seesaw Model,''
%  Phys.\ Rev.\ D {\bf 79}, 073003 (2009)
%  [arXiv:0902.1434 [hep-ph]].
%  %%CITATION = ARXIV:0902.1434;%%
%  %5 citations counted in INSPIRE as of 18 Aug 2014
% %\cite{Boucenna:2014zba}
% \bibitem{Boucenna:2014zba}
%   S.~M.~Boucenna, S.~Morisi and J.~W.~F.~Valle,
%   %``The low-scale approach to neutrino masses,''
%   arXiv:1404.3751 [hep-ph].
%   %%CITATION = ARXIV:1404.3751;%%
%   %4 citations counted in INSPIRE as of 18 Aug 2014

% %\cite{Ma:2007pi}
% \bibitem{Ma:2007pi} 
%   E.~Ma,
%   %``Multiplicative conservation of baryon number and baryogenesis,''
%   Phys.\ Lett.\ B {\bf 661}, 273 (2008)
%   [arXiv:0710.1102 [hep-ph]].
%   %%CITATION = ARXIV:0710.1102;%%
%   %6 citations counted in INSPIRE as of 18 Aug 2014


%\cite{AristizabalSierra:2007nf}
\bibitem{AristizabalSierra:2007nf} 
  D.~Aristizabal Sierra, M.~Hirsch and S.~G.~Kovalenko,
  %``Leptoquarks: Neutrino masses and accelerator phenomenology,''
  Phys.\ Rev.\ D {\bf 77}, 055011 (2008)
  [arXiv:0710.5699 [hep-ph]].
  %%CITATION = ARXIV:0710.5699;%%
  %26 citations counted in INSPIRE as of 13 Nov 2014

%\cite{Sierra:2008wj}
\bibitem{Sierra:2008wj} 
  D.~Aristizabal Sierra, J.~Kubo, D.~Restrepo, D.~Suematsu and O.~Zapata,
  % ``Radiative seesaw: Warm dark matter, collider and lepton flavour
  % violating signals,''
  Phys.\ Rev.\ D {\bf 79}, 013011 (2009)
  [arXiv:0808.3340 [hep-ph]].
  %%CITATION = ARXIV:0808.3340;%%
  %50 citations counted in INSPIRE as of 13 Nov 2014

%\cite{FileviezPerez:2009ud}
\bibitem{FileviezPerez:2009ud} 
  P.~Fileviez Perez and M.~B.~Wise,
  %``On the Origin of Neutrino Masses,''
  Phys.\ Rev.\ D {\bf 80}, 053006 (2009)
  [arXiv:0906.2950 [hep-ph]].
  %%CITATION = ARXIV:0906.2950;%%
  %42 citations counted in INSPIRE as of 13 Nov 2014

%\cite{Cai:2014kra}
\bibitem{Cai:2014kra} 
  Y.~Cai, J.~D.~Clarke, M.~A.~Schmidt and R.~R.~Volkas,
  %``Testing Radiative Neutrino Mass Models at the LHC,''
  arXiv:1410.0689 [hep-ph].
  %%CITATION = ARXIV:1410.0689;%%

%\cite{Restrepo:2013aga}
\bibitem{Restrepo:2013aga} 
  D.~Restrepo, O.~Zapata and C.~E.~Yaguna,
  % ``Models with radiative neutrino masses and viable dark matter
  % candidates,''
  JHEP {\bf 1311}, 011 (2013)
  [arXiv:1308.3655 [hep-ph]].
  %%CITATION = ARXIV:1308.3655;%%
  %10 citations counted in INSPIRE as of 10 Sep 2014


%\cite{McDonald:2003zj}
\bibitem{McDonald:2003zj} 
  K.~L.~McDonald and B.~H.~J.~McKellar,
  % ``Evaluating the two loop diagram responsible for neutrino mass in
  % Babu's model,''
  hep-ph/0309270.
  %%CITATION = HEP-PH/0309270;%%
  %30 citations counted in INSPIRE as of 09 Sep 2014

%\cite{vanderBij:1983bw}
\bibitem{vanderBij:1983bw} 
  J.~van der Bij and M.~J.~G.~Veltman,
  %``Two Loop Large Higgs Mass Correction to the rho Parameter,''
  Nucl.\ Phys.\ B {\bf 231}, 205 (1984).
  %%CITATION = NUPHA,B231,205;%%
  %342 citations counted in INSPIRE as of 09 Sep 2014

%\cite{Passarino:1978jh}
\bibitem{Passarino:1978jh} 
  G.~Passarino and M.~J.~G.~Veltman,
  % ``One Loop Corrections for e+ e- Annihilation Into mu+ mu- in the
  % Weinberg Model,''
  Nucl.\ Phys.\ B {\bf 160}, 151 (1979).
  %%CITATION = NUPHA,B160,151;%%
  %1790 citations counted in INSPIRE as of 12 Nov 2014
\end{thebibliography}
\end{document}